\documentclass[reprint,aps,prl,amsmath,amssymb,superscriptaddress,floatfix]{revtex4-2}

\usepackage{graphicx}
\usepackage{hyperref}
\usepackage{dcolumn}
\usepackage{bm}
\usepackage{tabularx}
\usepackage[english]{babel}
\usepackage{dcolumn}

\newcommand{\mytilde}{\raise.17ex\hbox{$\scriptstyle\mathtt{\sim}$}}

\newcommand{\ket}[1]{|#1\rangle}
\usepackage{xcolor}
\usepackage{multirow}
\usepackage{longtable}
\usepackage{cleveref}

\makeatletter
\def\nobreakhline{%
  \noalign{\ifnum0=`}\fi
    \penalty\@M
    \futurelet\@let@token\LT@@nobreakhline}
\def\LT@@nobreakhline{%
  \ifx\@let@token\hline
    \global\let\@gtempa\@gobble
    \gdef\LT@sep{\penalty\@M\vskip\doublerulesep}
  \else
    \global\let\@gtempa\@empty
    \gdef\LT@sep{\penalty\@M\vskip-\arrayrulewidth}
  \fi
  \ifnum0=`{\fi}%
  \multispan\LT@cols
     \unskip\leaders\hrule\@height\arrayrulewidth\hfill\cr
  \noalign{\LT@sep}%
  \multispan\LT@cols
     \unskip\leaders\hrule\@height\arrayrulewidth\hfill\cr
  \noalign{\penalty\@M}%
  \@gtempa}
\makeatother

\hypersetup{colorlinks=true,linkcolor=blue,citecolor=blue,urlcolor=blue}

\newcommand{\USASK}{Department of Physics \& Engineering Physics, University of Saskatchewan, Saskatoon, Canada S7N 5E2}
\newcommand{\UBC}{Stewart Blusson Quantum Matter Institute, University of British Columbia, Vancouver, Canada V6T 1Z1}

\begin{document}

\title{Negative Charge Transfer Energy in Correlated Compounds}

\author{Robert J. Green}
\email{robert.green@usask.ca}
\affiliation{\USASK}
\affiliation{\UBC}

\author{George A. Sawatzky}
\affiliation{\UBC}

\date{\today}

\begin{abstract}
In correlated compounds containing cations in high formal oxidation states (assigned by assuming that anions attain full valence shells), the energy of ligand to cation charge transfer can become small or even negative. This yields compounds with a high degree of covalence and can lead to a \emph{self-doping} of holes into the ligand states of the valence band. Such compounds are of particular topical interest, as highly studied perovskite oxides containing trivalent nickel or tetravalent iron are negative charge transfer systems, as are nickel-containing lithium ion battery cathode materials. In this report, we review the topic of negative charge transfer energy, with an emphasis on plots and diagrams as analysis tools, in the spirit of the celebrated Tanabe-Sugano diagrams which are the focus of this Special Topics Issue. 
\end{abstract}

\maketitle

\section{Introduction}

Seven decades ago, Tanabe and Sugano introduced their diagrams as a transformative concept in materials science \cite{1954_TSD, 1954_TSD2} . Tanabe-Sugano diagrams provide an excellent visualization and analysis tool for the multi-electron states in correlated compounds. Particularly illuminating are phenomena such as spin state transitions, which appear in the diagrams as level crossings resulting from the competition between crystal and ligand fields as well as Coulomb and exchange interactions.

While Tanabe-Sugano diagrams are incredibly useful in relatively ionic compounds, when covalence becomes strong it may become important to consider hybridization effects directly, rather than the typical indirect approach via effective ligand field splittings and reductions in atomic Coulomb and exchange interactions. Strong covalence can originate from large hopping integrals (i.e. valence orbital overlap of cations and ligands), but can also originate from small or negative charge transfer energies and can subsequently result in a breakdown of perturbative approaches. Here, charge transfer energy is defined as the energy required to transfer a ligand valence electron (e.g. oxygen $2p$ electron) to a cation valence shell (e.g. transition metal $d$ shell). In this work, we explore the field of negative charge transfer energy compounds in a similar spirit to Tanabe-Sugano diagrams, where plots of particular parameter series can provide crucial insight into the energetics and correlated states of such materials.

\section{Charge Transfer Energetics}

A suitable starting point to introduce charge transfer energetics is the single impurity Anderson model (SIAM) \cite{Anderson1961SIAM}, consisting of a single correlated impurity coupled to a non-interacting bath. While early studies of charge transfer energetics often used cluster models \cite{1984PRBFujimoriNi, 1984PRLSawatzkyAllenNiO}, by including the ligand bandwidth one can more clearly distinguish between different charge transfer classes. For the SIAM, assuming an impurity with a valence $d$ shell, we have a Hamiltonian
\begin{align}
\label{eq:SIAM}
H = H_I + H_B + H_V
\end{align}
which consists of terms for the impurity $H_I$, the bath $H_B$, and their hybridization interaction $H_V$. The impurity term includes the onsite energies of the $d$ shell and the local Coulomb interaction, 
\begin{align}
\label{eq:SIAMImp}
    H_I = \sum_{\tau} \epsilon_{d,\tau} \bm{d}^{\dag}_{\tau} \bm{d}^{\phantom{\dag}}_{\tau} +  \sum_{\tau_1,\tau_2,\tau_3,\tau_4} U_{\tau_1,\tau_2,\tau_3,\tau_4} \bm{d}^{\dag}_{\tau_1} \bm{d}^{\dag}_{\tau_2} \bm{d}^{\phantom{\dag}}_{\tau_3} \bm{d}^{\phantom{\dag}}_{\tau_4},
\end{align}
with $\tau$ labeling the 10 different $3d$ spin-orbitals and $\bm{d}^{\dag}_{\tau}$ ($\bm{d}^{\phantom{\dag}}_{\tau}$) the operator creating (annihilating) an electron in orbital $\tau$. The first term gives the onsite energy and the second gives the Coulomb interaction. Depending on the local point group symmetry, energy $\epsilon_{d,\tau}$ may include crystal field splitting, which for example in $O_h$ point group symmetry would give different onsite energies for the $e_g$ orbitals ($3z^2-r^2$, and $x^2-y^2$) than for the $t_{2g}$ orbitals ($yz$, $xz$, and $xy$). The Coulomb interaction can contain the typical monopole term (Racah $A$ parameter or Slater $F^0$) as well as higher order multipoles (Racah $B$ and $C$ parameters or Slater $F^2$ and $F^4$) which lead to atomic multiplet splitting.  

\begin{figure*}
\centering
\includegraphics[width=7in]{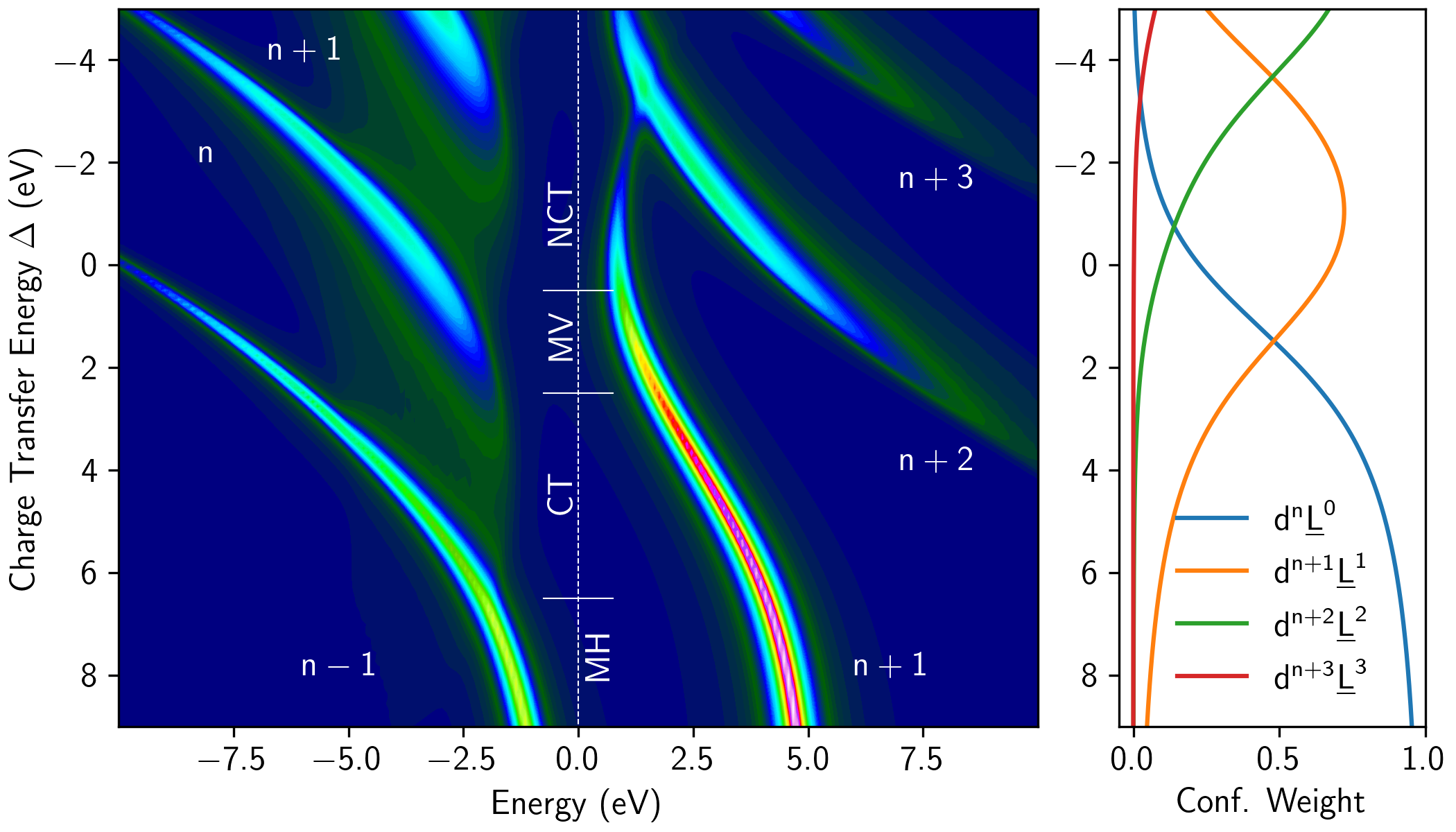}
\caption{(Color online) (Left) Electron removal and addition spectra for the Anderson impurity model at different charge transfer energies.  Addition and removal features are labelled according to the corresponding impurity electron counts, assuming a formal impurity occupation of $d^n$. Four regions are identified for specific charge transfer energy ranges: Mott-Hubbard (MH), positive charge transfer (CT), mixed-valence (MV), and negative charge transfer (NCT). (Right) Configuration weights in the ground state wavefunction over the same range of charge transfer energies.}
\label{fig:PESIPESMap}
\end{figure*}

The bath term of the Hamiltonian can be defined in various ways. In a conventional Anderson geometry \cite{Anderson1961SIAM}, the impurity couples to all bath sites independently, and there is no coupling between bath sites. Alternatively, an equivalent model can be constructed in what might be called a Wilson geometry \cite{Wilson1975}, where the impurity couples to one bath site, and all bath sites are coupled in a chain extending from this first site \cite{Haverkort_DMFT_2014}. In this geometry, the first bath site can correspond to nearest neighbours, and subsequent bath sites represent linear combinations of the shells from successively further neighbouring atoms. In this Wilson geometry we have for the bath 
\begin{equation}
\begin{aligned}
\label{eq:SIAMBath}
H_B = &\sum_{k=1}^{N_L} \sum_{\tau} \epsilon_{L,k,\tau} \cdot \left( \mathbf{L}^\dagger_{k,\tau} \mathbf{L}_{k,\tau} \right) \\ 
&+ \frac{W}{4}\sum_{k=1}^{N_L-1} \sum_{\tau} \left( \mathbf{L}^\dagger_{k,\tau} \mathbf{L}_{k+1,\tau} + \mathbf{L}^\dagger_{k+1,\tau} \mathbf{L}_{k,\tau} \right)
\end{aligned}
\end{equation}
which enumerates, over $k$, a set of $N_L$ ligand shells with creation (annihilation) operator $\mathbf{L^\dagger}_{k,\tau}$ ($\mathbf{L}_{k,\tau}$) at energy $\epsilon_{L,k,\tau}$ with a combined spin and orbital index $\tau$. The second term gives the hybridization of the chain of ligand atoms, yielding a band of finite width $W$. Note that for this bath we neglect all Coulomb and exchange interactions involving the bath atoms, assuming that they can be approximated by a one electron band structure which includes these interactions in a mean field average and exchange correlation fashion. This bath is coupled to the impurity via
\begin{align}
\label{eq:SIAMHyb}
H_V = \sum_{\tau} V_{\tau}\left(\mathbf{d}^\dagger_{\tau}\mathbf{L}_{1,\tau}+\mathbf{L}^\dagger_{1,\tau}\mathbf{d}_{\tau}\right)
\end{align}
where $V_{\tau}$ is the hopping integral from spin orbital $\tau$ of the impurity to first bath site (i.e. the nearest neighbours).

Note that this SIAM Hamiltonian easily reduces to the popular local cluster model when we truncate the bath at the first site.  Such single cluster models, typically referred to as multiplet ligand field theory (MLFT) \cite{2012PRBHaverkortMLFT} or charge transfer multiplet (CTM) models \cite{2021_JELSPEC_deGroot}, have been very successful in studying the electronic structure and core level spectroscopy of correlated compounds. Early applications of the cluster model for $3d$ oxides included studies of NiO \cite{1984PRBFujimoriNi, 1984PRLSawatzkyAllenNiO}  and CuO \cite{1990PRBEskesCuO}.

The onsite energies of the impurity and ligands can be defined in terms of charge transfer energy $\Delta$ and Coulomb interaction energy $U$, which can be more readily related to experiments (in particular core level spectroscopy). For an impurity with a formal number of $n$ electrons (e.g. $n=8$ for a Ni$^{2+}$ compound), we can define a reference energy for the $d^n$ configuration,
\begin{align}
n\epsilon_d + 10N_L\epsilon_L + n\left(n-1\right)U/2 = 0
\end{align}
and assign charge transfer energy $\Delta$ to the excited configuration where a ligand electron has transferred to the $d$ shell,
\begin{align}
\left(n+1\right)\epsilon_d + \left(10N_L-1\right)\epsilon_L + n\left(n+1\right)U/2 = \Delta
\end{align}
We can solve these equations for $\epsilon_d$ and $\epsilon_L$ to recast our Hamiltonian into these more common parameters of $U$ and $\Delta$. The Hamiltonian is then typically solved using a configuration interaction approach, expanding the wavefunction by configurations of the form $\ket{d^{n+m}\underline{L}^m}$, where $\underline{L}^m$ denotes a configuration with $m$ ligand holes. Using this approach Zaanen, Sawatzky, and Allen developed a classification of Mott-Hubbard (MH) and charge transfer (CT) insulators \cite{Zaanen_ZSA_PRL_1985}.  While in the former, the lowest energy charge fluctuations are governed by the Coulomb interaction energy $U$ and are of the form
\begin{align}
2d^n \rightarrow d^{n-1} + d^{n+1}
\end{align}
in the latter the fluctuations are governed by the charge transfer energy $\Delta$ and are of the type
\begin{align}
d^n \rightarrow d^{n+1}\underline{L}
\end{align}

This distinction between different types of compounds can be clearly demonstrated via plots of the impurity one electron removal and addition spectra.  In Fig. \ref{fig:PESIPESMap} we plot on the left these electron addition and removal spectra for the SIAM as a function of charge transfer energy. The SIAM was implemented and solved using the software QUANTY \cite{QuantyWeb,Haverkort_DMFT_2014, Haverkort_DMFTXAS_EPL2014}.  For large charge transfer energies, one clearly sees the upper and lower Mott-Hubbard bands (though not strictly \emph{bands} in the impurity model) which define the conductivity gap. This gap is roughly equivalent to $U$, and is verification of the ideas of Mott \cite{1937PRCMottPeierls,1949PPSAMott}, originally provided to explain the unexpected insulating nature of some transition metal compounds observed by de Boer and Verwey \cite{1937PRCDeBoerVerwey}. These ideas of Mott were later formalized by Hubbard \cite{1964PRSAHubbard1,1964PRSAHubbard2}.  In the diagram of Fig. \ref{fig:PESIPESMap}, we label the features according to the number of electrons -- in this Mott-Hubbard regime with large charge transfer energies, the ground state is primarily $\ket{d^n}$, so the removal and addition peaks have $n-1$ and $n+1$ electrons, respectively. The ground state character is shown in the right plot of Fig. \ref{fig:PESIPESMap}, where the wavefunction is decomposed into different configurations and their weight is plotted for the same values of $\Delta$.

For smaller charge transfer energies, we enter the charge transfer (CT) insulator regime, where now the lowest energy electron removal states (shown by the broad, weak, band-like feature at small removal energies near $\Delta=4~\mathrm{eV}$) are of $d^n\underline{L}$ character.  In the right panel for this charge transfer energy, one sees the configurations of the type $d^{n+1}\underline{L}^1$ gain more weight in the ground state.  

Continuing to smaller charge transfer energies, we enter the mixed-valence (M-V) regime, where the ground state is compose of near-equal weights of $d^n$ and $d^{n+1}\underline{L}$. Correspondingly one has two strong electron removal configurations ($d^{n-1}$ and $d^n\underline{L}$) and two strong electron addition configurations ($d^{n+1}$ and $d^{n+2}\underline{L}$). A clear example of such a compound is SmB$_6$, where the electron removal features are observed in photoemission spectra \cite{2016SawatzkyGreenJulich}. In this case, the weak hybridization interaction facilitates a clear distinction between the two valences present, and in particular there is a clear separation of valences in momentum space.  While some transition metal oxides have charge transfer energies that would place them in the mixed valence regime, their hybridization strengths are stronger, leading to a much more covalent wavefunction and not as clear separation of valences.

Finally, moving to even smaller charge transfer energies, we enter the negative charge transfer regime, where the ground state is dominated by the $d^{n+1}\underline{L}$ configuration (Fig. \ref{fig:PESIPESMap}, right panel).  Here, the removal spectra are primarily of nature $d^n\underline{L}$ and addition spectra of type $d^{n+1}$ and $d^{n+2}\underline{L}$. The strong ligand hole component in the ground state indicates a \emph{self-doping} \cite{1997LJPKhomskii} of the ligand band.  In examples such as the perovskite rare earth nickelates (e.g. NdNiO$_3$) and alkaline earth ferrates (e.g. SrFeO$_3$), this self-doping corresponds abundent oxygen holes in the ground state configuration, which have been verified by various spectroscopies \cite{2016NatComBisogniNNO, 2016PRBGreenDoubleCluster}.

\begin{figure}
\centering
\includegraphics[width=3.375in]{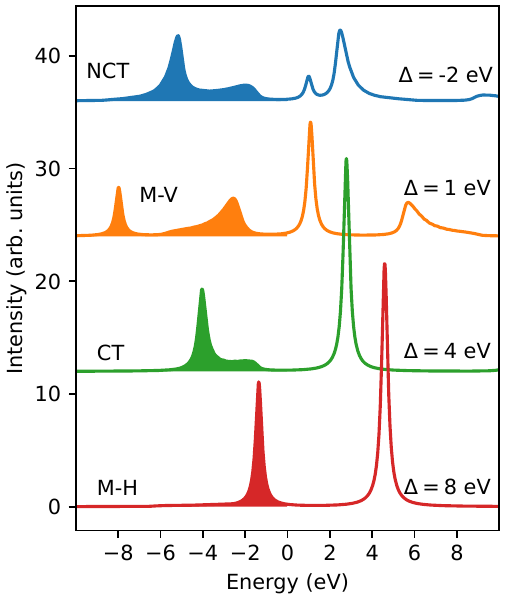}
\caption{(Color online) Selected electron removal (filled) and addition (unfilled) spectra extracted from Fig. \ref{fig:PESIPESMap}. The specific charge transfer energies are indicated, as are the corresponding classification of compounds. A Coulomb interaction energy of $U=5~\mathrm{eV}$ was used for the computation.}
\label{fig:PESIPES}
\end{figure}

\begin{figure*}
\centering
\includegraphics[width=7in]{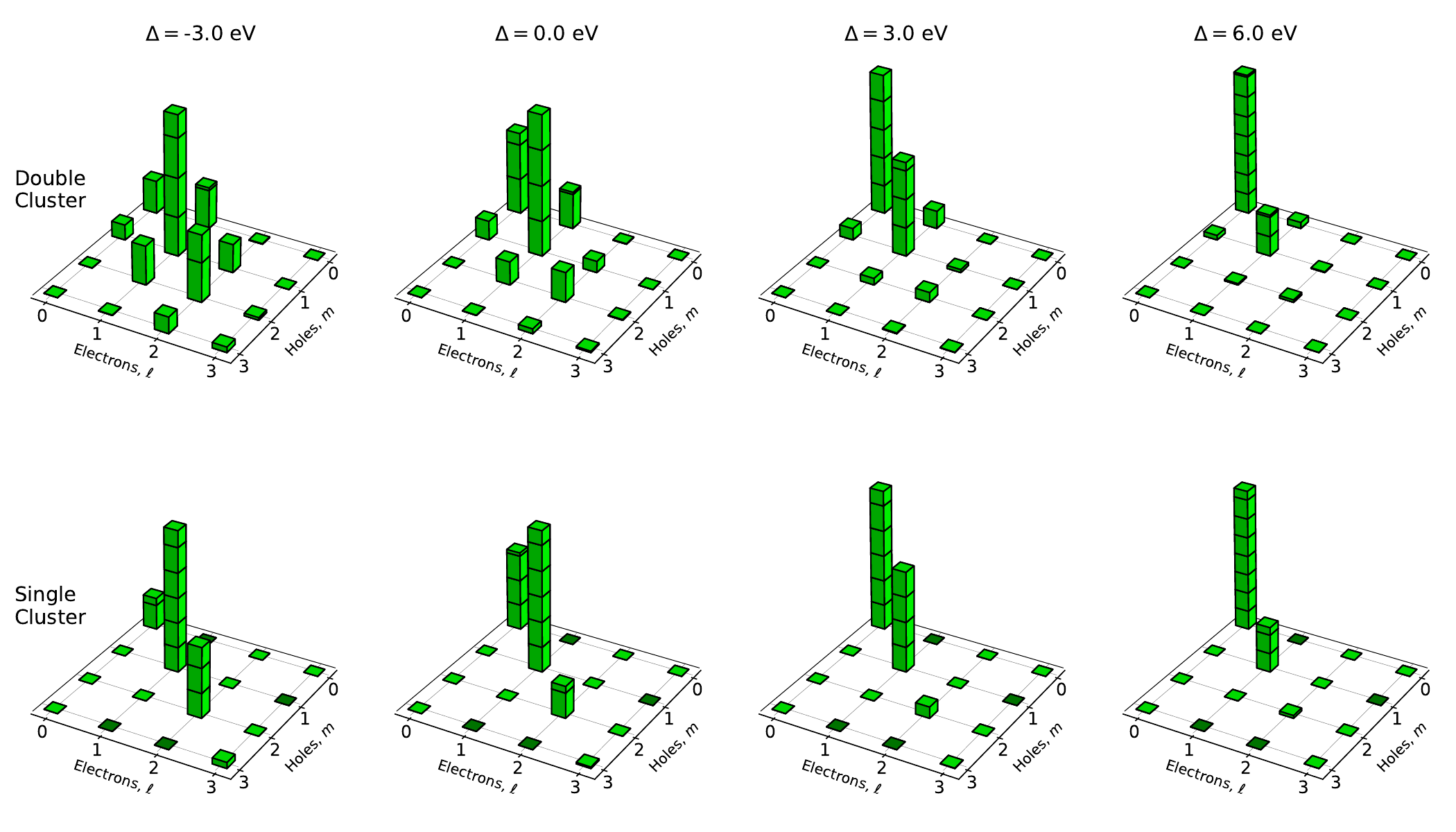}
\caption{(Color online) Comparison of the ground state wavefunctions obtained in the single cluster ligand field theory and double cluster models for various charge transfer energies. The bars show the weight of the $\ket{d^{n+\ell}\underline{L}^m}$ configurations. Scale markings on bars occur in intervals of 0.1, and the wavefunctions are normalized to a total weight of 1.}
\label{fig:CharPlot}
\end{figure*}

In Fig. \ref{fig:PESIPES}, we extract line cuts of the electron addition and removal spectra for charge transfer energies indicative of these four classes of compounds.  Here, the upper and lower Mott-Hubbard states are evident, separated by \mytilde 5 eV, which was the value of $U$ used in the SIAM calculation.  In the case of the charge transfer (CT) insulator, the ligand band is more clearly seen in the electron removal spectrum, demonstrating the the lowest energy removal states are of ligand character.  Again the four characteristic features are seen in the mixed valence spectrum, and the broad, strongly hybridized ligand states are evident in the negative charge transfer (NCT) case.

As shown above, the SIAM is a useful tool for demonstrating these four classes of compounds.  However, an impurity model cannot appropriately model self-doping and intersite cation fluctuations, as a single atom coupled to a continuum bath cannot affect the filling of the bath. To capture self-doping, and the nonlocal fluctuations introduced in such a case, one needs a more sophisticated method like dynamical mean field theory (DMFT). A computationally simpler alternative, particularly applicable to perovskite type of oxides, is the double cluster model \cite{2016PRBGreenDoubleCluster}. Here, one constructs two ligand field theory clusters which include a cation and nearest neighbour ligands, and then couples them together via symmetry-appropriate hybridization operators. The effect is a system with appropriate metal concentration, which can then lead to significant self-doping of ligands and subsequent intersite fluctuations.

\begin{figure*}[ht]
\centering
\includegraphics[width=7in]{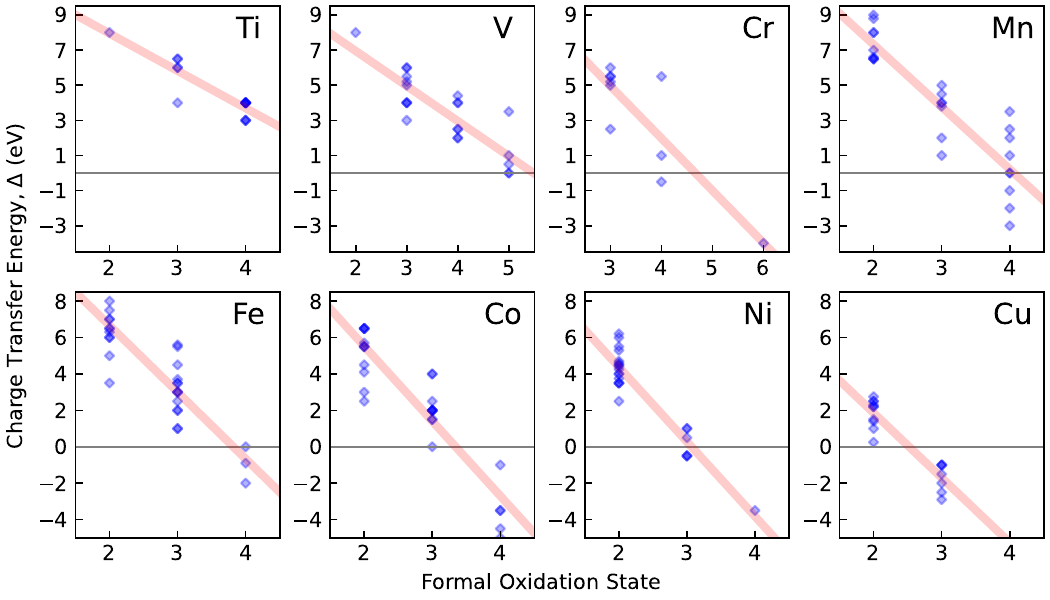}
\caption{(Color online) Charge transfer energies of $3d$ transition metal oxides compiled from published studies. The data points are plotted with transparency so that overlapping points are evident via higher opacity. Tables with references for all compounds can be found in the Appendix.}
\label{fig:CTLiterature}
\end{figure*}

In Fig. \ref{fig:CharPlot}, we compare ground state wavefunction decompositions for a single cluster (i.e. the SIAM with only nearest neighbours and thus zero ligand bandwidth) and the double cluster model, for values of charge transfer energy $\Delta$ spanning from the Mott-Hubbard to negative charge transfer regimes. Specifically, for the single cluster, we take the Hamiltonian from Eq. \ref{eq:SIAM}, but limit the bath term from Eq. \ref{eq:SIAMBath} to $k=1$ (nearest neighbours).  The result is a single cluster Hamiltonian, $H_{SC}$. For the double cluster model, we use two single cluster Hamiltonians, labelled $A$ and $B$, and introduce hybridization between them,
\begin{align}
    H_{DC} = H_{SC,A} + H_{SC,B} + H_{mix}
\end{align}
where the hybridization operator has the form
\begin{equation}
\begin{aligned}
H_{mix} = \sqrt{x}\sum_{\tau} V_{\tau}&\left(\mathbf{d_A}^\dagger_{\tau}\mathbf{L_B}_{\tau}+\mathbf{L_B}^\dagger_{\tau}\mathbf{d_A}_{\tau} \right. \\
&\left.+ \mathbf{d_B}^\dagger_{\tau}\mathbf{L_A}_{\tau}+\mathbf{L_A}^\dagger_{\tau}\mathbf{d_B}_{\tau}\right)
\end{aligned}
\end{equation}
and the original hybridization operator of Eq. \ref{eq:SIAMHyb} is correspondingly scaled by $\sqrt{1-x}$. In this formulation, the parameter $x$ scales the coupling between clusters. In perovskite oxides, a value of \mytilde 0.3 is used, which agrees well with the crystal structure symmetry \cite{2016PRBGreenDoubleCluster}. This value is used in Fig. \ref{fig:CharPlot}, where the double cluster ground states are compared to those of a single cluster. In the single cluster regime, beginning with a full ligand shell and a formal (stoichiometry defined) $d$ shell filling of $d^n$, one is restricted to configurations of the form $\ket{d^{n+m}\underline{L}^m}$ in the ground state wavefunction.  However, for the double cluster model, intersite fluctuations introduce more general configurations of the form $\ket{d^{n+\ell}\underline{L}^m}$.

For relatively large charge transfer energies in the Mott-Hubbard or positive charge transfer regimes (e.g. $\Delta$ of 3.0 or 6.0 eV in Fig. \ref{fig:CharPlot}), the two models exhibit similar wavefunctions.  The full ligand shell suppresses nonlocal charge fluctuations, and the configurations are primarily restricted to the $\ket{d^{n+m}\underline{L}^m}$ type, forming the diagonals of the plots in the figure.  However, for smaller charge transfer energies ($\Delta$ of 0.0 or -3.0 eV in Fig. \ref{fig:CharPlot}), the self-doping of significant ligand holes leads to more pronounced nonlocal charge flucutations, and the wavefunction weight spreads away from this diagonal to include significant character of the $\ket{d^{n+\ell}\underline{L}^m}$ type. It was shown that these intersite charge fluctuations are important for the perovskite nickelates \cite{2016PRBGreenDoubleCluster, 2018PRXYiDoubleCluster, 2019PRBYiDoubleCluster}.

\section{Trends in Charge Transfer Energy and Examples of Negative Charge Transfer Materials}

The charge transfer energy is determined by the cation electron affinity and the anion electronegativity.  Accordingly, compounds with small or negative charge transfer energy tend to contain cations from later in the transition element series with high formal oxidation states, and/or with heavier anions. Several previous works have investigated trends in the charge transfer energies of various compounds, which verified these general expectations \cite{1992PRBBocquetXPS, 1996PRBBocquetXPS, Fujimori1993JELSPEC_Trends}.  It is now roughly three decades since these illuminating works, so in Figure \ref{fig:CTLiterature} we compile charge transfer energies collected for $3d$ oxides from an extensive literature search up to the present day. We plot these charge transfer energies against formal oxidation state for oxides of the eight transition metals spanning from Ti to Cu. In each case we also plot in red a linear regression to the collected data to better visualize the trends in the charge transfer energy. The data and references for this plot are given in the tables of the Appendix.

\begin{figure}
\centering
\includegraphics[width=3.375in]{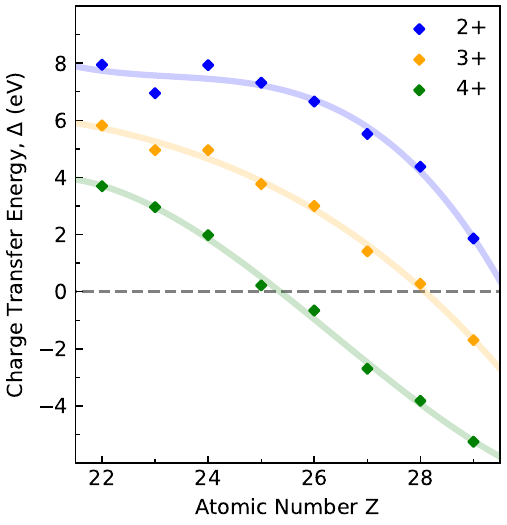}
\caption{(Color online) Trends of charge transfer energy $\Delta$ versus atomic number for several different formal oxidation states. The values plotted are extracted from the linear regressions to the data of Fig. \ref{fig:CTLiterature}. The lines are a third order polynomial fit to aid in visualizing the trends.}
\label{fig:CTTrend}
\end{figure}

While there is some scatter in the data (see the Appendix for further discussion), it is clear from Fig. \ref{fig:CTLiterature} that charge transfer energy decreases roughly linearly with formal oxidation state and generally decreases along the $3d$ series. From the linear regressions of Fig. \ref{fig:CTLiterature}, we extract an average charge transfer energy for common formal oxidation states of the $3d$ series, and plot these in Fig. \ref{fig:CTTrend}. Here a clear trend is present of decreasing charge transfer energy for higher atomic number in the series as demonstrated in previous works  \cite{1992PRBBocquetXPS, 1996PRBBocquetXPS, Fujimori1993JELSPEC_Trends}. We see that for oxides, negative charge transfer energies are generally present for tetravalent Fe-Cu ions or trivalent Ni and Cu ions. Apparent from Fig. \ref{fig:CTLiterature} is that Cr and V negative charge transfer oxides are also possible. For anions with higher electronegativities (sulfur for example) these series are expected to shift downward so that 3+ or even 2+ ions late in the series will be negative charge transfer compounds. We note that in Fig. \ref{fig:CTTrend}, anomalies for V and Cr may be due to a lack of data for these elements and that the data in Fig. \ref{fig:CTLiterature} for these elements spans more oxidation states than for other elements and therefore a linear regression may not be appropriate.

\begin{figure*}
\centering
\includegraphics[width=7in]{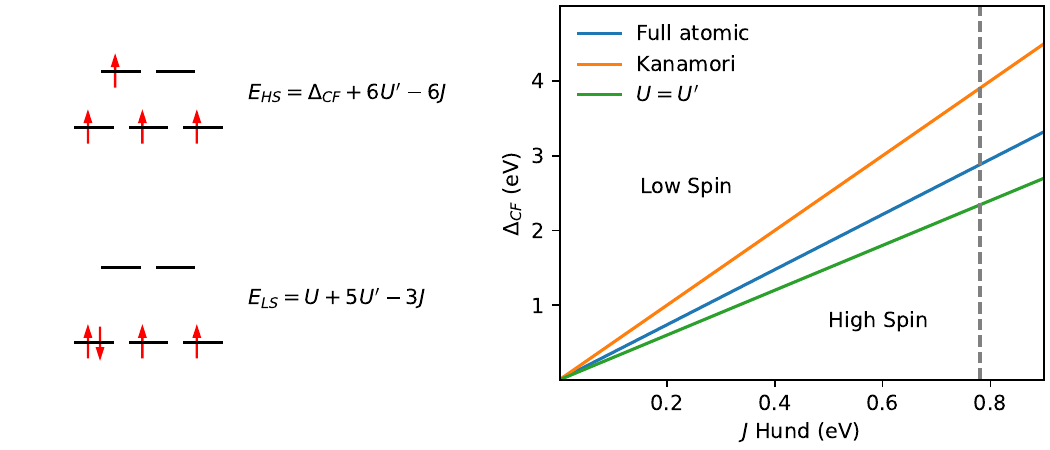}
\caption{(Color online) (Left) The energetics of high spin and low spin $3d^4$ configurations, as determined with the Kanamori approach. (Right) Comparison of the spin state crossover for the full atomic multiplet theory and the Kanamori approach, along with the simplified Kanamori approach with $U'=U$. The dashed grey line denotes the free ion value of $J$.}
\label{fig:HSLS}
\end{figure*}

The first report of a solid state compound with negative charge transfer energy was that of NaCuO$_2$, studied by Mizokawa \emph{et al} \cite{1991PRLMizokawaNaCuO2}. In this compound, the stoichiometry imposes a high 3+ oxidation state for the Cu, but a negative charge transfer energy yields a ground state that is primarily $d^9\underline{L}$. The material is insulating, but the authors determined that the gap is a $p-p$ type gap and therefore NaCuO$_2$ is neither a Mott-Hubbard or charge-transfer insulator. Follow up studies further clarified the nature of the $d^9\underline{L}$ ground state \cite{1994PRBMizokawaNaCuO2, 1991SSCSawatzkyNaCuO2, 1993PRBSarmaNaCuO2}.

More recently, the perovskite nickelates $R$NiO$_3$ (with $R$ a rare earth ion) have been extensively studied. With trivalent rare earths, formal oxidation state rules imply the presence of Ni$^{3+}$, or a $3d^7$ configuration. However, due to a negative charge transfer energy, the ground state is primarily $3d^8\underline{L}$. Note that the low spin $3d^7$ and the $3d^8\underline{L}$ are both $S=1/2$ due to the antiferromagnetic coupling of the ligand hole to the Ni. These compounds (except for LaNiO$_3$) exhibit a metal-insulator transition, with different transition temperatures for compounds having different rare-earth ion \cite{Medarde_JPCM_1997}. In the insulating phase, the NiO$_6$ octahedra expand and compress in a breathing distortion following a rocksalt lattice pattern. The compounds also transition into an antiferromagnetic ordered phase with a (1/4,1/4,1/4) ordering vector (in pseudocubic notation). 

Early theoretical studies of the perovskite nickelates predicted oxygen- or nickel-based charge ordering depending on the value of charge transfer energy \cite{Mizokawa_SDMI_PRB_2000}. Later computational studies expressed the ordering in terms of a bond disproportionation \cite{LauMillis_PRL_2013, Johnston_PRL_2014, 2016PRBGreenDoubleCluster}, where the negative charge transfer $d^8\underline{L}$ state disproportionates into $d^8\underline{L}^0$ and $d^8\underline{L}^2$. Experimental studies using core level spectroscopy verified the negative charge transfer nature of these compounds \cite{2016NatComBisogniNNO}. As nickelates can be grown epitaxially in thin film form, and the perovskite structure affords many different substrates which impose geometric and strain effects, numerous studies were carried out on nickelate heterostructures in recent years \cite{2018RPPCatalanoNickelates}. Examples include control of emergent properties via tensile and compressive strain \cite{2020PRBKimNickelates,2011NatMatBenckiser,2013NatCommChakhalian,2013PRBBenckiserNickelates,2013PRLFrano}, ultrathin films with control over the magnetic (non)-collinearity \cite{2018NatPhysHeptingNNO}, variations of the metal-insulator transition temperature for potential device applications \cite{2018PNASLiaoNickelates}, and control of the electronic structure and magnetic order via oxygen vacancy electron doping \cite{2021PRLJiaruiNickelates} or substitutional hole doping \cite{2022CommPhysMiddeyHoleDoping}.

The alkaline earth ferrates, $A$FeO$_3$, where Fe is in a high 4+ formal oxidation state, are also negative charge transfer compounds \cite{1992PRBBocquetSFO, 2002PRBFujimoriFerrates, 2002PRBAbbateFerrates, 2018PRMRoggeCaFeO3, 2018PRBRoggeCaFeO3Inv}. CaFeO$_3$ exhibits a breathing distortion and metal-insulator transition similar to the perovskite nickelates \cite{1998JPSJFujimorFerrates}. This is also found to coincide with a bond disproportionation, here into sites of the form $d^5\underline{L}^2$ and $d^5\underline{L}^0$ \cite{2002PRBFujimoriFerrates}. Also similar to the nickelates, both SrFeO$_3$ and CaFeO$_3$ exhibit antiferromagnetic ordering with an ordering vector parallel to the pseudocubic (111) direction.  However, different from the nickelates, the antiferromagnetic order is in the form of incommensurate helices \cite{1972JPSJTakedaSFO, 2000PRBWoowardCFO, 2019PRMRoggeSFO} which can form topological spin textures \cite{2020PRBIshiwataSFO}.

We have focused on oxide compounds up to this point, but the disulfide pyrites FeS$_2$, CoS$_2$, and NiS$_2$ are also negative charge transfer compounds. In these compounds, formal oxidation state assignments would suggest valences of 4- for the S$_2$ and therefore 4+ for the cations. However, it is instead found that the cations are 2+ and there are self-doped sulfur $3p$ holes \cite{1987JPCSawatzkyPyrites}. Additionally, the sulfur anions form pairs with full bonding and empty antibonding orbitals such that the S$_2$ molecule has a $2-$ configuration (similar to superoxides like KO$_2$). While this is indeed formally a negative charge transfer situation, often an effective positive charge transfer energy is defined for the divalent cations, relative to the full sulfur $3p$ valence band. The compounds have diverse and interesting properties--FeS$_2$ is a diamagnetic semiconductor, CoS$_2$ is a ferromagnetic metal, and NiS$_2$ is an antiferromagnetic Mott insulator \cite{1969JPSJAdachiPyrites, 1979JAPOgawaPyrites}. Spectroscopy studies revealed the self-doped nature of the compounds \cite{1996PRBFujimoriPyrites, 1996JPCMBocquetPyrites}, and an impurity model with an explicit self-doped conduction band was used to interpret photoemission spectra \cite{1996JPCMBocquetPyrites}.

\section{The Importance of Atomic Physics}

Atomic physics, i.e. multipole Coulomb interactions, generally remain very important for negative charge transfer compounds. In full atomic multiplet theory, the strength of the Coulomb interactions among (all pairs of) electrons are defined by either the radial Slater integrals $F^k$ (where $k=0,2,4$ for $d$ electrons) or the Racah parameters $A$, $B$, $C$.  These interactions are at times approximated--for example the Kanamori approach \cite{1963PTPKanamori} uses effective parameters $U$, $J$, and $U'$, where $U$ is the Coulomb repulsion of electrons in the same orbital, $J$ is the Hunds' rule exchange, and $U'=U-2J$ is the Coulomb interaction of electrons in different orbitals. The work of Tanabe and Sugano \cite{1954_TSD, 1954_TSD2} considered the full multiplet interactions, so this dedicated Special Issue is a good avenue to demonstrate the importance of these atomic effects. As an illustrative example, the energies of high spin and low spin configurations for a $3d^4$ ion with octahedral coordination within the Kanamori approximation are as given in Fig. \ref{fig:HSLS}. Equating these two energies gives us the condition of $\Delta_{CF}=5J$ for the spin state crossover. This line is compared to the crossover computed with full crystal field multiplet theory in the right panel of Figure \ref{fig:HSLS}, where we find the condition $\Delta_{CF}\approx 3.69J$. Also shown in the Figure is the simplification $U$=$U'$ in the Kanamori approach, where the crossover occurs at $\Delta_{CF}=3J$. Therefore we find the full multiplet result is between the two approximations. While the approximations are of course still very useful tools, care must be taken for compounds which are in close proximity to spin state transitions, as the full atomic approach becomes necessary.

\section{Conclusion}

We have reviewed how correlated compounds can be classified into four regimes, based on the relative values of charge transfer energy ($\Delta$) and Coulomb interaction $U$.  In addition to the more common Mott-Hubbard ($U<\Delta$) and charge transfer ($U>\Delta$) insulators, for small values of $\Delta$ one enters the mixed valence and then negative charge transfer regimes. The self-doping of the ligand band for these cases leads to new phenomena such as bond disproportionation, where ligand holes bond asymmetrically between sites. Clearly the anion states become very important in these compounds, and this can mean that non-conventional starting points might be necessary when constructing models to understand the electronic and magnetic structures of these compounds.

\section*{Acknowledgements}

We thank Daniel Khomskii for helpful discussions on the importance of atomic physics. This work was supported by the Natural Sciences and Engineering Research Council of Canada (NSERC). This research was undertaken thanks, in part, to funding from the Canada First Research Excellence Fund, Quantum Materials and Future Technologies Program.  We gratefully acknowledge the facilities of and assistance with the Plato computing cluster at the University of Saskatchewan.

\appendix
\section{Appendix: Oxide Charge Transfer Energy Data}

\renewcommand{\thetable}{A-\Roman{table}}  

This Appendix contains tables of charge transfer energies collected from a literature search of $3d$ transition metal oxides. Each table reports values for compounds containing a specific $3d$ cation, ranging from Ti to Cu in Tables \labelcref{Tab:Ti,Tab:V,Tab:Cr,Tab:Mn,Tab:Fe,Tab:Co,Tab:Ni,Tab:Cu}. The data from these tables was used to construct the plots in Fig. \ref{fig:CTLiterature}, which show clear trends in the charge transfer energies with different formal valences. It should be noted that generally relatively large uncertainties exist when extracting charge transfer energies using the typical core level spectroscopy approaches like x-ray photoelectron spectroscopy (XPS) and x-ray absorption spectroscopy (XAS).  Thus, there is some scatter in the data contained in these tables, even for different experiments on the same compounds. Further, different techniques can probe effectively different charge transfer energies--the value of $\Delta$ obtained by optimizing cluster model parameters against $2p$ XPS data is fundamentally different from that which would be extracted from oxygen x-ray emission and absorption spectra. Both values are valid, but relate to different defitions of $\Delta$. In compiling the data in the following tables, we have primarily focused on cluster-derived values so that they may be directly compared and thus the trends are meaningful. Of the various approaches to extract charge transfer energies, $2p3d$ resonant inelastic x-ray scattering (RIXS) seems to be especially powerful, as one typically sees a clear charge transfer band, separated or slightly overlapping with $dd$ orbital excitations \cite{2000JPSJKotaniTiO2, 2006PRBKotaniMnORIXS, 2008PRBKotaniCoORIXS, 2005JPCMGhiringhelliNiO, 2013JPCCGreenZnNiO, 2015PRLGreenIn2O3Fe, 2014JAPGreenSiO2MnCo, 2020PRBHariki2p3dRIXS}. When analyzed using a SIAM, the charge transfer energy can often be extracted in a direct and precise way.

\begin{table}
\caption{Collection of charge transfer energies for Ti oxides from literature.}
\label{Tab:Ti}
\begin{center}
\begin{tabular}{lccc}
\hline
Compound & Valence & ~$\Delta$ (eV)~ & Reference\\ 
\hline
TiO$_2$ & 4+ & 4.0 & \onlinecite{1996PRBBocquetXPS}\\ 
TiO$_2$ & 4+ & 4.0 & \onlinecite{1994JPSJOkadaTi}\\ 
TiO$_2$ & 4+ & 4.0 & \onlinecite{1993JPSJUozumiTiV}\\ 
TiO$_2$ & 4+ & 4.0 & \onlinecite{1994JPSJTanakaRPES}\\ 
TiO$_2$ & 4+ & 3.0 & \onlinecite{2000JPSJKotaniTiO2}\\ 
TiO$_2$ & 4+ & 4.0 & \onlinecite{1995JELSPECKotani}\\ 
TiO$_2$ & 4+ & 3.0 & \onlinecite{2022PRBHarikiTi}\\ 
TiO$_2$ & 4+ & 4.0 & \onlinecite{1999JPCMHufner}\\ 
SrTiO$_3$ & 4+ & 4.0 & \onlinecite{1996PRBBocquetXPS}\\ 
SrTiO$_3$ & 4+ & 3.0 & \onlinecite{2022PRBHarikiTi}\\ 
Ti$_2$O$_3$ & 3+ & 6.5 & \onlinecite{2018PRXTjengTi2O3}\\ 
Ti$_2$O$_3$ & 3+ & 6.5 & \onlinecite{1997JELSPECKotaniM2O3,1997JPSJKotaniTi2O3}\\ 
LaTiO$_3$ & 3+ & 6.0 & \onlinecite{1995PRBBocquet}\\ 
LaTiO$_3$ & 3+ & 4.0 & \onlinecite{2005PRLHaverkortLTO}\\ 
YTiO$_3$ & 3+ & 6.0 & \onlinecite{1996PRBBocquetXPS}\\ 
TiO & 2+ & 8.0 & \onlinecite{1996PRBBocquetXPS}\\ 
\hline
\end{tabular}
\end{center}
\end{table}

\begin{table}
\caption{Collection of charge transfer energies for V oxides from literature.}
\label{Tab:V}
\begin{center}
\begin{tabular}{lccc}
\hline
Compound & Valence & ~$\Delta$ (eV)~ & Reference\\ 
\hline
V$_2$O$_5$ & 5+ & 0.5 & \onlinecite{1996PRBBocquetXPS}\\ 
V$_2$O$_5$ & 5+ & 3.5 & \onlinecite{1999JPCMHufner}\\ 
Ba$_3$V$_2$S$_4$O$_3$ & 5+ & 1.0 & \onlinecite{2015ChemEJTjengBVSO}\\ 
NaV$_2$O$_5$ & 5+ & 0.0 & \onlinecite{2004JACKotaniV}\\ 
V$_6$O$_1$$_3$ & 5+ & 0.0 & \onlinecite{2004PRBKotaniV6O13}\\ 
VO$_2$ & 4+ & 2.0 & \onlinecite{1993JPSJUozumiTiV}\\ 
VO$_2$ & 4+ & 2.5 & \onlinecite{2005PRLHaverkortVO2}\\ 
VO$_2$ & 4+ & 2.0 & \onlinecite{2024PRBHarikiVCr}\\ 
VO$_2$ & 4+ & 4.4 & \onlinecite{1999JPCMHufner}\\ 
NaV$_2$O$_5$ & 4+ & 4.0 & \onlinecite{2004JACKotaniV}\\ 
V$_6$O$_1$$_3$(A) & 4+ & 4.0 & \onlinecite{2004PRBKotaniV6O13}\\ 
V$_6$O$_1$$_3$(B) & 4+ & 2.5 & \onlinecite{2004PRBKotaniV6O13}\\ 
LaVO$_3$ & 3+ & 4.0 & \onlinecite{1996PRBBocquetXPS}\\ 
LiVO$_2$ & 3+ & 4.0 & \onlinecite{1997PRBSawatzkyLiVO2}\\ 
V$_2$O$_3$ & 3+ & 4.0 & \onlinecite{1996PRBBocquetXPS}\\ 
V$_2$O$_3$ & 3+ & 6.0 & \onlinecite{1997JELSPECKotaniM2O3,2004JACKotaniV}\\ 
V$_2$O$_3$ & 3+ & 5.5 & \onlinecite{1999JPCMHufner}\\ 
V$_2$OPO$_4$ & 3+ & 6.0 & \onlinecite{2020PRBMizokawaV2OPO4}\\ 
ZnV$_2$O$_4$ & 3+ & 5.2 & \onlinecite{2006PRBTakuboV}\\ 
CdV$_2$O$_4$ & 3+ & 5.0 & \onlinecite{2006PRBTakuboV}\\ 
Ba$_3$V$_2$S$_4$O$_3$ & 3+ & 3.0 & \onlinecite{2015ChemEJTjengBVSO}\\ 
V$_2$OPO$_4$ & 2+ & 8.0 & \onlinecite{2020PRBMizokawaV2OPO4}\\ 
\hline
\end{tabular}
\end{center}
\end{table}

\begin{table}
\caption{Collection of charge transfer energies for Cr oxides from literature.}
\label{Tab:Cr}
\begin{center}
\begin{tabular}{lccc}
\hline
Compound & Valence & ~$\Delta$ (eV)~ & Reference\\ 
\hline
PbCrO$_3$ & 6+ & -4.0 & \onlinecite{2023PRBTjengPbCrO3}\\ 
CrO$_2$ & 4+ & -0.5\footnote{These references did not give values, but used words \emph{small} and \emph{negative}, so here we assign -0.5 eV} & \onlinecite{1998PRLSawatzkyCrO2,2003PRBTjengCrO2}\\ 
CrO$_2$ & 4+ & 1.0 & \onlinecite{2024PRBHarikiVCr}\\ 
CrO$_2$ & 4+ & 5.5 & \onlinecite{1999JPCMHufner}\\ 
LaCrO$_3$ & 3+ & 5.2 & \onlinecite{1995PRBBocquet}\\ 
LaCrO$_3$ & 3+ & 5.5 & \onlinecite{1996PRBSarmaCr}\\ 
YCrO$_3$ & 3+ & 2.5 & \onlinecite{2018APLChakYCO}\\ 
PbCrO$_3$ & 3+ & 5.0 & \onlinecite{2023PRBTjengPbCrO3}\\ 
Cr$_2$O$_3$ & 3+ & 5.5 & \onlinecite{1997JELSPECKotaniM2O3}\\ 
Cr$_2$O$_3$ & 3+ & 6.0 & \onlinecite{1999JPCMHufner}\\ 
\hline
\end{tabular}
\end{center}
\end{table}

\begin{table}
\caption{Collection of charge transfer energies for Mn oxides from literature.}
\label{Tab:Mn}
\begin{center}
\begin{tabular}{lccc}
\hline
Compound & Valence & ~$\Delta$ (eV)~ & Reference\\ 
\hline
Bi$_3$Mn$_4$O$_1$$_2$(NO$_3$) & 4+ & 1.0 & \onlinecite{2013SSCWadatiMn4}\\ 
La$_2$MnCoO$_6$ & 4+ & -3.0 & \onlinecite{2008PRBBurnusLMCO}\\ 
SrMnO$_3$ & 4+ & 2.0 & \onlinecite{1992PRBBocquetXPS}\\ 
CaMnO$_3$ & 4+ & 3.5 & \onlinecite{2002PhysBAbbateMn34}\\ 
LiMn$_2$O$_4$ & 4+ & -1.0 & \onlinecite{2017JPCCTjengLiMn2O4}\\ 
Li$_2$MnO$_3$ & 4+ & 0.0 & \onlinecite{2023JPSJMizokawaMn4}\\ 
LaSr$_3$Mn$_2$O$_4$ & 4+ & -2.0 & \onlinecite{2011PRBTjengLSMO}\\ 
TbSrMn$_2$O$_6$ & 4+ & 0.0 & \onlinecite{2021PRBdeGrootTSMO}\\ 
Pr$_2$MnNiO$_6$ & 4+ & 2.5 & \onlinecite{2018JPCMdeGrootMnNi}\\ 
LiMn$_2$O$_4$ & 3+ & 4.0 & \onlinecite{2017JPCCTjengLiMn2O4}\\ 
LaMnO$_3$ & 3+ & 4.5 & \onlinecite{1992PRBBocquetXPS}\\ 
CaMnO$_3$ & 3+ & 3.8 & \onlinecite{2002PhysBAbbateMn34}\\ 
DyMnO$_3$ & 3+ & 4.0 & \onlinecite{2010PRBTjengDyMnO3}\\ 
LaSr$_3$Mn$_2$O$_4$ & 3+ & 2.0 & \onlinecite{2011PRBTjengLSMO}\\ 
Mn$_2$O$_3$ & 3+ & 5.0 & \onlinecite{1997JELSPECKotaniM2O3}\\ 
TbSrMn$_2$O$_6$ & 3+ & 1.0 & \onlinecite{2021PRBdeGrootTSMO}\\ 
MnO & 2+ & 6.5 & \onlinecite{1992PRBBocquetXPS}\\ 
MnO & 2+ & 7.0 & \onlinecite{1990PRBFujimoriMnO}\\ 
MnO & 2+ & 9.0 & \onlinecite{1992JPSJKotaniFeMnXPS}\\ 
MnO & 2+ & 8.0 & \onlinecite{1997JPSJKotaniMnXPS,2003JPSJKotaniMnKEdge}\\ 
MnO & 2+ & 6.5 & \onlinecite{2006PRBKotaniMnORIXS}\\ 
MnO & 2+ & 8.8 & \onlinecite{1991PRBvanElpMnO}\\ 
Mn$_3$WO$_6$ & 2+ & 8.0 & \onlinecite{2020ChemMatTjengMn3WO6}\\ 
MnWO$_4$ & 2+ & 6.5 & \onlinecite{2010PRBTjengMnCoWO4}\\ 
SiO$_2$:Mn & 2+ & 6.6 & \onlinecite{2014JAPGreenSiO2MnCo}\\ 
\hline
\end{tabular}
\end{center}
\end{table}

\begin{table}
\caption{Collection of charge transfer energies for Fe oxides from literature.}
\label{Tab:Fe}
\begin{center}
\begin{tabular}{lccc}
\hline
Compound & Valence & ~$\Delta$ (eV)~ & Reference\\ 
\hline
SrFeO$_3$ & 4+ & 0.0 & \onlinecite{1992PRBBocquetXPS}\\ 
CaFeO$_3$ & 4+ & -2.0 & \onlinecite{2018PRBRoggeCaFeO3Inv}\\ 
BaFeO$_3$ & 4+ & -0.9 & \onlinecite{2015PRBWadatiBaFeO3}\\ 
LaFeO$_3$ & 3+ & 2.5 & \onlinecite{1992PRBBocquetXPS}\\ 
CoFe$_2$O$_4$ & 3+ & 2.0 & \onlinecite{2021JMMMTjengCoFe2O4}\\ 
FePO$_4$ & 3+ & 5.6 & \onlinecite{2018CPCHosonoLPFO}\\ 
Fe$_2$O$_3$ & 3+ & 3.5 & \onlinecite{1992PRBBocquetXPS}\\ 
Fe$_2$O$_3$ & 3+ & 3.0 & \onlinecite{1986PRBFujimoriFe2O3}\\ 
Fe$_2$O$_3$ & 3+ & 4.5 & \onlinecite{1997JELSPECKotaniM2O3}\\ 
Fe$_2$O$_3$ & 3+ & 2.0 & \onlinecite{2022PRBdeGrootFe2O3}\\ 
Fe$_2$O$_3$ & 3+ & 3.7 & \onlinecite{2019PRBHariki1sXPS}\\ 
Fe$_2$O$_3$ & 3+ & 5.5 & \onlinecite{1999JPCMHufner}\\ 
Fe$_3$O$_4$(Td) & 3+ & 3.5 & \onlinecite{2016PRXTjengFe3O4}\\ 
Fe$_3$O$_4$(Oh) & 3+ & 3.0 & \onlinecite{2016PRXTjengFe3O4}\\ 
SmFeO$_3$ & 3+ & 3.0 & \onlinecite{2014PRLTjengSmFeO3}\\ 
CaBaFe$_4$O$_7$ & 3+ & 1.0 & \onlinecite{2011PRBTjengCBFO}\\ 
(In,Fe)$_2$O$_3$ & 3+ & 1.0 & \onlinecite{2015PRLGreenIn2O3Fe}\\ 
Fe$_3$O$_4$ & 2+ & 7.0 & \onlinecite{2016PRXTjengFe3O4}\\ 
FeO & 2+ & 6.0 & \onlinecite{1992PRBBocquetXPS}\\ 
FeO & 2+ & 6.5 & \onlinecite{1987PRBFujimoriFeO}\\ 
FeO & 2+ & 7.0 & \onlinecite{1994JPSJTanakaRPES}\\ 
FeO & 2+ & 6.5 & \onlinecite{1992JPSJKotaniFeMnXPS}\\ 
FeO & 2+ & 6.3 & \onlinecite{1999JPCMHufner}\\ 
(Fe,Mg)O & 2+ & 7.5 & \onlinecite{2010PRBTjengMgOFe}\\ 
FeWO$_4$ & 2+ & 8.0 & \onlinecite{2023PRBTjengFeWO4}\\ 
FePO$_4$ & 2+ & 9.7 & \onlinecite{2018CPCHosonoLPFO}\\ 
CaBaFe$_4$O$_7$ & 2+ & 5.0 & \onlinecite{2011PRBTjengCBFO}\\ 
FeTiO$_3$ & 2+ & 3.5 & \onlinecite{2019PRBHariki1sXPS}\\ 
(In,Fe)$_2$O$_3$ & 2+ & 6.0 & \onlinecite{2015PRLGreenIn2O3Fe}\\ 
\hline
\end{tabular}
\end{center}
\end{table}

\begin{table}
\caption{Collection of charge transfer energies for Co oxides from literature.}
\label{Tab:Co}
\begin{center}
\begin{tabular}{lccc}
\hline
Compound & Valence & ~$\Delta$ (eV)~ & Reference\\ 
\hline
CaCu$_3$Co$_4$O$_1$$_2$ & 4+ & -4.5 & \onlinecite{20221PRBTjengCo4}\\ 
Ba$_2$CoO$_4$ & 4+ & -3.5 & \onlinecite{2023ZAACTjengBa2CoO4}\\ 
BaCoO$_3$ & 4+ & -3.5 & \onlinecite{2019PRBTjengBaCoO3}\\ 
SrCoO$_3$ & 4+ & -5.0 & \onlinecite{1995PRBSawatzkyCo4}\\ 
Na$_x$CoO$_2$ & 4+ & -1.0 & \onlinecite{2010PRBTjengNaCoO2}\\ 
Sr$_2$CoO$_3$Cl & 3+ & 2.5 & \onlinecite{2004PRLTjengSr2CoO3Cl}\\ 
LaCoO$_3$ & 3+ & 2.0 & \onlinecite{2006PRLHaverkortLCO}\\ 
LaCoO$_3$ & 3+ & 2.0 & \onlinecite{2023PRXTjengLCO}\\ 
BiCoO$_3$ & 3+ & 0.0 & \onlinecite{2011PRBShimakawaBiCoO3}\\ 
LiCoO$_2$ & 3+ & 4.0 & \onlinecite{1991PRBvanElpCo}\\ 
Sr$_4$CoIrO$_8$ & 3+ & 2.0 & \onlinecite{2017PRBTjengCo3}\\ 
NdCaCoO$_4$ & 3+ & 2.0 & \onlinecite{2017PRBTjengCo3}\\ 
Gd$_2$Ba$_2$Co$_4$O$_1$$_0$ & 3+ & 2.0 & \onlinecite{2012NJPTjengGdBaCoO}\\ 
Na$_x$CoO$_2$ & 3+ & 4.0 & \onlinecite{2010PRBTjengNaCoO2}\\ 
Ca$_3$Co$_2$O$_6$ & 3+ & 1.5 & \onlinecite{2006PRBTjengCCO}\\ 
Co$_3$O$_4$ & 3+ & 1.5 & \onlinecite{2022JPCCdeGrootCo3O4}\\ 
CoFe$_2$O$_4$ & 2+ & 6.5 & \onlinecite{2021JMMMTjengCoFe2O4}\\ 
La$_2$MnCoO$_6$ & 2+ & 5.5 & \onlinecite{2008PRBBurnusLMCO}\\ 
CoWO$_4$ & 2+ & 6.5 & \onlinecite{2010PRBTjengMnCoWO4}\\ 
CoV$_2$O$_6$ & 2+ & 6.5 & \onlinecite{2014PRBTjengCoV2O6}\\ 
CoO & 2+ & 6.5 & \onlinecite{2005PRLTjengCoO}\\ 
CoO & 2+ & 2.5 & \onlinecite{1992JPSJKotaniCoO}\\ 
CoO & 2+ & 3.0 & \onlinecite{2008PRBKotaniCoORIXS}\\ 
CoO & 2+ & 4.1 & \onlinecite{2019PRBHariki1sXPS}\\ 
CoO & 2+ & 5.7 & \onlinecite{1993CPLParmigiani}\\ 
CoO & 2+ & 5.5 & \onlinecite{1991PRBvanElpCo}\\ 
Co$_3$O$_4$ & 2+ & 4.5 & \onlinecite{2022JPCCdeGrootCo3O4}\\ 
SiO$_2$:Co & 2+ & 5.5 & \onlinecite{2014JAPGreenSiO2MnCo}\\ 
\hline
\end{tabular}
\end{center}
\end{table}

\begin{table}
\caption{Collection of charge transfer energies for Ni oxides from literature.}
\label{Tab:Ni}
\begin{center}
\begin{tabular}{lccc}
\hline
Compound & Valence & ~$\Delta$ (eV)~ & Reference\\ 
\hline
LiNiO$_2$ & 4+ & -3.5 & \onlinecite{2023NatCommTjengLiNiO2}\\ 
NdNiO$_3$ & 3+ & -0.5 & \onlinecite{2016PRBGreenDoubleCluster}\\ 
PrNiO$_3$ & 3+ & 1.0 & \onlinecite{1996JELSPECMizokawaPNO}\\ 
LaNiO$_3$ & 3+ & 1.0 & \onlinecite{2002PRBAbbateLaNiO3}\\ 
LiNiO$_2$ & 3+ & -0.5 & \onlinecite{2023NatCommTjengLiNiO2}\\ 
LiNiO$_2$ & 3+ & -0.5 & \onlinecite{1992PRBvanElpNi}\\ 
Nd$_4$LiNiO$_8$ & 3+ & 0.5 & \onlinecite{2017JPCCTjengLiMn2O4}\\ 
Li$_2$NiMn$_3$O$_8$ & 2+ & 6.0 & \onlinecite{2017JPCCTjengLiMn2O4}\\ 
Pr$_2$MnNiO$_6$ & 2+ & 3.5 & \onlinecite{2018JPCMdeGrootMnNi}\\ 
Lu$_2$BaNiO$_5$ & 2+ & 2.5 & \onlinecite{1999PRBSarmaNi}\\ 
Y$_2$BaNiO$_5$ & 2+ & 3.75 & \onlinecite{1999PRBSarmaNi}\\ 
Pr$_2$NiO$_4$ & 2+ & 3.5 & \onlinecite{1999PRBSarmaNi}\\ 
La$_2$NiO$_4$ & 2+ & 4.25 & \onlinecite{1999PRBSarmaNi}\\ 
NiO & 2+ & 5.5 & \onlinecite{1999PRBSarmaNi}\\ 
NiO & 2+ & 4.5 & \onlinecite{2017EPJSTTjengNiOReview}\\ 
NiO & 2+ & 4.5 & \onlinecite{1992PRBBocquetXPS}\\ 
NiO & 2+ & 4.6 & \onlinecite{1986PRBvanderlaanNi}\\ 
NiO & 2+ & 4.0 & \onlinecite{1984PRBFujimoriNi}\\ 
NiO & 2+ & 3.5 & \onlinecite{2005JPSJKotaniNiO,2005JPCMGhiringhelliNiO}\\ 
NiO & 2+ & 4.4 & \onlinecite{2019PRBHariki1sXPS}\\ 
NiO & 2+ & 5.3 & \onlinecite{1993CPLParmigiani}\\ 
NiO & 2+ & 6.2 & \onlinecite{1992PRBvanElpNi}\\ 
(Zn,Ni)O & 2+ & 4.0 & \onlinecite{2013JPCCGreenZnNiO}\\ 
(Mg,Ni)O & 2+ & 4.7 & \onlinecite{2000PRBTjengMgONi}\\ 
\hline
\end{tabular}
\end{center}
\end{table}

\begin{table}
\caption{Collection of charge transfer energies for Cu oxides from literature.}
\label{Tab:Cu}
\begin{center}
\begin{tabular}{lccc}
\hline
Compound & Valence & ~$\Delta$ (eV)~ & Reference\\ 
\hline
NaCuO$_2$ & 3+ & -2.0 & \onlinecite{1991PRLMizokawaNaCuO2}\\ 
NaCuO$_2$ & 3+ & -1.0 & \onlinecite{1994PRBMizokawaNaCuO2}\\ 
NaCuO$_2$ & 3+ & -1.0 & \onlinecite{1991SSCSawatzkyNaCuO2}\\ 
NaCuO$_2$ & 3+ & -2.5 & \onlinecite{1993PRBSarmaNaCuO2}\\ 
LaCuO$_3$ & 3+ & -1.0 & \onlinecite{1998PRBFujimoriLaCuO3}\\ 
LaCuO$_3$ & 3+ & -2.9 & \onlinecite{1999JPSJKotaniLaCuO3}\\ 
KCuO$_2$ & 3+ & -1.5 & \onlinecite{2015PRBChakCu3}\\ 
CuO & 2+ & 2.2 & \onlinecite{1988PRBTjengCuOxides}\\ 
CuO & 2+ & 2.75 & \onlinecite{1991JPSJKotaniCuOLCO}\\ 
CuO & 2+ & 1.5 & \onlinecite{1997EPLFujimoriCu}\\ 
CaCu$_3$Ru$_4$O$_1$$_2$ & 2+ & 2.2 & \onlinecite{2013PRBTjengCCRO}\\ 
Nd$_2$CuO$_4$ & 2+ & 2.5 & \onlinecite{2000PRBKotaniNCO,2000JPSJKotaniNCO}\\ 
La$_2$CuO$_4$ & 2+ & 1.0 & \onlinecite{1997EPLFujimoriCu}\\ 
La$_2$CuO$_4$ & 2+ & 1.4 & \onlinecite{1991SSCSawatzkyNaCuO2}\\ 
La$_2$CuO$_4$ & 2+ & 2.3 & \onlinecite{1991JPSJKotaniCuOLCO}\\ 
Sr$_2$CuO$_3$ & 2+ & 2.5 & \onlinecite{1996JPSJKotaniSCO}\\ 
Sr$_2$CuO$_3$ & 2+ & 0.25 & \onlinecite{1997EPLFujimoriCu}\\ 
\hline
\end{tabular}
\end{center}
\end{table}

\bibliography{Bibliography}

\begin{thebibliography}{158}%
\makeatletter
\providecommand \@ifxundefined [1]{%
 \@ifx{#1\undefined}
}%
\providecommand \@ifnum [1]{%
 \ifnum #1\expandafter \@firstoftwo
 \else \expandafter \@secondoftwo
 \fi
}%
\providecommand \@ifx [1]{%
 \ifx #1\expandafter \@firstoftwo
 \else \expandafter \@secondoftwo
 \fi
}%
\providecommand \natexlab [1]{#1}%
\providecommand \enquote  [1]{``#1''}%
\providecommand \bibnamefont  [1]{#1}%
\providecommand \bibfnamefont [1]{#1}%
\providecommand \citenamefont [1]{#1}%
\providecommand \href@noop [0]{\@secondoftwo}%
\providecommand \href [0]{\begingroup \@sanitize@url \@href}%
\providecommand \@href[1]{\@@startlink{#1}\@@href}%
\providecommand \@@href[1]{\endgroup#1\@@endlink}%
\providecommand \@sanitize@url [0]{\catcode `\\12\catcode `\$12\catcode `\&12\catcode `\#12\catcode `\^12\catcode `\_12\catcode `\%12\relax}%
\providecommand \@@startlink[1]{}%
\providecommand \@@endlink[0]{}%
\providecommand \url  [0]{\begingroup\@sanitize@url \@url }%
\providecommand \@url [1]{\endgroup\@href {#1}{\urlprefix }}%
\providecommand \urlprefix  [0]{URL }%
\providecommand \Eprint [0]{\href }%
\providecommand \doibase [0]{http://dx.doi.org/}%
\providecommand \selectlanguage [0]{\@gobble}%
\providecommand \bibinfo  [0]{\@secondoftwo}%
\providecommand \bibfield  [0]{\@secondoftwo}%
\providecommand \translation [1]{[#1]}%
\providecommand \BibitemOpen [0]{}%
\providecommand \bibitemStop [0]{}%
\providecommand \bibitemNoStop [0]{.\EOS\space}%
\providecommand \EOS [0]{\spacefactor3000\relax}%
\providecommand \BibitemShut  [1]{\csname bibitem#1\endcsname}%
\let\auto@bib@innerbib\@empty
\bibitem [{\citenamefont {Tanabe}\ and\ \citenamefont {Sugano}(1954{\natexlab{a}})}]{1954_TSD}%
  \BibitemOpen
  \bibfield  {author} {\bibinfo {author} {\bibfnamefont {Y.}~\bibnamefont {Tanabe}}\ and\ \bibinfo {author} {\bibfnamefont {S.}~\bibnamefont {Sugano}},\ }\href {\doibase 10.1143/JPSJ.9.753} {\bibfield  {journal} {\bibinfo  {journal} {Journal of the Physical Society of Japan}\ }\textbf {\bibinfo {volume} {9}},\ \bibinfo {pages} {753} (\bibinfo {year} {1954}{\natexlab{a}})}\BibitemShut {NoStop}%
\bibitem [{\citenamefont {Tanabe}\ and\ \citenamefont {Sugano}(1954{\natexlab{b}})}]{1954_TSD2}%
  \BibitemOpen
  \bibfield  {author} {\bibinfo {author} {\bibfnamefont {Y.}~\bibnamefont {Tanabe}}\ and\ \bibinfo {author} {\bibfnamefont {S.}~\bibnamefont {Sugano}},\ }\href {\doibase 10.1143/JPSJ.9.766} {\bibfield  {journal} {\bibinfo  {journal} {Journal of the Physical Society of Japan}\ }\textbf {\bibinfo {volume} {9}},\ \bibinfo {pages} {766} (\bibinfo {year} {1954}{\natexlab{b}})}\BibitemShut {NoStop}%
\bibitem [{\citenamefont {Anderson}(1961)}]{Anderson1961SIAM}%
  \BibitemOpen
  \bibfield  {author} {\bibinfo {author} {\bibfnamefont {P.~W.}\ \bibnamefont {Anderson}},\ }\href {\doibase 10.1103/PhysRev.124.41} {\bibfield  {journal} {\bibinfo  {journal} {Phys. Rev.}\ }\textbf {\bibinfo {volume} {124}},\ \bibinfo {pages} {41} (\bibinfo {year} {1961})}\BibitemShut {NoStop}%
\bibitem [{\citenamefont {Fujimori}\ and\ \citenamefont {Minami}(1984)}]{1984PRBFujimoriNi}%
  \BibitemOpen
  \bibfield  {author} {\bibinfo {author} {\bibfnamefont {A.}~\bibnamefont {Fujimori}}\ and\ \bibinfo {author} {\bibfnamefont {F.}~\bibnamefont {Minami}},\ }\href {\doibase 10.1103/PhysRevB.30.957} {\bibfield  {journal} {\bibinfo  {journal} {Phys. Rev. B}\ }\textbf {\bibinfo {volume} {30}},\ \bibinfo {pages} {957} (\bibinfo {year} {1984})}\BibitemShut {NoStop}%
\bibitem [{\citenamefont {Sawatzky}\ and\ \citenamefont {Allen}(1984)}]{1984PRLSawatzkyAllenNiO}%
  \BibitemOpen
  \bibfield  {author} {\bibinfo {author} {\bibfnamefont {G.~A.}\ \bibnamefont {Sawatzky}}\ and\ \bibinfo {author} {\bibfnamefont {J.~W.}\ \bibnamefont {Allen}},\ }\href {\doibase 10.1103/PhysRevLett.53.2339} {\bibfield  {journal} {\bibinfo  {journal} {Phys. Rev. Lett.}\ }\textbf {\bibinfo {volume} {53}},\ \bibinfo {pages} {2339} (\bibinfo {year} {1984})}\BibitemShut {NoStop}%
\bibitem [{\citenamefont {Wilson}(1975)}]{Wilson1975}%
  \BibitemOpen
  \bibfield  {author} {\bibinfo {author} {\bibfnamefont {K.~G.}\ \bibnamefont {Wilson}},\ }\href {\doibase 10.1103/RevModPhys.47.773} {\bibfield  {journal} {\bibinfo  {journal} {Rev. Mod. Phys.}\ }\textbf {\bibinfo {volume} {47}},\ \bibinfo {pages} {773} (\bibinfo {year} {1975})}\BibitemShut {NoStop}%
\bibitem [{\citenamefont {Lu}\ \emph {et~al.}(2014)\citenamefont {Lu}, \citenamefont {H\"oppner}, \citenamefont {Gunnarsson},\ and\ \citenamefont {Haverkort}}]{Haverkort_DMFT_2014}%
  \BibitemOpen
  \bibfield  {author} {\bibinfo {author} {\bibfnamefont {Y.}~\bibnamefont {Lu}}, \bibinfo {author} {\bibfnamefont {M.}~\bibnamefont {H\"oppner}}, \bibinfo {author} {\bibfnamefont {O.}~\bibnamefont {Gunnarsson}}, \ and\ \bibinfo {author} {\bibfnamefont {M.~W.}\ \bibnamefont {Haverkort}},\ }\href {\doibase 10.1103/PhysRevB.90.085102} {\bibfield  {journal} {\bibinfo  {journal} {Phys. Rev. B}\ }\textbf {\bibinfo {volume} {90}},\ \bibinfo {pages} {085102} (\bibinfo {year} {2014})}\BibitemShut {NoStop}%
\bibitem [{\citenamefont {Haverkort}\ \emph {et~al.}(2012)\citenamefont {Haverkort}, \citenamefont {Zwierzycki},\ and\ \citenamefont {Andersen}}]{2012PRBHaverkortMLFT}%
  \BibitemOpen
  \bibfield  {author} {\bibinfo {author} {\bibfnamefont {M.~W.}\ \bibnamefont {Haverkort}}, \bibinfo {author} {\bibfnamefont {M.}~\bibnamefont {Zwierzycki}}, \ and\ \bibinfo {author} {\bibfnamefont {O.~K.}\ \bibnamefont {Andersen}},\ }\href {\doibase 10.1103/PhysRevB.85.165113} {\bibfield  {journal} {\bibinfo  {journal} {Phys. Rev. B}\ }\textbf {\bibinfo {volume} {85}},\ \bibinfo {pages} {165113} (\bibinfo {year} {2012})}\BibitemShut {NoStop}%
\bibitem [{\citenamefont {{de Groot}}\ \emph {et~al.}(2021)\citenamefont {{de Groot}}, \citenamefont {Elnaggar}, \citenamefont {Frati}, \citenamefont {pan Wang}, \citenamefont {Delgado-Jaime}, \citenamefont {{van Veenendaal}}, \citenamefont {Fernandez-Rodriguez}, \citenamefont {Haverkort}, \citenamefont {Green}, \citenamefont {{van der Laan}}, \citenamefont {Kvashnin}, \citenamefont {Hariki}, \citenamefont {Ikeno}, \citenamefont {Ramanantoanina}, \citenamefont {Daul}, \citenamefont {Delley}, \citenamefont {Odelius}, \citenamefont {Lundberg}, \citenamefont {Kuhn}, \citenamefont {Bokarev}, \citenamefont {Shirley}, \citenamefont {Vinson}, \citenamefont {Gilmore}, \citenamefont {Stener}, \citenamefont {Fronzoni}, \citenamefont {Decleva}, \citenamefont {Kruger}, \citenamefont {Retegan}, \citenamefont {Joly}, \citenamefont {Vorwerk}, \citenamefont {Draxl}, \citenamefont {Rehr},\ and\ \citenamefont {Tanaka}}]{2021_JELSPEC_deGroot}%
  \BibitemOpen
  \bibfield  {author} {\bibinfo {author} {\bibfnamefont {F.~M.}\ \bibnamefont {{de Groot}}}, \bibinfo {author} {\bibfnamefont {H.}~\bibnamefont {Elnaggar}}, \bibinfo {author} {\bibfnamefont {F.}~\bibnamefont {Frati}}, \bibinfo {author} {\bibfnamefont {R.}~\bibnamefont {pan Wang}}, \bibinfo {author} {\bibfnamefont {M.~U.}\ \bibnamefont {Delgado-Jaime}}, \bibinfo {author} {\bibfnamefont {M.}~\bibnamefont {{van Veenendaal}}}, \bibinfo {author} {\bibfnamefont {J.}~\bibnamefont {Fernandez-Rodriguez}}, \bibinfo {author} {\bibfnamefont {M.~W.}\ \bibnamefont {Haverkort}}, \bibinfo {author} {\bibfnamefont {R.~J.}\ \bibnamefont {Green}}, \bibinfo {author} {\bibfnamefont {G.}~\bibnamefont {{van der Laan}}}, \bibinfo {author} {\bibfnamefont {Y.}~\bibnamefont {Kvashnin}}, \bibinfo {author} {\bibfnamefont {A.}~\bibnamefont {Hariki}}, \bibinfo {author} {\bibfnamefont {H.}~\bibnamefont {Ikeno}}, \bibinfo {author} {\bibfnamefont {H.}~\bibnamefont {Ramanantoanina}}, \bibinfo {author} {\bibfnamefont {C.}~\bibnamefont {Daul}},
  \bibinfo {author} {\bibfnamefont {B.}~\bibnamefont {Delley}}, \bibinfo {author} {\bibfnamefont {M.}~\bibnamefont {Odelius}}, \bibinfo {author} {\bibfnamefont {M.}~\bibnamefont {Lundberg}}, \bibinfo {author} {\bibfnamefont {O.}~\bibnamefont {Kuhn}}, \bibinfo {author} {\bibfnamefont {S.~I.}\ \bibnamefont {Bokarev}}, \bibinfo {author} {\bibfnamefont {E.}~\bibnamefont {Shirley}}, \bibinfo {author} {\bibfnamefont {J.}~\bibnamefont {Vinson}}, \bibinfo {author} {\bibfnamefont {K.}~\bibnamefont {Gilmore}}, \bibinfo {author} {\bibfnamefont {M.}~\bibnamefont {Stener}}, \bibinfo {author} {\bibfnamefont {G.}~\bibnamefont {Fronzoni}}, \bibinfo {author} {\bibfnamefont {P.}~\bibnamefont {Decleva}}, \bibinfo {author} {\bibfnamefont {P.}~\bibnamefont {Kruger}}, \bibinfo {author} {\bibfnamefont {M.}~\bibnamefont {Retegan}}, \bibinfo {author} {\bibfnamefont {Y.}~\bibnamefont {Joly}}, \bibinfo {author} {\bibfnamefont {C.}~\bibnamefont {Vorwerk}}, \bibinfo {author} {\bibfnamefont {C.}~\bibnamefont {Draxl}}, \bibinfo {author}
  {\bibfnamefont {J.}~\bibnamefont {Rehr}}, \ and\ \bibinfo {author} {\bibfnamefont {A.}~\bibnamefont {Tanaka}},\ }\href {\doibase https://doi.org/10.1016/j.elspec.2021.147061} {\bibfield  {journal} {\bibinfo  {journal} {Journal of Electron Spectroscopy and Related Phenomena}\ }\textbf {\bibinfo {volume} {249}},\ \bibinfo {pages} {147061} (\bibinfo {year} {2021})}\BibitemShut {NoStop}%
\bibitem [{\citenamefont {Eskes}\ \emph {et~al.}(1990)\citenamefont {Eskes}, \citenamefont {Tjeng},\ and\ \citenamefont {Sawatzky}}]{1990PRBEskesCuO}%
  \BibitemOpen
  \bibfield  {author} {\bibinfo {author} {\bibfnamefont {H.}~\bibnamefont {Eskes}}, \bibinfo {author} {\bibfnamefont {L.~H.}\ \bibnamefont {Tjeng}}, \ and\ \bibinfo {author} {\bibfnamefont {G.~A.}\ \bibnamefont {Sawatzky}},\ }\href {\doibase 10.1103/PhysRevB.41.288} {\bibfield  {journal} {\bibinfo  {journal} {Phys. Rev. B}\ }\textbf {\bibinfo {volume} {41}},\ \bibinfo {pages} {288} (\bibinfo {year} {1990})}\BibitemShut {NoStop}%
\bibitem [{\citenamefont {Zaanen}\ \emph {et~al.}(1985)\citenamefont {Zaanen}, \citenamefont {Sawatzky},\ and\ \citenamefont {Allen}}]{Zaanen_ZSA_PRL_1985}%
  \BibitemOpen
  \bibfield  {author} {\bibinfo {author} {\bibfnamefont {J.}~\bibnamefont {Zaanen}}, \bibinfo {author} {\bibfnamefont {G.~A.}\ \bibnamefont {Sawatzky}}, \ and\ \bibinfo {author} {\bibfnamefont {J.~W.}\ \bibnamefont {Allen}},\ }\href {\doibase 10.1103/PhysRevLett.55.418} {\bibfield  {journal} {\bibinfo  {journal} {Phys. Rev. Lett.}\ }\textbf {\bibinfo {volume} {55}},\ \bibinfo {pages} {418} (\bibinfo {year} {1985})}\BibitemShut {NoStop}%
\bibitem [{\citenamefont {Haverkort}()}]{QuantyWeb}%
  \BibitemOpen
  \bibfield  {author} {\bibinfo {author} {\bibfnamefont {M.~W.}\ \bibnamefont {Haverkort}},\ }\href@noop {} {}\bibinfo {note} {\url{http://www.quanty.org}}\BibitemShut {NoStop}%
\bibitem [{\citenamefont {Haverkort}\ \emph {et~al.}(2014)\citenamefont {Haverkort}, \citenamefont {Sangiovanni}, \citenamefont {Hansmann}, \citenamefont {Toschi}, \citenamefont {Lu},\ and\ \citenamefont {Macke}}]{Haverkort_DMFTXAS_EPL2014}%
  \BibitemOpen
  \bibfield  {author} {\bibinfo {author} {\bibfnamefont {M.~W.}\ \bibnamefont {Haverkort}}, \bibinfo {author} {\bibfnamefont {G.}~\bibnamefont {Sangiovanni}}, \bibinfo {author} {\bibfnamefont {P.}~\bibnamefont {Hansmann}}, \bibinfo {author} {\bibfnamefont {A.}~\bibnamefont {Toschi}}, \bibinfo {author} {\bibfnamefont {Y.}~\bibnamefont {Lu}}, \ and\ \bibinfo {author} {\bibfnamefont {S.}~\bibnamefont {Macke}},\ }\href {\doibase 10.1209/0295-5075/108/57004} {\bibfield  {journal} {\bibinfo  {journal} {Europhysics Letters}\ }\textbf {\bibinfo {volume} {108}},\ \bibinfo {pages} {57004} (\bibinfo {year} {2014})}\BibitemShut {NoStop}%
\bibitem [{\citenamefont {Mott}\ and\ \citenamefont {Peierls}(1937)}]{1937PRCMottPeierls}%
  \BibitemOpen
  \bibfield  {author} {\bibinfo {author} {\bibfnamefont {N.~F.}\ \bibnamefont {Mott}}\ and\ \bibinfo {author} {\bibfnamefont {R.}~\bibnamefont {Peierls}},\ }\href {\doibase 10.1088/0959-5309/49/4S/308} {\bibfield  {journal} {\bibinfo  {journal} {Proceedings of the Physical Society}\ }\textbf {\bibinfo {volume} {49}},\ \bibinfo {pages} {72} (\bibinfo {year} {1937})}\BibitemShut {NoStop}%
\bibitem [{\citenamefont {Mott}(1949)}]{1949PPSAMott}%
  \BibitemOpen
  \bibfield  {author} {\bibinfo {author} {\bibfnamefont {N.~F.}\ \bibnamefont {Mott}},\ }\href {\doibase 10.1088/0370-1298/62/7/303} {\bibfield  {journal} {\bibinfo  {journal} {Proceedings of the Physical Society. Section A}\ }\textbf {\bibinfo {volume} {62}},\ \bibinfo {pages} {416} (\bibinfo {year} {1949})}\BibitemShut {NoStop}%
\bibitem [{\citenamefont {de~Boer}\ and\ \citenamefont {Verwey}(1937)}]{1937PRCDeBoerVerwey}%
  \BibitemOpen
  \bibfield  {author} {\bibinfo {author} {\bibfnamefont {J.~H.}\ \bibnamefont {de~Boer}}\ and\ \bibinfo {author} {\bibfnamefont {E.~J.~W.}\ \bibnamefont {Verwey}},\ }\href {\doibase 10.1088/0959-5309/49/4S/307} {\bibfield  {journal} {\bibinfo  {journal} {Proceedings of the Physical Society}\ }\textbf {\bibinfo {volume} {49}},\ \bibinfo {pages} {59} (\bibinfo {year} {1937})}\BibitemShut {NoStop}%
\bibitem [{\citenamefont {Hubbard}(1964{\natexlab{a}})}]{1964PRSAHubbard1}%
  \BibitemOpen
  \bibfield  {author} {\bibinfo {author} {\bibfnamefont {J.}~\bibnamefont {Hubbard}},\ }\href {\doibase 10.1098/rspa.1964.0019} {\bibfield  {journal} {\bibinfo  {journal} {Proceedings of the Royal Society of London. Series A. Mathematical and Physical Sciences}\ }\textbf {\bibinfo {volume} {277}},\ \bibinfo {pages} {237} (\bibinfo {year} {1964}{\natexlab{a}})}\BibitemShut {NoStop}%
\bibitem [{\citenamefont {Hubbard}(1964{\natexlab{b}})}]{1964PRSAHubbard2}%
  \BibitemOpen
  \bibfield  {author} {\bibinfo {author} {\bibfnamefont {J.}~\bibnamefont {Hubbard}},\ }\href {\doibase 10.1098/rspa.1964.0190} {\bibfield  {journal} {\bibinfo  {journal} {Proceedings of the Royal Society of London. Series A. Mathematical and Physical Sciences}\ }\textbf {\bibinfo {volume} {281}},\ \bibinfo {pages} {401} (\bibinfo {year} {1964}{\natexlab{b}})}\BibitemShut {NoStop}%
\bibitem [{\citenamefont {Sawatzky}\ and\ \citenamefont {Green}(2016)}]{2016SawatzkyGreenJulich}%
  \BibitemOpen
  \bibfield  {author} {\bibinfo {author} {\bibfnamefont {G.~A.}\ \bibnamefont {Sawatzky}}\ and\ \bibinfo {author} {\bibfnamefont {R.~J.}\ \bibnamefont {Green}},\ }in\ \href {https://juser.fz-juelich.de/record/819465} {\emph {\bibinfo {booktitle} {Quantum Materials: Experiments and Theory}}},\ \bibinfo {editor} {edited by\ \bibinfo {editor} {\bibfnamefont {J.~v. d.~B.}\ \bibnamefont {E.~Pavarini}, \bibfnamefont {E.~Koch}}\ and\ \bibinfo {editor} {\bibfnamefont {G.}~\bibnamefont {Sawatzky}}}\ (\bibinfo  {publisher} {Forschungszentrum Julich},\ \bibinfo {year} {2016})\BibitemShut {NoStop}%
\bibitem [{\citenamefont {Khomskii}(1997)}]{1997LJPKhomskii}%
  \BibitemOpen
  \bibfield  {author} {\bibinfo {author} {\bibfnamefont {D.~I.}\ \bibnamefont {Khomskii}},\ }\href {https://link.aps.org/doi/10.1103/PhysRevB.33.4253} {\bibfield  {journal} {\bibinfo  {journal} {Lithuanian J. Phys.}\ }\textbf {\bibinfo {volume} {37}},\ \bibinfo {pages} {65} (\bibinfo {year} {1997})}\BibitemShut {NoStop}%
\bibitem [{\citenamefont {Bisogni}\ \emph {et~al.}(2016)\citenamefont {Bisogni}, \citenamefont {Catalano}, \citenamefont {Green}, \citenamefont {Gibert}, \citenamefont {Scherwitzl}, \citenamefont {Huang}, \citenamefont {Strocov}, \citenamefont {Zubko}, \citenamefont {Balandeh}, \citenamefont {Triscone}, \citenamefont {Sawatzky},\ and\ \citenamefont {Schmitt}}]{2016NatComBisogniNNO}%
  \BibitemOpen
  \bibfield  {author} {\bibinfo {author} {\bibfnamefont {V.}~\bibnamefont {Bisogni}}, \bibinfo {author} {\bibfnamefont {S.}~\bibnamefont {Catalano}}, \bibinfo {author} {\bibfnamefont {R.~J.}\ \bibnamefont {Green}}, \bibinfo {author} {\bibfnamefont {M.}~\bibnamefont {Gibert}}, \bibinfo {author} {\bibfnamefont {R.}~\bibnamefont {Scherwitzl}}, \bibinfo {author} {\bibfnamefont {Y.}~\bibnamefont {Huang}}, \bibinfo {author} {\bibfnamefont {V.~N.}\ \bibnamefont {Strocov}}, \bibinfo {author} {\bibfnamefont {P.}~\bibnamefont {Zubko}}, \bibinfo {author} {\bibfnamefont {S.}~\bibnamefont {Balandeh}}, \bibinfo {author} {\bibfnamefont {J.-M.}\ \bibnamefont {Triscone}}, \bibinfo {author} {\bibfnamefont {G.}~\bibnamefont {Sawatzky}}, \ and\ \bibinfo {author} {\bibfnamefont {T.}~\bibnamefont {Schmitt}},\ }\href {\doibase 10.1038/ncomms13017} {\bibfield  {journal} {\bibinfo  {journal} {Nature Communications}\ }\textbf {\bibinfo {volume} {7}},\ \bibinfo {pages} {13017} (\bibinfo {year} {2016})}\BibitemShut {NoStop}%
\bibitem [{\citenamefont {Green}\ \emph {et~al.}(2016)\citenamefont {Green}, \citenamefont {Haverkort},\ and\ \citenamefont {Sawatzky}}]{2016PRBGreenDoubleCluster}%
  \BibitemOpen
  \bibfield  {author} {\bibinfo {author} {\bibfnamefont {R.~J.}\ \bibnamefont {Green}}, \bibinfo {author} {\bibfnamefont {M.~W.}\ \bibnamefont {Haverkort}}, \ and\ \bibinfo {author} {\bibfnamefont {G.~A.}\ \bibnamefont {Sawatzky}},\ }\href {\doibase 10.1103/PhysRevB.94.195127} {\bibfield  {journal} {\bibinfo  {journal} {Phys. Rev. B}\ }\textbf {\bibinfo {volume} {94}},\ \bibinfo {pages} {195127} (\bibinfo {year} {2016})}\BibitemShut {NoStop}%
\bibitem [{\citenamefont {Lu}\ \emph {et~al.}(2018)\citenamefont {Lu}, \citenamefont {Betto}, \citenamefont {F\"ursich}, \citenamefont {Suzuki}, \citenamefont {Kim}, \citenamefont {Cristiani}, \citenamefont {Logvenov}, \citenamefont {Brookes}, \citenamefont {Benckiser}, \citenamefont {Haverkort}, \citenamefont {Khaliullin}, \citenamefont {Le~Tacon}, \citenamefont {Minola},\ and\ \citenamefont {Keimer}}]{2018PRXYiDoubleCluster}%
  \BibitemOpen
  \bibfield  {author} {\bibinfo {author} {\bibfnamefont {Y.}~\bibnamefont {Lu}}, \bibinfo {author} {\bibfnamefont {D.}~\bibnamefont {Betto}}, \bibinfo {author} {\bibfnamefont {K.}~\bibnamefont {F\"ursich}}, \bibinfo {author} {\bibfnamefont {H.}~\bibnamefont {Suzuki}}, \bibinfo {author} {\bibfnamefont {H.-H.}\ \bibnamefont {Kim}}, \bibinfo {author} {\bibfnamefont {G.}~\bibnamefont {Cristiani}}, \bibinfo {author} {\bibfnamefont {G.}~\bibnamefont {Logvenov}}, \bibinfo {author} {\bibfnamefont {N.~B.}\ \bibnamefont {Brookes}}, \bibinfo {author} {\bibfnamefont {E.}~\bibnamefont {Benckiser}}, \bibinfo {author} {\bibfnamefont {M.~W.}\ \bibnamefont {Haverkort}}, \bibinfo {author} {\bibfnamefont {G.}~\bibnamefont {Khaliullin}}, \bibinfo {author} {\bibfnamefont {M.}~\bibnamefont {Le~Tacon}}, \bibinfo {author} {\bibfnamefont {M.}~\bibnamefont {Minola}}, \ and\ \bibinfo {author} {\bibfnamefont {B.}~\bibnamefont {Keimer}},\ }\href {\doibase 10.1103/PhysRevX.8.031014} {\bibfield  {journal} {\bibinfo  {journal} {Phys. Rev.
  X}\ }\textbf {\bibinfo {volume} {8}},\ \bibinfo {pages} {031014} (\bibinfo {year} {2018})}\BibitemShut {NoStop}%
\bibitem [{\citenamefont {F\"ursich}\ \emph {et~al.}(2019)\citenamefont {F\"ursich}, \citenamefont {Lu}, \citenamefont {Betto}, \citenamefont {Bluschke}, \citenamefont {Porras}, \citenamefont {Schierle}, \citenamefont {Ortiz}, \citenamefont {Suzuki}, \citenamefont {Cristiani}, \citenamefont {Logvenov}, \citenamefont {Brookes}, \citenamefont {Haverkort}, \citenamefont {Le~Tacon}, \citenamefont {Benckiser}, \citenamefont {Minola},\ and\ \citenamefont {Keimer}}]{2019PRBYiDoubleCluster}%
  \BibitemOpen
  \bibfield  {author} {\bibinfo {author} {\bibfnamefont {K.}~\bibnamefont {F\"ursich}}, \bibinfo {author} {\bibfnamefont {Y.}~\bibnamefont {Lu}}, \bibinfo {author} {\bibfnamefont {D.}~\bibnamefont {Betto}}, \bibinfo {author} {\bibfnamefont {M.}~\bibnamefont {Bluschke}}, \bibinfo {author} {\bibfnamefont {J.}~\bibnamefont {Porras}}, \bibinfo {author} {\bibfnamefont {E.}~\bibnamefont {Schierle}}, \bibinfo {author} {\bibfnamefont {R.}~\bibnamefont {Ortiz}}, \bibinfo {author} {\bibfnamefont {H.}~\bibnamefont {Suzuki}}, \bibinfo {author} {\bibfnamefont {G.}~\bibnamefont {Cristiani}}, \bibinfo {author} {\bibfnamefont {G.}~\bibnamefont {Logvenov}}, \bibinfo {author} {\bibfnamefont {N.~B.}\ \bibnamefont {Brookes}}, \bibinfo {author} {\bibfnamefont {M.~W.}\ \bibnamefont {Haverkort}}, \bibinfo {author} {\bibfnamefont {M.}~\bibnamefont {Le~Tacon}}, \bibinfo {author} {\bibfnamefont {E.}~\bibnamefont {Benckiser}}, \bibinfo {author} {\bibfnamefont {M.}~\bibnamefont {Minola}}, \ and\ \bibinfo {author} {\bibfnamefont
  {B.}~\bibnamefont {Keimer}},\ }\href {\doibase 10.1103/PhysRevB.99.165124} {\bibfield  {journal} {\bibinfo  {journal} {Phys. Rev. B}\ }\textbf {\bibinfo {volume} {99}},\ \bibinfo {pages} {165124} (\bibinfo {year} {2019})}\BibitemShut {NoStop}%
\bibitem [{\citenamefont {Bocquet}\ \emph {et~al.}(1992{\natexlab{a}})\citenamefont {Bocquet}, \citenamefont {Mizokawa}, \citenamefont {Saitoh}, \citenamefont {Namatame},\ and\ \citenamefont {Fujimori}}]{1992PRBBocquetXPS}%
  \BibitemOpen
  \bibfield  {author} {\bibinfo {author} {\bibfnamefont {A.~E.}\ \bibnamefont {Bocquet}}, \bibinfo {author} {\bibfnamefont {T.}~\bibnamefont {Mizokawa}}, \bibinfo {author} {\bibfnamefont {T.}~\bibnamefont {Saitoh}}, \bibinfo {author} {\bibfnamefont {H.}~\bibnamefont {Namatame}}, \ and\ \bibinfo {author} {\bibfnamefont {A.}~\bibnamefont {Fujimori}},\ }\href {\doibase 10.1103/PhysRevB.46.3771} {\bibfield  {journal} {\bibinfo  {journal} {Phys. Rev. B}\ }\textbf {\bibinfo {volume} {46}},\ \bibinfo {pages} {3771} (\bibinfo {year} {1992}{\natexlab{a}})}\BibitemShut {NoStop}%
\bibitem [{\citenamefont {Bocquet}\ \emph {et~al.}(1996{\natexlab{a}})\citenamefont {Bocquet}, \citenamefont {Mizokawa}, \citenamefont {Morikawa}, \citenamefont {Fujimori}, \citenamefont {Barman}, \citenamefont {Maiti}, \citenamefont {Sarma}, \citenamefont {Tokura},\ and\ \citenamefont {Onoda}}]{1996PRBBocquetXPS}%
  \BibitemOpen
  \bibfield  {author} {\bibinfo {author} {\bibfnamefont {A.~E.}\ \bibnamefont {Bocquet}}, \bibinfo {author} {\bibfnamefont {T.}~\bibnamefont {Mizokawa}}, \bibinfo {author} {\bibfnamefont {K.}~\bibnamefont {Morikawa}}, \bibinfo {author} {\bibfnamefont {A.}~\bibnamefont {Fujimori}}, \bibinfo {author} {\bibfnamefont {S.~R.}\ \bibnamefont {Barman}}, \bibinfo {author} {\bibfnamefont {K.}~\bibnamefont {Maiti}}, \bibinfo {author} {\bibfnamefont {D.~D.}\ \bibnamefont {Sarma}}, \bibinfo {author} {\bibfnamefont {Y.}~\bibnamefont {Tokura}}, \ and\ \bibinfo {author} {\bibfnamefont {M.}~\bibnamefont {Onoda}},\ }\href {\doibase 10.1103/PhysRevB.53.1161} {\bibfield  {journal} {\bibinfo  {journal} {Phys. Rev. B}\ }\textbf {\bibinfo {volume} {53}},\ \bibinfo {pages} {1161} (\bibinfo {year} {1996}{\natexlab{a}})}\BibitemShut {NoStop}%
\bibitem [{\citenamefont {Fujimori}\ \emph {et~al.}(1993)\citenamefont {Fujimori}, \citenamefont {Bocquet}, \citenamefont {Saitoh},\ and\ \citenamefont {Mizokawa}}]{Fujimori1993JELSPEC_Trends}%
  \BibitemOpen
  \bibfield  {author} {\bibinfo {author} {\bibfnamefont {A.}~\bibnamefont {Fujimori}}, \bibinfo {author} {\bibfnamefont {A.}~\bibnamefont {Bocquet}}, \bibinfo {author} {\bibfnamefont {T.}~\bibnamefont {Saitoh}}, \ and\ \bibinfo {author} {\bibfnamefont {T.}~\bibnamefont {Mizokawa}},\ }\href {\doibase https://doi.org/10.1016/0368-2048(93)80011-A} {\bibfield  {journal} {\bibinfo  {journal} {Journal of Electron Spectroscopy and Related Phenomena}\ }\textbf {\bibinfo {volume} {62}},\ \bibinfo {pages} {141} (\bibinfo {year} {1993})}\BibitemShut {NoStop}%
\bibitem [{\citenamefont {Mizokawa}\ \emph {et~al.}(1991)\citenamefont {Mizokawa}, \citenamefont {Namatame}, \citenamefont {Fujimori}, \citenamefont {Akeyama}, \citenamefont {Kondoh}, \citenamefont {Kuroda},\ and\ \citenamefont {Kosugi}}]{1991PRLMizokawaNaCuO2}%
  \BibitemOpen
  \bibfield  {author} {\bibinfo {author} {\bibfnamefont {T.}~\bibnamefont {Mizokawa}}, \bibinfo {author} {\bibfnamefont {H.}~\bibnamefont {Namatame}}, \bibinfo {author} {\bibfnamefont {A.}~\bibnamefont {Fujimori}}, \bibinfo {author} {\bibfnamefont {K.}~\bibnamefont {Akeyama}}, \bibinfo {author} {\bibfnamefont {H.}~\bibnamefont {Kondoh}}, \bibinfo {author} {\bibfnamefont {H.}~\bibnamefont {Kuroda}}, \ and\ \bibinfo {author} {\bibfnamefont {N.}~\bibnamefont {Kosugi}},\ }\href {\doibase 10.1103/PhysRevLett.67.1638} {\bibfield  {journal} {\bibinfo  {journal} {Phys. Rev. Lett.}\ }\textbf {\bibinfo {volume} {67}},\ \bibinfo {pages} {1638} (\bibinfo {year} {1991})}\BibitemShut {NoStop}%
\bibitem [{\citenamefont {Mizokawa}\ \emph {et~al.}(1994)\citenamefont {Mizokawa}, \citenamefont {Fujimori}, \citenamefont {Namatame}, \citenamefont {Akeyama},\ and\ \citenamefont {Kosugi}}]{1994PRBMizokawaNaCuO2}%
  \BibitemOpen
  \bibfield  {author} {\bibinfo {author} {\bibfnamefont {T.}~\bibnamefont {Mizokawa}}, \bibinfo {author} {\bibfnamefont {A.}~\bibnamefont {Fujimori}}, \bibinfo {author} {\bibfnamefont {H.}~\bibnamefont {Namatame}}, \bibinfo {author} {\bibfnamefont {K.}~\bibnamefont {Akeyama}}, \ and\ \bibinfo {author} {\bibfnamefont {N.}~\bibnamefont {Kosugi}},\ }\href {\doibase 10.1103/PhysRevB.49.7193} {\bibfield  {journal} {\bibinfo  {journal} {Phys. Rev. B}\ }\textbf {\bibinfo {volume} {49}},\ \bibinfo {pages} {7193} (\bibinfo {year} {1994})}\BibitemShut {NoStop}%
\bibitem [{\citenamefont {Okada}\ \emph {et~al.}(1991)\citenamefont {Okada}, \citenamefont {Kotani}, \citenamefont {Thole},\ and\ \citenamefont {Sawatzky}}]{1991SSCSawatzkyNaCuO2}%
  \BibitemOpen
  \bibfield  {author} {\bibinfo {author} {\bibfnamefont {K.}~\bibnamefont {Okada}}, \bibinfo {author} {\bibfnamefont {A.}~\bibnamefont {Kotani}}, \bibinfo {author} {\bibfnamefont {B.}~\bibnamefont {Thole}}, \ and\ \bibinfo {author} {\bibfnamefont {G.}~\bibnamefont {Sawatzky}},\ }\href {\doibase https://doi.org/10.1016/0038-1098(91)90815-D} {\bibfield  {journal} {\bibinfo  {journal} {Solid State Communications}\ }\textbf {\bibinfo {volume} {77}},\ \bibinfo {pages} {835} (\bibinfo {year} {1991})}\BibitemShut {NoStop}%
\bibitem [{\citenamefont {Nimkar}\ \emph {et~al.}(1993)\citenamefont {Nimkar}, \citenamefont {Sarma},\ and\ \citenamefont {Krishnamurthy}}]{1993PRBSarmaNaCuO2}%
  \BibitemOpen
  \bibfield  {author} {\bibinfo {author} {\bibfnamefont {S.}~\bibnamefont {Nimkar}}, \bibinfo {author} {\bibfnamefont {D.~D.}\ \bibnamefont {Sarma}}, \ and\ \bibinfo {author} {\bibfnamefont {H.~R.}\ \bibnamefont {Krishnamurthy}},\ }\href {\doibase 10.1103/PhysRevB.47.10927} {\bibfield  {journal} {\bibinfo  {journal} {Phys. Rev. B}\ }\textbf {\bibinfo {volume} {47}},\ \bibinfo {pages} {10927} (\bibinfo {year} {1993})}\BibitemShut {NoStop}%
\bibitem [{\citenamefont {Medarde}(1997)}]{Medarde_JPCM_1997}%
  \BibitemOpen
  \bibfield  {author} {\bibinfo {author} {\bibfnamefont {M.}~\bibnamefont {Medarde}},\ }\href {http://stacks.iop.org/0953-8984/9/i=8/a=003} {\bibfield  {journal} {\bibinfo  {journal} {Journal of Physics: Condensed Matter}\ }\textbf {\bibinfo {volume} {9}},\ \bibinfo {pages} {1679} (\bibinfo {year} {1997})}\BibitemShut {NoStop}%
\bibitem [{\citenamefont {Mizokawa}\ \emph {et~al.}(2000)\citenamefont {Mizokawa}, \citenamefont {Khomskii},\ and\ \citenamefont {Sawatzky}}]{Mizokawa_SDMI_PRB_2000}%
  \BibitemOpen
  \bibfield  {author} {\bibinfo {author} {\bibfnamefont {T.}~\bibnamefont {Mizokawa}}, \bibinfo {author} {\bibfnamefont {D.~I.}\ \bibnamefont {Khomskii}}, \ and\ \bibinfo {author} {\bibfnamefont {G.~A.}\ \bibnamefont {Sawatzky}},\ }\href {\doibase 10.1103/PhysRevB.61.11263} {\bibfield  {journal} {\bibinfo  {journal} {Phys. Rev. B}\ }\textbf {\bibinfo {volume} {61}},\ \bibinfo {pages} {11263} (\bibinfo {year} {2000})}\BibitemShut {NoStop}%
\bibitem [{\citenamefont {Lau}\ and\ \citenamefont {Millis}(2013)}]{LauMillis_PRL_2013}%
  \BibitemOpen
  \bibfield  {author} {\bibinfo {author} {\bibfnamefont {B.}~\bibnamefont {Lau}}\ and\ \bibinfo {author} {\bibfnamefont {A.~J.}\ \bibnamefont {Millis}},\ }\href {\doibase 10.1103/PhysRevLett.110.126404} {\bibfield  {journal} {\bibinfo  {journal} {Phys. Rev. Lett.}\ }\textbf {\bibinfo {volume} {110}},\ \bibinfo {pages} {126404} (\bibinfo {year} {2013})}\BibitemShut {NoStop}%
\bibitem [{\citenamefont {Johnston}\ \emph {et~al.}(2014)\citenamefont {Johnston}, \citenamefont {Mukherjee}, \citenamefont {Elfimov}, \citenamefont {Berciu},\ and\ \citenamefont {Sawatzky}}]{Johnston_PRL_2014}%
  \BibitemOpen
  \bibfield  {author} {\bibinfo {author} {\bibfnamefont {S.}~\bibnamefont {Johnston}}, \bibinfo {author} {\bibfnamefont {A.}~\bibnamefont {Mukherjee}}, \bibinfo {author} {\bibfnamefont {I.}~\bibnamefont {Elfimov}}, \bibinfo {author} {\bibfnamefont {M.}~\bibnamefont {Berciu}}, \ and\ \bibinfo {author} {\bibfnamefont {G.~A.}\ \bibnamefont {Sawatzky}},\ }\href {\doibase 10.1103/PhysRevLett.112.106404} {\bibfield  {journal} {\bibinfo  {journal} {Phys. Rev. Lett.}\ }\textbf {\bibinfo {volume} {112}},\ \bibinfo {pages} {106404} (\bibinfo {year} {2014})}\BibitemShut {NoStop}%
\bibitem [{\citenamefont {Catalano}\ \emph {et~al.}(2018)\citenamefont {Catalano}, \citenamefont {Gibert}, \citenamefont {Fowlie}, \citenamefont {Íñiguez}, \citenamefont {Triscone},\ and\ \citenamefont {Kreisel}}]{2018RPPCatalanoNickelates}%
  \BibitemOpen
  \bibfield  {author} {\bibinfo {author} {\bibfnamefont {S.}~\bibnamefont {Catalano}}, \bibinfo {author} {\bibfnamefont {M.}~\bibnamefont {Gibert}}, \bibinfo {author} {\bibfnamefont {J.}~\bibnamefont {Fowlie}}, \bibinfo {author} {\bibfnamefont {J.}~\bibnamefont {Íñiguez}}, \bibinfo {author} {\bibfnamefont {J.-M.}\ \bibnamefont {Triscone}}, \ and\ \bibinfo {author} {\bibfnamefont {J.}~\bibnamefont {Kreisel}},\ }\href {\doibase 10.1088/1361-6633/aaa37a} {\bibfield  {journal} {\bibinfo  {journal} {Reports on Progress in Physics}\ }\textbf {\bibinfo {volume} {81}},\ \bibinfo {pages} {046501} (\bibinfo {year} {2018})}\BibitemShut {NoStop}%
\bibitem [{\citenamefont {Kim}\ \emph {et~al.}(2020)\citenamefont {Kim}, \citenamefont {Paudel}, \citenamefont {Green}, \citenamefont {Song}, \citenamefont {Lee}, \citenamefont {Choi}, \citenamefont {Irwin}, \citenamefont {Noesges}, \citenamefont {Brillson}, \citenamefont {Rzchowski}, \citenamefont {Sawatzky}, \citenamefont {Tsymbal},\ and\ \citenamefont {Eom}}]{2020PRBKimNickelates}%
  \BibitemOpen
  \bibfield  {author} {\bibinfo {author} {\bibfnamefont {T.~H.}\ \bibnamefont {Kim}}, \bibinfo {author} {\bibfnamefont {T.~R.}\ \bibnamefont {Paudel}}, \bibinfo {author} {\bibfnamefont {R.~J.}\ \bibnamefont {Green}}, \bibinfo {author} {\bibfnamefont {K.}~\bibnamefont {Song}}, \bibinfo {author} {\bibfnamefont {H.-S.}\ \bibnamefont {Lee}}, \bibinfo {author} {\bibfnamefont {S.-Y.}\ \bibnamefont {Choi}}, \bibinfo {author} {\bibfnamefont {J.}~\bibnamefont {Irwin}}, \bibinfo {author} {\bibfnamefont {B.}~\bibnamefont {Noesges}}, \bibinfo {author} {\bibfnamefont {L.~J.}\ \bibnamefont {Brillson}}, \bibinfo {author} {\bibfnamefont {M.~S.}\ \bibnamefont {Rzchowski}}, \bibinfo {author} {\bibfnamefont {G.~A.}\ \bibnamefont {Sawatzky}}, \bibinfo {author} {\bibfnamefont {E.~Y.}\ \bibnamefont {Tsymbal}}, \ and\ \bibinfo {author} {\bibfnamefont {C.~B.}\ \bibnamefont {Eom}},\ }\href {\doibase 10.1103/PhysRevB.101.121105} {\bibfield  {journal} {\bibinfo  {journal} {Phys. Rev. B}\ }\textbf {\bibinfo {volume} {101}},\ \bibinfo
  {pages} {121105} (\bibinfo {year} {2020})}\BibitemShut {NoStop}%
\bibitem [{\citenamefont {Benckiser}\ \emph {et~al.}(2011)\citenamefont {Benckiser}, \citenamefont {Haverkort}, \citenamefont {Brueck}, \citenamefont {Goering}, \citenamefont {Macke}, \citenamefont {Frano}, \citenamefont {Yang}, \citenamefont {Andersen}, \citenamefont {Cristiani}, \citenamefont {Habermeier}, \citenamefont {Boris}, \citenamefont {Zegkinoglou}, \citenamefont {Wochner}, \citenamefont {Kim}, \citenamefont {Hinkov},\ and\ \citenamefont {Keimer}}]{2011NatMatBenckiser}%
  \BibitemOpen
  \bibfield  {author} {\bibinfo {author} {\bibfnamefont {E.}~\bibnamefont {Benckiser}}, \bibinfo {author} {\bibfnamefont {M.~W.}\ \bibnamefont {Haverkort}}, \bibinfo {author} {\bibfnamefont {S.}~\bibnamefont {Brueck}}, \bibinfo {author} {\bibfnamefont {E.}~\bibnamefont {Goering}}, \bibinfo {author} {\bibfnamefont {S.}~\bibnamefont {Macke}}, \bibinfo {author} {\bibfnamefont {A.}~\bibnamefont {Frano}}, \bibinfo {author} {\bibfnamefont {X.}~\bibnamefont {Yang}}, \bibinfo {author} {\bibfnamefont {O.~K.}\ \bibnamefont {Andersen}}, \bibinfo {author} {\bibfnamefont {G.}~\bibnamefont {Cristiani}}, \bibinfo {author} {\bibfnamefont {H.-U.}\ \bibnamefont {Habermeier}}, \bibinfo {author} {\bibfnamefont {A.~V.}\ \bibnamefont {Boris}}, \bibinfo {author} {\bibfnamefont {I.}~\bibnamefont {Zegkinoglou}}, \bibinfo {author} {\bibfnamefont {P.}~\bibnamefont {Wochner}}, \bibinfo {author} {\bibfnamefont {H.-J.}\ \bibnamefont {Kim}}, \bibinfo {author} {\bibfnamefont {V.}~\bibnamefont {Hinkov}}, \ and\ \bibinfo {author}
  {\bibfnamefont {B.}~\bibnamefont {Keimer}},\ }\href {\doibase 10.1038/NMAT2958} {\bibfield  {journal} {\bibinfo  {journal} {Nature Materials}\ }\textbf {\bibinfo {volume} {10}},\ \bibinfo {pages} {189} (\bibinfo {year} {2011})}\BibitemShut {NoStop}%
\bibitem [{\citenamefont {Liu}\ \emph {et~al.}(2013)\citenamefont {Liu}, \citenamefont {Kargarian}, \citenamefont {Kareev}, \citenamefont {Gray}, \citenamefont {Ryan}, \citenamefont {Cruz}, \citenamefont {Tahir}, \citenamefont {Chuang}, \citenamefont {Guo}, \citenamefont {Rondinelli}, \citenamefont {Freeland}, \citenamefont {Fiete},\ and\ \citenamefont {Chakhalian}}]{2013NatCommChakhalian}%
  \BibitemOpen
  \bibfield  {author} {\bibinfo {author} {\bibfnamefont {J.}~\bibnamefont {Liu}}, \bibinfo {author} {\bibfnamefont {M.}~\bibnamefont {Kargarian}}, \bibinfo {author} {\bibfnamefont {M.}~\bibnamefont {Kareev}}, \bibinfo {author} {\bibfnamefont {B.}~\bibnamefont {Gray}}, \bibinfo {author} {\bibfnamefont {P.~J.}\ \bibnamefont {Ryan}}, \bibinfo {author} {\bibfnamefont {A.}~\bibnamefont {Cruz}}, \bibinfo {author} {\bibfnamefont {N.}~\bibnamefont {Tahir}}, \bibinfo {author} {\bibfnamefont {Y.-D.}\ \bibnamefont {Chuang}}, \bibinfo {author} {\bibfnamefont {J.}~\bibnamefont {Guo}}, \bibinfo {author} {\bibfnamefont {J.~M.}\ \bibnamefont {Rondinelli}}, \bibinfo {author} {\bibfnamefont {J.~W.}\ \bibnamefont {Freeland}}, \bibinfo {author} {\bibfnamefont {G.~A.}\ \bibnamefont {Fiete}}, \ and\ \bibinfo {author} {\bibfnamefont {J.}~\bibnamefont {Chakhalian}},\ }\href {\doibase 10.1038/ncomms3714} {\bibfield  {journal} {\bibinfo  {journal} {Nature Communications}\ }\textbf {\bibinfo {volume} {4}},\ \bibinfo {pages} {2714}
  (\bibinfo {year} {2013})}\BibitemShut {NoStop}%
\bibitem [{\citenamefont {Wu}\ \emph {et~al.}(2013)\citenamefont {Wu}, \citenamefont {Benckiser}, \citenamefont {Haverkort}, \citenamefont {Frano}, \citenamefont {Lu}, \citenamefont {Nwankwo}, \citenamefont {Br\"uck}, \citenamefont {Audehm}, \citenamefont {Goering}, \citenamefont {Macke}, \citenamefont {Hinkov}, \citenamefont {Wochner}, \citenamefont {Christiani}, \citenamefont {Heinze}, \citenamefont {Logvenov}, \citenamefont {Habermeier},\ and\ \citenamefont {Keimer}}]{2013PRBBenckiserNickelates}%
  \BibitemOpen
  \bibfield  {author} {\bibinfo {author} {\bibfnamefont {M.}~\bibnamefont {Wu}}, \bibinfo {author} {\bibfnamefont {E.}~\bibnamefont {Benckiser}}, \bibinfo {author} {\bibfnamefont {M.~W.}\ \bibnamefont {Haverkort}}, \bibinfo {author} {\bibfnamefont {A.}~\bibnamefont {Frano}}, \bibinfo {author} {\bibfnamefont {Y.}~\bibnamefont {Lu}}, \bibinfo {author} {\bibfnamefont {U.}~\bibnamefont {Nwankwo}}, \bibinfo {author} {\bibfnamefont {S.}~\bibnamefont {Br\"uck}}, \bibinfo {author} {\bibfnamefont {P.}~\bibnamefont {Audehm}}, \bibinfo {author} {\bibfnamefont {E.}~\bibnamefont {Goering}}, \bibinfo {author} {\bibfnamefont {S.}~\bibnamefont {Macke}}, \bibinfo {author} {\bibfnamefont {V.}~\bibnamefont {Hinkov}}, \bibinfo {author} {\bibfnamefont {P.}~\bibnamefont {Wochner}}, \bibinfo {author} {\bibfnamefont {G.}~\bibnamefont {Christiani}}, \bibinfo {author} {\bibfnamefont {S.}~\bibnamefont {Heinze}}, \bibinfo {author} {\bibfnamefont {G.}~\bibnamefont {Logvenov}}, \bibinfo {author} {\bibfnamefont {H.-U.}\ \bibnamefont
  {Habermeier}}, \ and\ \bibinfo {author} {\bibfnamefont {B.}~\bibnamefont {Keimer}},\ }\href {\doibase 10.1103/PhysRevB.88.125124} {\bibfield  {journal} {\bibinfo  {journal} {Phys. Rev. B}\ }\textbf {\bibinfo {volume} {88}},\ \bibinfo {pages} {125124} (\bibinfo {year} {2013})}\BibitemShut {NoStop}%
\bibitem [{\citenamefont {Frano}\ \emph {et~al.}(2013)\citenamefont {Frano}, \citenamefont {Schierle}, \citenamefont {Haverkort}, \citenamefont {Lu}, \citenamefont {Wu}, \citenamefont {Blanco-Canosa}, \citenamefont {Nwankwo}, \citenamefont {Boris}, \citenamefont {Wochner}, \citenamefont {Cristiani}, \citenamefont {Habermeier}, \citenamefont {Logvenov}, \citenamefont {Hinkov}, \citenamefont {Benckiser}, \citenamefont {Weschke},\ and\ \citenamefont {Keimer}}]{2013PRLFrano}%
  \BibitemOpen
  \bibfield  {author} {\bibinfo {author} {\bibfnamefont {A.}~\bibnamefont {Frano}}, \bibinfo {author} {\bibfnamefont {E.}~\bibnamefont {Schierle}}, \bibinfo {author} {\bibfnamefont {M.~W.}\ \bibnamefont {Haverkort}}, \bibinfo {author} {\bibfnamefont {Y.}~\bibnamefont {Lu}}, \bibinfo {author} {\bibfnamefont {M.}~\bibnamefont {Wu}}, \bibinfo {author} {\bibfnamefont {S.}~\bibnamefont {Blanco-Canosa}}, \bibinfo {author} {\bibfnamefont {U.}~\bibnamefont {Nwankwo}}, \bibinfo {author} {\bibfnamefont {A.~V.}\ \bibnamefont {Boris}}, \bibinfo {author} {\bibfnamefont {P.}~\bibnamefont {Wochner}}, \bibinfo {author} {\bibfnamefont {G.}~\bibnamefont {Cristiani}}, \bibinfo {author} {\bibfnamefont {H.~U.}\ \bibnamefont {Habermeier}}, \bibinfo {author} {\bibfnamefont {G.}~\bibnamefont {Logvenov}}, \bibinfo {author} {\bibfnamefont {V.}~\bibnamefont {Hinkov}}, \bibinfo {author} {\bibfnamefont {E.}~\bibnamefont {Benckiser}}, \bibinfo {author} {\bibfnamefont {E.}~\bibnamefont {Weschke}}, \ and\ \bibinfo {author} {\bibfnamefont
  {B.}~\bibnamefont {Keimer}},\ }\href {\doibase 10.1103/PhysRevLett.111.106804} {\bibfield  {journal} {\bibinfo  {journal} {Phys. Rev. Lett.}\ }\textbf {\bibinfo {volume} {111}},\ \bibinfo {pages} {106804} (\bibinfo {year} {2013})}\BibitemShut {NoStop}%
\bibitem [{\citenamefont {Hepting}\ \emph {et~al.}(2018)\citenamefont {Hepting}, \citenamefont {Green}, \citenamefont {Zhong}, \citenamefont {Bluschke}, \citenamefont {Suyolcu}, \citenamefont {Macke}, \citenamefont {Frano}, \citenamefont {Catalano}, \citenamefont {Gibert}, \citenamefont {Sutarto}, \citenamefont {He}, \citenamefont {Cristiani}, \citenamefont {Logvenov}, \citenamefont {Wang}, \citenamefont {van Aken}, \citenamefont {Hansmann}, \citenamefont {Le~Tacon}, \citenamefont {Triscone}, \citenamefont {Sawatzky}, \citenamefont {Keimer},\ and\ \citenamefont {Benckiser}}]{2018NatPhysHeptingNNO}%
  \BibitemOpen
  \bibfield  {author} {\bibinfo {author} {\bibfnamefont {M.}~\bibnamefont {Hepting}}, \bibinfo {author} {\bibfnamefont {R.~J.}\ \bibnamefont {Green}}, \bibinfo {author} {\bibfnamefont {Z.}~\bibnamefont {Zhong}}, \bibinfo {author} {\bibfnamefont {M.}~\bibnamefont {Bluschke}}, \bibinfo {author} {\bibfnamefont {Y.~E.}\ \bibnamefont {Suyolcu}}, \bibinfo {author} {\bibfnamefont {S.}~\bibnamefont {Macke}}, \bibinfo {author} {\bibfnamefont {A.}~\bibnamefont {Frano}}, \bibinfo {author} {\bibfnamefont {S.}~\bibnamefont {Catalano}}, \bibinfo {author} {\bibfnamefont {M.}~\bibnamefont {Gibert}}, \bibinfo {author} {\bibfnamefont {R.}~\bibnamefont {Sutarto}}, \bibinfo {author} {\bibfnamefont {F.}~\bibnamefont {He}}, \bibinfo {author} {\bibfnamefont {G.}~\bibnamefont {Cristiani}}, \bibinfo {author} {\bibfnamefont {G.}~\bibnamefont {Logvenov}}, \bibinfo {author} {\bibfnamefont {Y.}~\bibnamefont {Wang}}, \bibinfo {author} {\bibfnamefont {P.~A.}\ \bibnamefont {van Aken}}, \bibinfo {author} {\bibfnamefont {P.}~\bibnamefont
  {Hansmann}}, \bibinfo {author} {\bibfnamefont {M.}~\bibnamefont {Le~Tacon}}, \bibinfo {author} {\bibfnamefont {J.~M.}\ \bibnamefont {Triscone}}, \bibinfo {author} {\bibfnamefont {G.~A.}\ \bibnamefont {Sawatzky}}, \bibinfo {author} {\bibfnamefont {B.}~\bibnamefont {Keimer}}, \ and\ \bibinfo {author} {\bibfnamefont {E.}~\bibnamefont {Benckiser}},\ }\href {\doibase 10.1038/s41567-018-0218-5} {\bibfield  {journal} {\bibinfo  {journal} {Nature Physics}\ }\textbf {\bibinfo {volume} {14}},\ \bibinfo {pages} {1097+} (\bibinfo {year} {2018})}\BibitemShut {NoStop}%
\bibitem [{\citenamefont {Liao}\ \emph {et~al.}(2018)\citenamefont {Liao}, \citenamefont {Gauquelin}, \citenamefont {Green}, \citenamefont {M{\"u}ller-Caspary}, \citenamefont {Lobato}, \citenamefont {Li}, \citenamefont {Van~Aert}, \citenamefont {Verbeeck}, \citenamefont {Huijben}, \citenamefont {Grisolia}, \citenamefont {Rouco}, \citenamefont {Hage}, \citenamefont {Villegas}, \citenamefont {Mercy}, \citenamefont {Bibes}, \citenamefont {Ghosez}, \citenamefont {Sawatzky}, \citenamefont {Rijnders},\ and\ \citenamefont {Koster}}]{2018PNASLiaoNickelates}%
  \BibitemOpen
  \bibfield  {author} {\bibinfo {author} {\bibfnamefont {Z.}~\bibnamefont {Liao}}, \bibinfo {author} {\bibfnamefont {N.}~\bibnamefont {Gauquelin}}, \bibinfo {author} {\bibfnamefont {R.~J.}\ \bibnamefont {Green}}, \bibinfo {author} {\bibfnamefont {K.}~\bibnamefont {M{\"u}ller-Caspary}}, \bibinfo {author} {\bibfnamefont {I.}~\bibnamefont {Lobato}}, \bibinfo {author} {\bibfnamefont {L.}~\bibnamefont {Li}}, \bibinfo {author} {\bibfnamefont {S.}~\bibnamefont {Van~Aert}}, \bibinfo {author} {\bibfnamefont {J.}~\bibnamefont {Verbeeck}}, \bibinfo {author} {\bibfnamefont {M.}~\bibnamefont {Huijben}}, \bibinfo {author} {\bibfnamefont {M.~N.}\ \bibnamefont {Grisolia}}, \bibinfo {author} {\bibfnamefont {V.}~\bibnamefont {Rouco}}, \bibinfo {author} {\bibfnamefont {R.~E.}\ \bibnamefont {Hage}}, \bibinfo {author} {\bibfnamefont {J.~E.}\ \bibnamefont {Villegas}}, \bibinfo {author} {\bibfnamefont {A.}~\bibnamefont {Mercy}}, \bibinfo {author} {\bibfnamefont {M.}~\bibnamefont {Bibes}}, \bibinfo {author} {\bibfnamefont
  {P.}~\bibnamefont {Ghosez}}, \bibinfo {author} {\bibfnamefont {G.~A.}\ \bibnamefont {Sawatzky}}, \bibinfo {author} {\bibfnamefont {G.}~\bibnamefont {Rijnders}}, \ and\ \bibinfo {author} {\bibfnamefont {G.}~\bibnamefont {Koster}},\ }\href {\doibase 10.1073/pnas.1807457115} {\bibfield  {journal} {\bibinfo  {journal} {Proceedings of the National Academy of Sciences}\ }\textbf {\bibinfo {volume} {115}},\ \bibinfo {pages} {9515} (\bibinfo {year} {2018})}\BibitemShut {NoStop}%
\bibitem [{\citenamefont {Li}\ \emph {et~al.}(2021)\citenamefont {Li}, \citenamefont {Green}, \citenamefont {Zhang}, \citenamefont {Sutarto}, \citenamefont {Sadowski}, \citenamefont {Zhu}, \citenamefont {Zhang}, \citenamefont {Zhou}, \citenamefont {Sun}, \citenamefont {He}, \citenamefont {Ramanathan},\ and\ \citenamefont {Comin}}]{2021PRLJiaruiNickelates}%
  \BibitemOpen
  \bibfield  {author} {\bibinfo {author} {\bibfnamefont {J.}~\bibnamefont {Li}}, \bibinfo {author} {\bibfnamefont {R.~J.}\ \bibnamefont {Green}}, \bibinfo {author} {\bibfnamefont {Z.}~\bibnamefont {Zhang}}, \bibinfo {author} {\bibfnamefont {R.}~\bibnamefont {Sutarto}}, \bibinfo {author} {\bibfnamefont {J.~T.}\ \bibnamefont {Sadowski}}, \bibinfo {author} {\bibfnamefont {Z.}~\bibnamefont {Zhu}}, \bibinfo {author} {\bibfnamefont {G.}~\bibnamefont {Zhang}}, \bibinfo {author} {\bibfnamefont {D.}~\bibnamefont {Zhou}}, \bibinfo {author} {\bibfnamefont {Y.}~\bibnamefont {Sun}}, \bibinfo {author} {\bibfnamefont {F.}~\bibnamefont {He}}, \bibinfo {author} {\bibfnamefont {S.}~\bibnamefont {Ramanathan}}, \ and\ \bibinfo {author} {\bibfnamefont {R.}~\bibnamefont {Comin}},\ }\href {\doibase 10.1103/PhysRevLett.126.187602} {\bibfield  {journal} {\bibinfo  {journal} {Phys. Rev. Lett.}\ }\textbf {\bibinfo {volume} {126}},\ \bibinfo {pages} {187602} (\bibinfo {year} {2021})}\BibitemShut {NoStop}%
\bibitem [{\citenamefont {Patel}\ \emph {et~al.}(2022)\citenamefont {Patel}, \citenamefont {Patra}, \citenamefont {Ojha}, \citenamefont {Kumar}, \citenamefont {Sarkar}, \citenamefont {Saha}, \citenamefont {Bhattacharya}, \citenamefont {Freeland}, \citenamefont {Kim}, \citenamefont {Ryan},\ and\ \citenamefont {Middey}}]{2022CommPhysMiddeyHoleDoping}%
  \BibitemOpen
  \bibfield  {author} {\bibinfo {author} {\bibfnamefont {R.~K.}\ \bibnamefont {Patel}}, \bibinfo {author} {\bibfnamefont {K.}~\bibnamefont {Patra}}, \bibinfo {author} {\bibfnamefont {S.~K.}\ \bibnamefont {Ojha}}, \bibinfo {author} {\bibfnamefont {S.}~\bibnamefont {Kumar}}, \bibinfo {author} {\bibfnamefont {S.}~\bibnamefont {Sarkar}}, \bibinfo {author} {\bibfnamefont {A.}~\bibnamefont {Saha}}, \bibinfo {author} {\bibfnamefont {N.}~\bibnamefont {Bhattacharya}}, \bibinfo {author} {\bibfnamefont {J.~W.}\ \bibnamefont {Freeland}}, \bibinfo {author} {\bibfnamefont {J.-W.}\ \bibnamefont {Kim}}, \bibinfo {author} {\bibfnamefont {P.}~\bibnamefont {Ryan}, \bibfnamefont {Philip J~Mahadevan}}, \ and\ \bibinfo {author} {\bibfnamefont {S.}~\bibnamefont {Middey}},\ }\href {\doibase 10.1038/s42005-022-00993-1} {\bibfield  {journal} {\bibinfo  {journal} {Communications Physics}\ }\textbf {\bibinfo {volume} {5}},\ \bibinfo {pages} {216} (\bibinfo {year} {2022})}\BibitemShut {NoStop}%
\bibitem [{\citenamefont {Bocquet}\ \emph {et~al.}(1992{\natexlab{b}})\citenamefont {Bocquet}, \citenamefont {Fujimori}, \citenamefont {Mizokawa}, \citenamefont {Saitoh}, \citenamefont {Namatame}, \citenamefont {Suga}, \citenamefont {Kimizuka}, \citenamefont {Takeda},\ and\ \citenamefont {Takano}}]{1992PRBBocquetSFO}%
  \BibitemOpen
  \bibfield  {author} {\bibinfo {author} {\bibfnamefont {A.~E.}\ \bibnamefont {Bocquet}}, \bibinfo {author} {\bibfnamefont {A.}~\bibnamefont {Fujimori}}, \bibinfo {author} {\bibfnamefont {T.}~\bibnamefont {Mizokawa}}, \bibinfo {author} {\bibfnamefont {T.}~\bibnamefont {Saitoh}}, \bibinfo {author} {\bibfnamefont {H.}~\bibnamefont {Namatame}}, \bibinfo {author} {\bibfnamefont {S.}~\bibnamefont {Suga}}, \bibinfo {author} {\bibfnamefont {N.}~\bibnamefont {Kimizuka}}, \bibinfo {author} {\bibfnamefont {Y.}~\bibnamefont {Takeda}}, \ and\ \bibinfo {author} {\bibfnamefont {M.}~\bibnamefont {Takano}},\ }\href {\doibase 10.1103/PhysRevB.45.1561} {\bibfield  {journal} {\bibinfo  {journal} {Phys. Rev. B}\ }\textbf {\bibinfo {volume} {45}},\ \bibinfo {pages} {1561} (\bibinfo {year} {1992}{\natexlab{b}})}\BibitemShut {NoStop}%
\bibitem [{\citenamefont {Matsuno}\ \emph {et~al.}(2002)\citenamefont {Matsuno}, \citenamefont {Mizokawa}, \citenamefont {Fujimori}, \citenamefont {Takeda}, \citenamefont {Kawasaki},\ and\ \citenamefont {Takano}}]{2002PRBFujimoriFerrates}%
  \BibitemOpen
  \bibfield  {author} {\bibinfo {author} {\bibfnamefont {J.}~\bibnamefont {Matsuno}}, \bibinfo {author} {\bibfnamefont {T.}~\bibnamefont {Mizokawa}}, \bibinfo {author} {\bibfnamefont {A.}~\bibnamefont {Fujimori}}, \bibinfo {author} {\bibfnamefont {Y.}~\bibnamefont {Takeda}}, \bibinfo {author} {\bibfnamefont {S.}~\bibnamefont {Kawasaki}}, \ and\ \bibinfo {author} {\bibfnamefont {M.}~\bibnamefont {Takano}},\ }\href {\doibase 10.1103/PhysRevB.66.193103} {\bibfield  {journal} {\bibinfo  {journal} {Phys. Rev. B}\ }\textbf {\bibinfo {volume} {66}},\ \bibinfo {pages} {193103} (\bibinfo {year} {2002})}\BibitemShut {NoStop}%
\bibitem [{\citenamefont {Abbate}\ \emph {et~al.}(2002{\natexlab{a}})\citenamefont {Abbate}, \citenamefont {Zampieri}, \citenamefont {Okamoto}, \citenamefont {Fujimori}, \citenamefont {Kawasaki},\ and\ \citenamefont {Takano}}]{2002PRBAbbateFerrates}%
  \BibitemOpen
  \bibfield  {author} {\bibinfo {author} {\bibfnamefont {M.}~\bibnamefont {Abbate}}, \bibinfo {author} {\bibfnamefont {G.}~\bibnamefont {Zampieri}}, \bibinfo {author} {\bibfnamefont {J.}~\bibnamefont {Okamoto}}, \bibinfo {author} {\bibfnamefont {A.}~\bibnamefont {Fujimori}}, \bibinfo {author} {\bibfnamefont {S.}~\bibnamefont {Kawasaki}}, \ and\ \bibinfo {author} {\bibfnamefont {M.}~\bibnamefont {Takano}},\ }\href {\doibase 10.1103/PhysRevB.65.165120} {\bibfield  {journal} {\bibinfo  {journal} {Phys. Rev. B}\ }\textbf {\bibinfo {volume} {65}},\ \bibinfo {pages} {165120} (\bibinfo {year} {2002}{\natexlab{a}})}\BibitemShut {NoStop}%
\bibitem [{\citenamefont {Rogge}\ \emph {et~al.}(2018{\natexlab{a}})\citenamefont {Rogge}, \citenamefont {Chandrasena}, \citenamefont {Cammarata}, \citenamefont {Green}, \citenamefont {Shafer}, \citenamefont {Lefler}, \citenamefont {Huon}, \citenamefont {Arab}, \citenamefont {Arenholz}, \citenamefont {Lee}, \citenamefont {Lee}, \citenamefont {Nem\ifmmode~\check{s}\else \v{s}\fi{}\'ak}, \citenamefont {Rondinelli}, \citenamefont {Gray},\ and\ \citenamefont {May}}]{2018PRMRoggeCaFeO3}%
  \BibitemOpen
  \bibfield  {author} {\bibinfo {author} {\bibfnamefont {P.~C.}\ \bibnamefont {Rogge}}, \bibinfo {author} {\bibfnamefont {R.~U.}\ \bibnamefont {Chandrasena}}, \bibinfo {author} {\bibfnamefont {A.}~\bibnamefont {Cammarata}}, \bibinfo {author} {\bibfnamefont {R.~J.}\ \bibnamefont {Green}}, \bibinfo {author} {\bibfnamefont {P.}~\bibnamefont {Shafer}}, \bibinfo {author} {\bibfnamefont {B.~M.}\ \bibnamefont {Lefler}}, \bibinfo {author} {\bibfnamefont {A.}~\bibnamefont {Huon}}, \bibinfo {author} {\bibfnamefont {A.}~\bibnamefont {Arab}}, \bibinfo {author} {\bibfnamefont {E.}~\bibnamefont {Arenholz}}, \bibinfo {author} {\bibfnamefont {H.~N.}\ \bibnamefont {Lee}}, \bibinfo {author} {\bibfnamefont {T.-L.}\ \bibnamefont {Lee}}, \bibinfo {author} {\bibfnamefont {S.}~\bibnamefont {Nem\ifmmode~\check{s}\else \v{s}\fi{}\'ak}}, \bibinfo {author} {\bibfnamefont {J.~M.}\ \bibnamefont {Rondinelli}}, \bibinfo {author} {\bibfnamefont {A.~X.}\ \bibnamefont {Gray}}, \ and\ \bibinfo {author} {\bibfnamefont {S.~J.}\ \bibnamefont
  {May}},\ }\href {\doibase 10.1103/PhysRevMaterials.2.015002} {\bibfield  {journal} {\bibinfo  {journal} {Phys. Rev. Mater.}\ }\textbf {\bibinfo {volume} {2}},\ \bibinfo {pages} {015002} (\bibinfo {year} {2018}{\natexlab{a}})}\BibitemShut {NoStop}%
\bibitem [{\citenamefont {Rogge}\ \emph {et~al.}(2018{\natexlab{b}})\citenamefont {Rogge}, \citenamefont {Green}, \citenamefont {Shafer}, \citenamefont {Fabbris}, \citenamefont {Barbour}, \citenamefont {Lefler}, \citenamefont {Arenholz}, \citenamefont {Dean},\ and\ \citenamefont {May}}]{2018PRBRoggeCaFeO3Inv}%
  \BibitemOpen
  \bibfield  {author} {\bibinfo {author} {\bibfnamefont {P.~C.}\ \bibnamefont {Rogge}}, \bibinfo {author} {\bibfnamefont {R.~J.}\ \bibnamefont {Green}}, \bibinfo {author} {\bibfnamefont {P.}~\bibnamefont {Shafer}}, \bibinfo {author} {\bibfnamefont {G.}~\bibnamefont {Fabbris}}, \bibinfo {author} {\bibfnamefont {A.~M.}\ \bibnamefont {Barbour}}, \bibinfo {author} {\bibfnamefont {B.~M.}\ \bibnamefont {Lefler}}, \bibinfo {author} {\bibfnamefont {E.}~\bibnamefont {Arenholz}}, \bibinfo {author} {\bibfnamefont {M.~P.~M.}\ \bibnamefont {Dean}}, \ and\ \bibinfo {author} {\bibfnamefont {S.~J.}\ \bibnamefont {May}},\ }\href {\doibase 10.1103/PhysRevB.98.201115} {\bibfield  {journal} {\bibinfo  {journal} {Phys. Rev. B}\ }\textbf {\bibinfo {volume} {98}},\ \bibinfo {pages} {201115} (\bibinfo {year} {2018}{\natexlab{b}})}\BibitemShut {NoStop}%
\bibitem [{\citenamefont {Kawasaki}\ \emph {et~al.}(1998)\citenamefont {Kawasaki}, \citenamefont {Takano}, \citenamefont {Kanno}, \citenamefont {Takeda},\ and\ \citenamefont {Fujimori}}]{1998JPSJFujimorFerrates}%
  \BibitemOpen
  \bibfield  {author} {\bibinfo {author} {\bibfnamefont {S.}~\bibnamefont {Kawasaki}}, \bibinfo {author} {\bibfnamefont {M.}~\bibnamefont {Takano}}, \bibinfo {author} {\bibfnamefont {R.}~\bibnamefont {Kanno}}, \bibinfo {author} {\bibfnamefont {T.}~\bibnamefont {Takeda}}, \ and\ \bibinfo {author} {\bibfnamefont {A.}~\bibnamefont {Fujimori}},\ }\href {\doibase 10.1143/JPSJ.67.1529} {\bibfield  {journal} {\bibinfo  {journal} {Journal of the Physical Society of Japan}\ }\textbf {\bibinfo {volume} {67}},\ \bibinfo {pages} {1529} (\bibinfo {year} {1998})}\BibitemShut {NoStop}%
\bibitem [{\citenamefont {Takeda}\ \emph {et~al.}(1972)\citenamefont {Takeda}, \citenamefont {Yamaguchi},\ and\ \citenamefont {Watanabe}}]{1972JPSJTakedaSFO}%
  \BibitemOpen
  \bibfield  {author} {\bibinfo {author} {\bibfnamefont {T.}~\bibnamefont {Takeda}}, \bibinfo {author} {\bibfnamefont {Y.}~\bibnamefont {Yamaguchi}}, \ and\ \bibinfo {author} {\bibfnamefont {H.}~\bibnamefont {Watanabe}},\ }\href {\doibase 10.1143/JPSJ.33.967} {\bibfield  {journal} {\bibinfo  {journal} {Journal of the Physical Society of Japan}\ }\textbf {\bibinfo {volume} {33}},\ \bibinfo {pages} {967} (\bibinfo {year} {1972})}\BibitemShut {NoStop}%
\bibitem [{\citenamefont {Woodward}\ \emph {et~al.}(2000)\citenamefont {Woodward}, \citenamefont {Cox}, \citenamefont {Moshopoulou}, \citenamefont {Sleight},\ and\ \citenamefont {Morimoto}}]{2000PRBWoowardCFO}%
  \BibitemOpen
  \bibfield  {author} {\bibinfo {author} {\bibfnamefont {P.~M.}\ \bibnamefont {Woodward}}, \bibinfo {author} {\bibfnamefont {D.~E.}\ \bibnamefont {Cox}}, \bibinfo {author} {\bibfnamefont {E.}~\bibnamefont {Moshopoulou}}, \bibinfo {author} {\bibfnamefont {A.~W.}\ \bibnamefont {Sleight}}, \ and\ \bibinfo {author} {\bibfnamefont {S.}~\bibnamefont {Morimoto}},\ }\href {\doibase 10.1103/PhysRevB.62.844} {\bibfield  {journal} {\bibinfo  {journal} {Phys. Rev. B}\ }\textbf {\bibinfo {volume} {62}},\ \bibinfo {pages} {844} (\bibinfo {year} {2000})}\BibitemShut {NoStop}%
\bibitem [{\citenamefont {Rogge}\ \emph {et~al.}(2019)\citenamefont {Rogge}, \citenamefont {Green}, \citenamefont {Sutarto},\ and\ \citenamefont {May}}]{2019PRMRoggeSFO}%
  \BibitemOpen
  \bibfield  {author} {\bibinfo {author} {\bibfnamefont {P.~C.}\ \bibnamefont {Rogge}}, \bibinfo {author} {\bibfnamefont {R.~J.}\ \bibnamefont {Green}}, \bibinfo {author} {\bibfnamefont {R.}~\bibnamefont {Sutarto}}, \ and\ \bibinfo {author} {\bibfnamefont {S.~J.}\ \bibnamefont {May}},\ }\href {\doibase 10.1103/PhysRevMaterials.3.084404} {\bibfield  {journal} {\bibinfo  {journal} {Phys. Rev. Mater.}\ }\textbf {\bibinfo {volume} {3}},\ \bibinfo {pages} {084404} (\bibinfo {year} {2019})}\BibitemShut {NoStop}%
\bibitem [{\citenamefont {Ishiwata}\ \emph {et~al.}(2020)\citenamefont {Ishiwata}, \citenamefont {Nakajima}, \citenamefont {Kim}, \citenamefont {Inosov}, \citenamefont {Kanazawa}, \citenamefont {White}, \citenamefont {Gavilano}, \citenamefont {Georgii}, \citenamefont {Seemann}, \citenamefont {Brandl}, \citenamefont {Manuel}, \citenamefont {Khalyavin}, \citenamefont {Seki}, \citenamefont {Tokunaga}, \citenamefont {Kinoshita}, \citenamefont {Long}, \citenamefont {Kaneko}, \citenamefont {Taguchi}, \citenamefont {Arima}, \citenamefont {Keimer},\ and\ \citenamefont {Tokura}}]{2020PRBIshiwataSFO}%
  \BibitemOpen
  \bibfield  {author} {\bibinfo {author} {\bibfnamefont {S.}~\bibnamefont {Ishiwata}}, \bibinfo {author} {\bibfnamefont {T.}~\bibnamefont {Nakajima}}, \bibinfo {author} {\bibfnamefont {J.-H.}\ \bibnamefont {Kim}}, \bibinfo {author} {\bibfnamefont {D.~S.}\ \bibnamefont {Inosov}}, \bibinfo {author} {\bibfnamefont {N.}~\bibnamefont {Kanazawa}}, \bibinfo {author} {\bibfnamefont {J.~S.}\ \bibnamefont {White}}, \bibinfo {author} {\bibfnamefont {J.~L.}\ \bibnamefont {Gavilano}}, \bibinfo {author} {\bibfnamefont {R.}~\bibnamefont {Georgii}}, \bibinfo {author} {\bibfnamefont {K.~M.}\ \bibnamefont {Seemann}}, \bibinfo {author} {\bibfnamefont {G.}~\bibnamefont {Brandl}}, \bibinfo {author} {\bibfnamefont {P.}~\bibnamefont {Manuel}}, \bibinfo {author} {\bibfnamefont {D.~D.}\ \bibnamefont {Khalyavin}}, \bibinfo {author} {\bibfnamefont {S.}~\bibnamefont {Seki}}, \bibinfo {author} {\bibfnamefont {Y.}~\bibnamefont {Tokunaga}}, \bibinfo {author} {\bibfnamefont {M.}~\bibnamefont {Kinoshita}}, \bibinfo {author} {\bibfnamefont
  {Y.~W.}\ \bibnamefont {Long}}, \bibinfo {author} {\bibfnamefont {Y.}~\bibnamefont {Kaneko}}, \bibinfo {author} {\bibfnamefont {Y.}~\bibnamefont {Taguchi}}, \bibinfo {author} {\bibfnamefont {T.}~\bibnamefont {Arima}}, \bibinfo {author} {\bibfnamefont {B.}~\bibnamefont {Keimer}}, \ and\ \bibinfo {author} {\bibfnamefont {Y.}~\bibnamefont {Tokura}},\ }\href {\doibase 10.1103/PhysRevB.101.134406} {\bibfield  {journal} {\bibinfo  {journal} {Phys. Rev. B}\ }\textbf {\bibinfo {volume} {101}},\ \bibinfo {pages} {134406} (\bibinfo {year} {2020})}\BibitemShut {NoStop}%
\bibitem [{\citenamefont {Folkerts}\ \emph {et~al.}(1987)\citenamefont {Folkerts}, \citenamefont {Sawatzky}, \citenamefont {Haas}, \citenamefont {de~Groot},\ and\ \citenamefont {Hillebrecht}}]{1987JPCSawatzkyPyrites}%
  \BibitemOpen
  \bibfield  {author} {\bibinfo {author} {\bibfnamefont {W.}~\bibnamefont {Folkerts}}, \bibinfo {author} {\bibfnamefont {G.~A.}\ \bibnamefont {Sawatzky}}, \bibinfo {author} {\bibfnamefont {C.}~\bibnamefont {Haas}}, \bibinfo {author} {\bibfnamefont {R.~A.}\ \bibnamefont {de~Groot}}, \ and\ \bibinfo {author} {\bibfnamefont {F.~U.}\ \bibnamefont {Hillebrecht}},\ }\href {\doibase 10.1088/0022-3719/20/26/015} {\bibfield  {journal} {\bibinfo  {journal} {Journal of Physics C: Solid State Physics}\ }\textbf {\bibinfo {volume} {20}},\ \bibinfo {pages} {4135} (\bibinfo {year} {1987})}\BibitemShut {NoStop}%
\bibitem [{\citenamefont {Adachi}\ \emph {et~al.}(1969)\citenamefont {Adachi}, \citenamefont {Sato},\ and\ \citenamefont {Takeda}}]{1969JPSJAdachiPyrites}%
  \BibitemOpen
  \bibfield  {author} {\bibinfo {author} {\bibfnamefont {K.}~\bibnamefont {Adachi}}, \bibinfo {author} {\bibfnamefont {K.}~\bibnamefont {Sato}}, \ and\ \bibinfo {author} {\bibfnamefont {M.}~\bibnamefont {Takeda}},\ }\href {\doibase 10.1143/JPSJ.26.631} {\bibfield  {journal} {\bibinfo  {journal} {Journal of the Physical Society of Japan}\ }\textbf {\bibinfo {volume} {26}},\ \bibinfo {pages} {631} (\bibinfo {year} {1969})}\BibitemShut {NoStop}%
\bibitem [{\citenamefont {Ogawa}(1979)}]{1979JAPOgawaPyrites}%
  \BibitemOpen
  \bibfield  {author} {\bibinfo {author} {\bibfnamefont {S.}~\bibnamefont {Ogawa}},\ }\href {\doibase 10.1063/1.327037} {\bibfield  {journal} {\bibinfo  {journal} {Journal of Applied Physics}\ }\textbf {\bibinfo {volume} {50}},\ \bibinfo {pages} {2308} (\bibinfo {year} {1979})}\BibitemShut {NoStop}%
\bibitem [{\citenamefont {Fujimori}\ \emph {et~al.}(1996)\citenamefont {Fujimori}, \citenamefont {Mamiya}, \citenamefont {Mizokawa}, \citenamefont {Miyadai}, \citenamefont {Sekiguchi}, \citenamefont {Takahashi}, \citenamefont {M\^ori},\ and\ \citenamefont {Suga}}]{1996PRBFujimoriPyrites}%
  \BibitemOpen
  \bibfield  {author} {\bibinfo {author} {\bibfnamefont {A.}~\bibnamefont {Fujimori}}, \bibinfo {author} {\bibfnamefont {K.}~\bibnamefont {Mamiya}}, \bibinfo {author} {\bibfnamefont {T.}~\bibnamefont {Mizokawa}}, \bibinfo {author} {\bibfnamefont {T.}~\bibnamefont {Miyadai}}, \bibinfo {author} {\bibfnamefont {T.}~\bibnamefont {Sekiguchi}}, \bibinfo {author} {\bibfnamefont {H.}~\bibnamefont {Takahashi}}, \bibinfo {author} {\bibfnamefont {N.}~\bibnamefont {M\^ori}}, \ and\ \bibinfo {author} {\bibfnamefont {S.}~\bibnamefont {Suga}},\ }\href {\doibase 10.1103/PhysRevB.54.16329} {\bibfield  {journal} {\bibinfo  {journal} {Phys. Rev. B}\ }\textbf {\bibinfo {volume} {54}},\ \bibinfo {pages} {16329} (\bibinfo {year} {1996})}\BibitemShut {NoStop}%
\bibitem [{\citenamefont {Bocquet}\ \emph {et~al.}(1996{\natexlab{b}})\citenamefont {Bocquet}, \citenamefont {Mamiya}, \citenamefont {Mizokawa}, \citenamefont {Fujimori}, \citenamefont {Miyadai}, \citenamefont {Takahashi}, \citenamefont {Môri},\ and\ \citenamefont {Suga}}]{1996JPCMBocquetPyrites}%
  \BibitemOpen
  \bibfield  {author} {\bibinfo {author} {\bibfnamefont {A.~E.}\ \bibnamefont {Bocquet}}, \bibinfo {author} {\bibfnamefont {K.}~\bibnamefont {Mamiya}}, \bibinfo {author} {\bibfnamefont {T.}~\bibnamefont {Mizokawa}}, \bibinfo {author} {\bibfnamefont {A.}~\bibnamefont {Fujimori}}, \bibinfo {author} {\bibfnamefont {T.}~\bibnamefont {Miyadai}}, \bibinfo {author} {\bibfnamefont {H.}~\bibnamefont {Takahashi}}, \bibinfo {author} {\bibfnamefont {M.}~\bibnamefont {Môri}}, \ and\ \bibinfo {author} {\bibfnamefont {S.}~\bibnamefont {Suga}},\ }\href {\doibase 10.1088/0953-8984/8/14/013} {\bibfield  {journal} {\bibinfo  {journal} {Journal of Physics: Condensed Matter}\ }\textbf {\bibinfo {volume} {8}},\ \bibinfo {pages} {2389} (\bibinfo {year} {1996}{\natexlab{b}})}\BibitemShut {NoStop}%
\bibitem [{\citenamefont {Kanamori}(1963)}]{1963PTPKanamori}%
  \BibitemOpen
  \bibfield  {author} {\bibinfo {author} {\bibfnamefont {J.}~\bibnamefont {Kanamori}},\ }\href {\doibase 10.1143/PTP.30.275} {\bibfield  {journal} {\bibinfo  {journal} {Progress of Theoretical Physics}\ }\textbf {\bibinfo {volume} {30}},\ \bibinfo {pages} {275} (\bibinfo {year} {1963})}\BibitemShut {NoStop}%
\bibitem [{\citenamefont {Matsubara}\ \emph {et~al.}(2000)\citenamefont {Matsubara}, \citenamefont {Uozumi}, \citenamefont {Kotani}, \citenamefont {Harada},\ and\ \citenamefont {Shin}}]{2000JPSJKotaniTiO2}%
  \BibitemOpen
  \bibfield  {author} {\bibinfo {author} {\bibfnamefont {M.}~\bibnamefont {Matsubara}}, \bibinfo {author} {\bibfnamefont {T.}~\bibnamefont {Uozumi}}, \bibinfo {author} {\bibfnamefont {A.}~\bibnamefont {Kotani}}, \bibinfo {author} {\bibfnamefont {Y.}~\bibnamefont {Harada}}, \ and\ \bibinfo {author} {\bibfnamefont {S.}~\bibnamefont {Shin}},\ }\href {\doibase 10.1143/JPSJ.69.1558} {\bibfield  {journal} {\bibinfo  {journal} {Journal of the Physical Society of Japan}\ }\textbf {\bibinfo {volume} {69}},\ \bibinfo {pages} {1558} (\bibinfo {year} {2000})}\BibitemShut {NoStop}%
\bibitem [{\citenamefont {Ghiringhelli}\ \emph {et~al.}(2006)\citenamefont {Ghiringhelli}, \citenamefont {Matsubara}, \citenamefont {Dallera}, \citenamefont {Fracassi}, \citenamefont {Tagliaferri}, \citenamefont {Brookes}, \citenamefont {Kotani},\ and\ \citenamefont {Braicovich}}]{2006PRBKotaniMnORIXS}%
  \BibitemOpen
  \bibfield  {author} {\bibinfo {author} {\bibfnamefont {G.}~\bibnamefont {Ghiringhelli}}, \bibinfo {author} {\bibfnamefont {M.}~\bibnamefont {Matsubara}}, \bibinfo {author} {\bibfnamefont {C.}~\bibnamefont {Dallera}}, \bibinfo {author} {\bibfnamefont {F.}~\bibnamefont {Fracassi}}, \bibinfo {author} {\bibfnamefont {A.}~\bibnamefont {Tagliaferri}}, \bibinfo {author} {\bibfnamefont {N.~B.}\ \bibnamefont {Brookes}}, \bibinfo {author} {\bibfnamefont {A.}~\bibnamefont {Kotani}}, \ and\ \bibinfo {author} {\bibfnamefont {L.}~\bibnamefont {Braicovich}},\ }\href {\doibase 10.1103/PhysRevB.73.035111} {\bibfield  {journal} {\bibinfo  {journal} {Phys. Rev. B}\ }\textbf {\bibinfo {volume} {73}},\ \bibinfo {pages} {035111} (\bibinfo {year} {2006})}\BibitemShut {NoStop}%
\bibitem [{\citenamefont {Chiuzb\ifmmode~\u{a}\else \u{a}\fi{}ian}\ \emph {et~al.}(2008)\citenamefont {Chiuzb\ifmmode~\u{a}\else \u{a}\fi{}ian}, \citenamefont {Schmitt}, \citenamefont {Matsubara}, \citenamefont {Kotani}, \citenamefont {Ghiringhelli}, \citenamefont {Dallera}, \citenamefont {Tagliaferri}, \citenamefont {Braicovich}, \citenamefont {Scagnoli}, \citenamefont {Brookes}, \citenamefont {Staub},\ and\ \citenamefont {Patthey}}]{2008PRBKotaniCoORIXS}%
  \BibitemOpen
  \bibfield  {author} {\bibinfo {author} {\bibfnamefont {S.~G.}\ \bibnamefont {Chiuzb\ifmmode~\u{a}\else \u{a}\fi{}ian}}, \bibinfo {author} {\bibfnamefont {T.}~\bibnamefont {Schmitt}}, \bibinfo {author} {\bibfnamefont {M.}~\bibnamefont {Matsubara}}, \bibinfo {author} {\bibfnamefont {A.}~\bibnamefont {Kotani}}, \bibinfo {author} {\bibfnamefont {G.}~\bibnamefont {Ghiringhelli}}, \bibinfo {author} {\bibfnamefont {C.}~\bibnamefont {Dallera}}, \bibinfo {author} {\bibfnamefont {A.}~\bibnamefont {Tagliaferri}}, \bibinfo {author} {\bibfnamefont {L.}~\bibnamefont {Braicovich}}, \bibinfo {author} {\bibfnamefont {V.}~\bibnamefont {Scagnoli}}, \bibinfo {author} {\bibfnamefont {N.~B.}\ \bibnamefont {Brookes}}, \bibinfo {author} {\bibfnamefont {U.}~\bibnamefont {Staub}}, \ and\ \bibinfo {author} {\bibfnamefont {L.}~\bibnamefont {Patthey}},\ }\href {\doibase 10.1103/PhysRevB.78.245102} {\bibfield  {journal} {\bibinfo  {journal} {Phys. Rev. B}\ }\textbf {\bibinfo {volume} {78}},\ \bibinfo {pages} {245102} (\bibinfo {year}
  {2008})}\BibitemShut {NoStop}%
\bibitem [{\citenamefont {Ghiringhelli}\ \emph {et~al.}(2005)\citenamefont {Ghiringhelli}, \citenamefont {Matsubara}, \citenamefont {Dallera}, \citenamefont {Fracassi}, \citenamefont {Gusmeroli}, \citenamefont {Piazzalunga}, \citenamefont {Tagliaferri}, \citenamefont {Brookes}, \citenamefont {Kotani},\ and\ \citenamefont {Braicovich}}]{2005JPCMGhiringhelliNiO}%
  \BibitemOpen
  \bibfield  {author} {\bibinfo {author} {\bibfnamefont {G.}~\bibnamefont {Ghiringhelli}}, \bibinfo {author} {\bibfnamefont {M.}~\bibnamefont {Matsubara}}, \bibinfo {author} {\bibfnamefont {C.}~\bibnamefont {Dallera}}, \bibinfo {author} {\bibfnamefont {F.}~\bibnamefont {Fracassi}}, \bibinfo {author} {\bibfnamefont {R.}~\bibnamefont {Gusmeroli}}, \bibinfo {author} {\bibfnamefont {A.}~\bibnamefont {Piazzalunga}}, \bibinfo {author} {\bibfnamefont {A.}~\bibnamefont {Tagliaferri}}, \bibinfo {author} {\bibfnamefont {N.~B.}\ \bibnamefont {Brookes}}, \bibinfo {author} {\bibfnamefont {A.}~\bibnamefont {Kotani}}, \ and\ \bibinfo {author} {\bibfnamefont {L.}~\bibnamefont {Braicovich}},\ }\href {\doibase 10.1088/0953-8984/17/35/007} {\bibfield  {journal} {\bibinfo  {journal} {Journal of Physics: Condensed Matter}\ }\textbf {\bibinfo {volume} {17}},\ \bibinfo {pages} {5397} (\bibinfo {year} {2005})}\BibitemShut {NoStop}%
\bibitem [{\citenamefont {Das}\ \emph {et~al.}(2013)\citenamefont {Das}, \citenamefont {Green}, \citenamefont {Podder}, \citenamefont {Regier}, \citenamefont {Chang},\ and\ \citenamefont {Moewes}}]{2013JPCCGreenZnNiO}%
  \BibitemOpen
  \bibfield  {author} {\bibinfo {author} {\bibfnamefont {S.~C.}\ \bibnamefont {Das}}, \bibinfo {author} {\bibfnamefont {R.~J.}\ \bibnamefont {Green}}, \bibinfo {author} {\bibfnamefont {J.}~\bibnamefont {Podder}}, \bibinfo {author} {\bibfnamefont {T.~Z.}\ \bibnamefont {Regier}}, \bibinfo {author} {\bibfnamefont {G.~S.}\ \bibnamefont {Chang}}, \ and\ \bibinfo {author} {\bibfnamefont {A.}~\bibnamefont {Moewes}},\ }\href {\doibase 10.1021/jp3126329} {\bibfield  {journal} {\bibinfo  {journal} {The Journal of Physical Chemistry C}\ }\textbf {\bibinfo {volume} {117}},\ \bibinfo {pages} {12745} (\bibinfo {year} {2013})}\BibitemShut {NoStop}%
\bibitem [{\citenamefont {Green}\ \emph {et~al.}(2015)\citenamefont {Green}, \citenamefont {Regier}, \citenamefont {Leedahl}, \citenamefont {McLeod}, \citenamefont {Xu}, \citenamefont {Chang}, \citenamefont {Kurmaev},\ and\ \citenamefont {Moewes}}]{2015PRLGreenIn2O3Fe}%
  \BibitemOpen
  \bibfield  {author} {\bibinfo {author} {\bibfnamefont {R.~J.}\ \bibnamefont {Green}}, \bibinfo {author} {\bibfnamefont {T.~Z.}\ \bibnamefont {Regier}}, \bibinfo {author} {\bibfnamefont {B.}~\bibnamefont {Leedahl}}, \bibinfo {author} {\bibfnamefont {J.~A.}\ \bibnamefont {McLeod}}, \bibinfo {author} {\bibfnamefont {X.~H.}\ \bibnamefont {Xu}}, \bibinfo {author} {\bibfnamefont {G.~S.}\ \bibnamefont {Chang}}, \bibinfo {author} {\bibfnamefont {E.~Z.}\ \bibnamefont {Kurmaev}}, \ and\ \bibinfo {author} {\bibfnamefont {A.}~\bibnamefont {Moewes}},\ }\href {\doibase 10.1103/PhysRevLett.115.167401} {\bibfield  {journal} {\bibinfo  {journal} {Phys. Rev. Lett.}\ }\textbf {\bibinfo {volume} {115}},\ \bibinfo {pages} {167401} (\bibinfo {year} {2015})}\BibitemShut {NoStop}%
\bibitem [{\citenamefont {Green}\ \emph {et~al.}(2014)\citenamefont {Green}, \citenamefont {Zatsepin}, \citenamefont {St.~Onge}, \citenamefont {Kurmaev}, \citenamefont {Gavrilov}, \citenamefont {Zatsepin},\ and\ \citenamefont {Moewes}}]{2014JAPGreenSiO2MnCo}%
  \BibitemOpen
  \bibfield  {author} {\bibinfo {author} {\bibfnamefont {R.~J.}\ \bibnamefont {Green}}, \bibinfo {author} {\bibfnamefont {D.~A.}\ \bibnamefont {Zatsepin}}, \bibinfo {author} {\bibfnamefont {D.~J.}\ \bibnamefont {St.~Onge}}, \bibinfo {author} {\bibfnamefont {E.~Z.}\ \bibnamefont {Kurmaev}}, \bibinfo {author} {\bibfnamefont {N.~V.}\ \bibnamefont {Gavrilov}}, \bibinfo {author} {\bibfnamefont {A.~F.}\ \bibnamefont {Zatsepin}}, \ and\ \bibinfo {author} {\bibfnamefont {A.}~\bibnamefont {Moewes}},\ }\href {\doibase 10.1063/1.4868297} {\bibfield  {journal} {\bibinfo  {journal} {Journal of Applied Physics}\ }\textbf {\bibinfo {volume} {115}},\ \bibinfo {pages} {103708} (\bibinfo {year} {2014})}\BibitemShut {NoStop}%
\bibitem [{\citenamefont {Hariki}\ \emph {et~al.}(2020)\citenamefont {Hariki}, \citenamefont {Winder}, \citenamefont {Uozumi},\ and\ \citenamefont {Kune\ifmmode~\check{s}\else \v{s}\fi{}}}]{2020PRBHariki2p3dRIXS}%
  \BibitemOpen
  \bibfield  {author} {\bibinfo {author} {\bibfnamefont {A.}~\bibnamefont {Hariki}}, \bibinfo {author} {\bibfnamefont {M.}~\bibnamefont {Winder}}, \bibinfo {author} {\bibfnamefont {T.}~\bibnamefont {Uozumi}}, \ and\ \bibinfo {author} {\bibfnamefont {J.}~\bibnamefont {Kune\ifmmode~\check{s}\else \v{s}\fi{}}},\ }\href {\doibase 10.1103/PhysRevB.101.115130} {\bibfield  {journal} {\bibinfo  {journal} {Phys. Rev. B}\ }\textbf {\bibinfo {volume} {101}},\ \bibinfo {pages} {115130} (\bibinfo {year} {2020})}\BibitemShut {NoStop}%
\bibitem [{\citenamefont {Okada}\ \emph {et~al.}(1994)\citenamefont {Okada}, \citenamefont {Uozumi},\ and\ \citenamefont {Kotani}}]{1994JPSJOkadaTi}%
  \BibitemOpen
  \bibfield  {author} {\bibinfo {author} {\bibfnamefont {K.}~\bibnamefont {Okada}}, \bibinfo {author} {\bibfnamefont {T.}~\bibnamefont {Uozumi}}, \ and\ \bibinfo {author} {\bibfnamefont {A.}~\bibnamefont {Kotani}},\ }\href {\doibase 10.1143/JPSJ.63.3176} {\bibfield  {journal} {\bibinfo  {journal} {Journal of the Physical Society of Japan}\ }\textbf {\bibinfo {volume} {63}},\ \bibinfo {pages} {3176} (\bibinfo {year} {1994})}\BibitemShut {NoStop}%
\bibitem [{\citenamefont {Uozumi}\ \emph {et~al.}(1993)\citenamefont {Uozumi}, \citenamefont {Okada},\ and\ \citenamefont {Kotani}}]{1993JPSJUozumiTiV}%
  \BibitemOpen
  \bibfield  {author} {\bibinfo {author} {\bibfnamefont {T.}~\bibnamefont {Uozumi}}, \bibinfo {author} {\bibfnamefont {K.}~\bibnamefont {Okada}}, \ and\ \bibinfo {author} {\bibfnamefont {A.}~\bibnamefont {Kotani}},\ }\href {\doibase 10.1143/JPSJ.62.2595} {\bibfield  {journal} {\bibinfo  {journal} {Journal of the Physical Society of Japan}\ }\textbf {\bibinfo {volume} {62}},\ \bibinfo {pages} {2595} (\bibinfo {year} {1993})}\BibitemShut {NoStop}%
\bibitem [{\citenamefont {Tanaka}\ and\ \citenamefont {Jo}(1994)}]{1994JPSJTanakaRPES}%
  \BibitemOpen
  \bibfield  {author} {\bibinfo {author} {\bibfnamefont {A.}~\bibnamefont {Tanaka}}\ and\ \bibinfo {author} {\bibfnamefont {T.}~\bibnamefont {Jo}},\ }\href {\doibase 10.1143/JPSJ.63.2788} {\bibfield  {journal} {\bibinfo  {journal} {Journal of the Physical Society of Japan}\ }\textbf {\bibinfo {volume} {63}},\ \bibinfo {pages} {2788} (\bibinfo {year} {1994})}\BibitemShut {NoStop}%
\bibitem [{\citenamefont {Parlebas}\ \emph {et~al.}(1995)\citenamefont {Parlebas}, \citenamefont {Khan}, \citenamefont {Uozumi}, \citenamefont {Okada},\ and\ \citenamefont {Kotani}}]{1995JELSPECKotani}%
  \BibitemOpen
  \bibfield  {author} {\bibinfo {author} {\bibfnamefont {J.}~\bibnamefont {Parlebas}}, \bibinfo {author} {\bibfnamefont {M.}~\bibnamefont {Khan}}, \bibinfo {author} {\bibfnamefont {T.}~\bibnamefont {Uozumi}}, \bibinfo {author} {\bibfnamefont {K.}~\bibnamefont {Okada}}, \ and\ \bibinfo {author} {\bibfnamefont {A.}~\bibnamefont {Kotani}},\ }\href {\doibase https://doi.org/10.1016/0368-2048(94)02262-3} {\bibfield  {journal} {\bibinfo  {journal} {Journal of Electron Spectroscopy and Related Phenomena}\ }\textbf {\bibinfo {volume} {71}},\ \bibinfo {pages} {117} (\bibinfo {year} {1995})}\BibitemShut {NoStop}%
\bibitem [{\citenamefont {Hariki}\ \emph {et~al.}(2022)\citenamefont {Hariki}, \citenamefont {Higashi}, \citenamefont {Yamaguchi}, \citenamefont {Li}, \citenamefont {Kalha}, \citenamefont {Mascheck}, \citenamefont {Eriksson}, \citenamefont {Wiell}, \citenamefont {de~Groot},\ and\ \citenamefont {Regoutz}}]{2022PRBHarikiTi}%
  \BibitemOpen
  \bibfield  {author} {\bibinfo {author} {\bibfnamefont {A.}~\bibnamefont {Hariki}}, \bibinfo {author} {\bibfnamefont {K.}~\bibnamefont {Higashi}}, \bibinfo {author} {\bibfnamefont {T.}~\bibnamefont {Yamaguchi}}, \bibinfo {author} {\bibfnamefont {J.}~\bibnamefont {Li}}, \bibinfo {author} {\bibfnamefont {C.}~\bibnamefont {Kalha}}, \bibinfo {author} {\bibfnamefont {M.}~\bibnamefont {Mascheck}}, \bibinfo {author} {\bibfnamefont {S.~K.}\ \bibnamefont {Eriksson}}, \bibinfo {author} {\bibfnamefont {T.}~\bibnamefont {Wiell}}, \bibinfo {author} {\bibfnamefont {F.~M.~F.}\ \bibnamefont {de~Groot}}, \ and\ \bibinfo {author} {\bibfnamefont {A.}~\bibnamefont {Regoutz}},\ }\href {\doibase 10.1103/PhysRevB.106.205138} {\bibfield  {journal} {\bibinfo  {journal} {Phys. Rev. B}\ }\textbf {\bibinfo {volume} {106}},\ \bibinfo {pages} {205138} (\bibinfo {year} {2022})}\BibitemShut {NoStop}%
\bibitem [{\citenamefont {Zimmermann}\ \emph {et~al.}(1999)\citenamefont {Zimmermann}, \citenamefont {Steiner}, \citenamefont {Claessen}, \citenamefont {Reinert}, \citenamefont {Hüfner}, \citenamefont {Blaha},\ and\ \citenamefont {Dufek}}]{1999JPCMHufner}%
  \BibitemOpen
  \bibfield  {author} {\bibinfo {author} {\bibfnamefont {R.}~\bibnamefont {Zimmermann}}, \bibinfo {author} {\bibfnamefont {P.}~\bibnamefont {Steiner}}, \bibinfo {author} {\bibfnamefont {R.}~\bibnamefont {Claessen}}, \bibinfo {author} {\bibfnamefont {F.}~\bibnamefont {Reinert}}, \bibinfo {author} {\bibfnamefont {S.}~\bibnamefont {Hüfner}}, \bibinfo {author} {\bibfnamefont {P.}~\bibnamefont {Blaha}}, \ and\ \bibinfo {author} {\bibfnamefont {P.}~\bibnamefont {Dufek}},\ }\href {\doibase 10.1088/0953-8984/11/7/002} {\bibfield  {journal} {\bibinfo  {journal} {Journal of Physics: Condensed Matter}\ }\textbf {\bibinfo {volume} {11}},\ \bibinfo {pages} {1657} (\bibinfo {year} {1999})}\BibitemShut {NoStop}%
\bibitem [{\citenamefont {Chang}\ \emph {et~al.}(2018)\citenamefont {Chang}, \citenamefont {Koethe}, \citenamefont {Hu}, \citenamefont {Weinen}, \citenamefont {Agrestini}, \citenamefont {Zhao}, \citenamefont {Gegner}, \citenamefont {Ott}, \citenamefont {Panaccione}, \citenamefont {Wu}, \citenamefont {Haverkort}, \citenamefont {Roth}, \citenamefont {Komarek}, \citenamefont {Offi}, \citenamefont {Monaco}, \citenamefont {Liao}, \citenamefont {Tsuei}, \citenamefont {Lin}, \citenamefont {Chen}, \citenamefont {Tanaka},\ and\ \citenamefont {Tjeng}}]{2018PRXTjengTi2O3}%
  \BibitemOpen
  \bibfield  {author} {\bibinfo {author} {\bibfnamefont {C.~F.}\ \bibnamefont {Chang}}, \bibinfo {author} {\bibfnamefont {T.~C.}\ \bibnamefont {Koethe}}, \bibinfo {author} {\bibfnamefont {Z.}~\bibnamefont {Hu}}, \bibinfo {author} {\bibfnamefont {J.}~\bibnamefont {Weinen}}, \bibinfo {author} {\bibfnamefont {S.}~\bibnamefont {Agrestini}}, \bibinfo {author} {\bibfnamefont {L.}~\bibnamefont {Zhao}}, \bibinfo {author} {\bibfnamefont {J.}~\bibnamefont {Gegner}}, \bibinfo {author} {\bibfnamefont {H.}~\bibnamefont {Ott}}, \bibinfo {author} {\bibfnamefont {G.}~\bibnamefont {Panaccione}}, \bibinfo {author} {\bibfnamefont {H.}~\bibnamefont {Wu}}, \bibinfo {author} {\bibfnamefont {M.~W.}\ \bibnamefont {Haverkort}}, \bibinfo {author} {\bibfnamefont {H.}~\bibnamefont {Roth}}, \bibinfo {author} {\bibfnamefont {A.~C.}\ \bibnamefont {Komarek}}, \bibinfo {author} {\bibfnamefont {F.}~\bibnamefont {Offi}}, \bibinfo {author} {\bibfnamefont {G.}~\bibnamefont {Monaco}}, \bibinfo {author} {\bibfnamefont {Y.-F.}\ \bibnamefont
  {Liao}}, \bibinfo {author} {\bibfnamefont {K.-D.}\ \bibnamefont {Tsuei}}, \bibinfo {author} {\bibfnamefont {H.-J.}\ \bibnamefont {Lin}}, \bibinfo {author} {\bibfnamefont {C.~T.}\ \bibnamefont {Chen}}, \bibinfo {author} {\bibfnamefont {A.}~\bibnamefont {Tanaka}}, \ and\ \bibinfo {author} {\bibfnamefont {L.~H.}\ \bibnamefont {Tjeng}},\ }\href {\doibase 10.1103/PhysRevX.8.021004} {\bibfield  {journal} {\bibinfo  {journal} {Phys. Rev. X}\ }\textbf {\bibinfo {volume} {8}},\ \bibinfo {pages} {021004} (\bibinfo {year} {2018})}\BibitemShut {NoStop}%
\bibitem [{\citenamefont {Uozumi}\ \emph {et~al.}(1997)\citenamefont {Uozumi}, \citenamefont {Okada}, \citenamefont {Kotani}, \citenamefont {Zimmermann}, \citenamefont {Steiner}, \citenamefont {Hüfner}, \citenamefont {Tezuka},\ and\ \citenamefont {Shin}}]{1997JELSPECKotaniM2O3}%
  \BibitemOpen
  \bibfield  {author} {\bibinfo {author} {\bibfnamefont {T.}~\bibnamefont {Uozumi}}, \bibinfo {author} {\bibfnamefont {K.}~\bibnamefont {Okada}}, \bibinfo {author} {\bibfnamefont {A.}~\bibnamefont {Kotani}}, \bibinfo {author} {\bibfnamefont {R.}~\bibnamefont {Zimmermann}}, \bibinfo {author} {\bibfnamefont {P.}~\bibnamefont {Steiner}}, \bibinfo {author} {\bibfnamefont {S.}~\bibnamefont {Hüfner}}, \bibinfo {author} {\bibfnamefont {Y.}~\bibnamefont {Tezuka}}, \ and\ \bibinfo {author} {\bibfnamefont {S.}~\bibnamefont {Shin}},\ }\href {\doibase https://doi.org/10.1016/S0368-2048(96)03063-0} {\bibfield  {journal} {\bibinfo  {journal} {Journal of Electron Spectroscopy and Related Phenomena}\ }\textbf {\bibinfo {volume} {83}},\ \bibinfo {pages} {9} (\bibinfo {year} {1997})}\BibitemShut {NoStop}%
\bibitem [{\citenamefont {Tezuka}\ \emph {et~al.}(1997)\citenamefont {Tezuka}, \citenamefont {Shin}, \citenamefont {Uozumi},\ and\ \citenamefont {Kotani}}]{1997JPSJKotaniTi2O3}%
  \BibitemOpen
  \bibfield  {author} {\bibinfo {author} {\bibfnamefont {Y.}~\bibnamefont {Tezuka}}, \bibinfo {author} {\bibfnamefont {S.}~\bibnamefont {Shin}}, \bibinfo {author} {\bibfnamefont {T.}~\bibnamefont {Uozumi}}, \ and\ \bibinfo {author} {\bibfnamefont {A.}~\bibnamefont {Kotani}},\ }\href {\doibase 10.1143/JPSJ.66.3153} {\bibfield  {journal} {\bibinfo  {journal} {Journal of the Physical Society of Japan}\ }\textbf {\bibinfo {volume} {66}},\ \bibinfo {pages} {3153} (\bibinfo {year} {1997})}\BibitemShut {NoStop}%
\bibitem [{\citenamefont {Saitoh}\ \emph {et~al.}(1995)\citenamefont {Saitoh}, \citenamefont {Bocquet}, \citenamefont {Mizokawa},\ and\ \citenamefont {Fujimori}}]{1995PRBBocquet}%
  \BibitemOpen
  \bibfield  {author} {\bibinfo {author} {\bibfnamefont {T.}~\bibnamefont {Saitoh}}, \bibinfo {author} {\bibfnamefont {A.~E.}\ \bibnamefont {Bocquet}}, \bibinfo {author} {\bibfnamefont {T.}~\bibnamefont {Mizokawa}}, \ and\ \bibinfo {author} {\bibfnamefont {A.}~\bibnamefont {Fujimori}},\ }\href {\doibase 10.1103/PhysRevB.52.7934} {\bibfield  {journal} {\bibinfo  {journal} {Phys. Rev. B}\ }\textbf {\bibinfo {volume} {52}},\ \bibinfo {pages} {7934} (\bibinfo {year} {1995})}\BibitemShut {NoStop}%
\bibitem [{\citenamefont {Haverkort}\ \emph {et~al.}(2005{\natexlab{a}})\citenamefont {Haverkort}, \citenamefont {Hu}, \citenamefont {Tanaka}, \citenamefont {Ghiringhelli}, \citenamefont {Roth}, \citenamefont {Cwik}, \citenamefont {Lorenz}, \citenamefont {Sch\"u\ss{}ler-Langeheine}, \citenamefont {Streltsov}, \citenamefont {Mylnikova}, \citenamefont {Anisimov}, \citenamefont {de~Nadai}, \citenamefont {Brookes}, \citenamefont {Hsieh}, \citenamefont {Lin}, \citenamefont {Chen}, \citenamefont {Mizokawa}, \citenamefont {Taguchi}, \citenamefont {Tokura}, \citenamefont {Khomskii},\ and\ \citenamefont {Tjeng}}]{2005PRLHaverkortLTO}%
  \BibitemOpen
  \bibfield  {author} {\bibinfo {author} {\bibfnamefont {M.~W.}\ \bibnamefont {Haverkort}}, \bibinfo {author} {\bibfnamefont {Z.}~\bibnamefont {Hu}}, \bibinfo {author} {\bibfnamefont {A.}~\bibnamefont {Tanaka}}, \bibinfo {author} {\bibfnamefont {G.}~\bibnamefont {Ghiringhelli}}, \bibinfo {author} {\bibfnamefont {H.}~\bibnamefont {Roth}}, \bibinfo {author} {\bibfnamefont {M.}~\bibnamefont {Cwik}}, \bibinfo {author} {\bibfnamefont {T.}~\bibnamefont {Lorenz}}, \bibinfo {author} {\bibfnamefont {C.}~\bibnamefont {Sch\"u\ss{}ler-Langeheine}}, \bibinfo {author} {\bibfnamefont {S.~V.}\ \bibnamefont {Streltsov}}, \bibinfo {author} {\bibfnamefont {A.~S.}\ \bibnamefont {Mylnikova}}, \bibinfo {author} {\bibfnamefont {V.~I.}\ \bibnamefont {Anisimov}}, \bibinfo {author} {\bibfnamefont {C.}~\bibnamefont {de~Nadai}}, \bibinfo {author} {\bibfnamefont {N.~B.}\ \bibnamefont {Brookes}}, \bibinfo {author} {\bibfnamefont {H.~H.}\ \bibnamefont {Hsieh}}, \bibinfo {author} {\bibfnamefont {H.-J.}\ \bibnamefont {Lin}}, \bibinfo
  {author} {\bibfnamefont {C.~T.}\ \bibnamefont {Chen}}, \bibinfo {author} {\bibfnamefont {T.}~\bibnamefont {Mizokawa}}, \bibinfo {author} {\bibfnamefont {Y.}~\bibnamefont {Taguchi}}, \bibinfo {author} {\bibfnamefont {Y.}~\bibnamefont {Tokura}}, \bibinfo {author} {\bibfnamefont {D.~I.}\ \bibnamefont {Khomskii}}, \ and\ \bibinfo {author} {\bibfnamefont {L.~H.}\ \bibnamefont {Tjeng}},\ }\href {\doibase 10.1103/PhysRevLett.94.056401} {\bibfield  {journal} {\bibinfo  {journal} {Phys. Rev. Lett.}\ }\textbf {\bibinfo {volume} {94}},\ \bibinfo {pages} {056401} (\bibinfo {year} {2005}{\natexlab{a}})}\BibitemShut {NoStop}%
\bibitem [{\citenamefont {Hopkins}\ \emph {et~al.}(2015)\citenamefont {Hopkins}, \citenamefont {Prots}, \citenamefont {Burkhardt}, \citenamefont {Watier}, \citenamefont {Hu}, \citenamefont {Kuo}, \citenamefont {Chiang}, \citenamefont {Pi}, \citenamefont {Tanaka}, \citenamefont {Tjeng},\ and\ \citenamefont {Valldor}}]{2015ChemEJTjengBVSO}%
  \BibitemOpen
  \bibfield  {author} {\bibinfo {author} {\bibfnamefont {E.~J.}\ \bibnamefont {Hopkins}}, \bibinfo {author} {\bibfnamefont {Y.}~\bibnamefont {Prots}}, \bibinfo {author} {\bibfnamefont {U.}~\bibnamefont {Burkhardt}}, \bibinfo {author} {\bibfnamefont {Y.}~\bibnamefont {Watier}}, \bibinfo {author} {\bibfnamefont {Z.}~\bibnamefont {Hu}}, \bibinfo {author} {\bibfnamefont {C.-Y.}\ \bibnamefont {Kuo}}, \bibinfo {author} {\bibfnamefont {J.-C.}\ \bibnamefont {Chiang}}, \bibinfo {author} {\bibfnamefont {T.-W.}\ \bibnamefont {Pi}}, \bibinfo {author} {\bibfnamefont {A.}~\bibnamefont {Tanaka}}, \bibinfo {author} {\bibfnamefont {L.~H.}\ \bibnamefont {Tjeng}}, \ and\ \bibinfo {author} {\bibfnamefont {M.}~\bibnamefont {Valldor}},\ }\href {\doibase https://doi.org/10.1002/chem.201406511} {\bibfield  {journal} {\bibinfo  {journal} {Chemistry – A European Journal}\ }\textbf {\bibinfo {volume} {21}},\ \bibinfo {pages} {7938} (\bibinfo {year} {2015})}\BibitemShut {NoStop}%
\bibitem [{\citenamefont {Schmitt}\ \emph {et~al.}(2004{\natexlab{a}})\citenamefont {Schmitt}, \citenamefont {Duda}, \citenamefont {Matsubara}, \citenamefont {Augustsson}, \citenamefont {Trif}, \citenamefont {Guo}, \citenamefont {Gridneva}, \citenamefont {Uozumi}, \citenamefont {Kotani},\ and\ \citenamefont {Nordgren}}]{2004JACKotaniV}%
  \BibitemOpen
  \bibfield  {author} {\bibinfo {author} {\bibfnamefont {T.}~\bibnamefont {Schmitt}}, \bibinfo {author} {\bibfnamefont {L.-C.}\ \bibnamefont {Duda}}, \bibinfo {author} {\bibfnamefont {M.}~\bibnamefont {Matsubara}}, \bibinfo {author} {\bibfnamefont {A.}~\bibnamefont {Augustsson}}, \bibinfo {author} {\bibfnamefont {F.}~\bibnamefont {Trif}}, \bibinfo {author} {\bibfnamefont {J.-H.}\ \bibnamefont {Guo}}, \bibinfo {author} {\bibfnamefont {L.}~\bibnamefont {Gridneva}}, \bibinfo {author} {\bibfnamefont {T.}~\bibnamefont {Uozumi}}, \bibinfo {author} {\bibfnamefont {A.}~\bibnamefont {Kotani}}, \ and\ \bibinfo {author} {\bibfnamefont {J.}~\bibnamefont {Nordgren}},\ }\href {\doibase https://doi.org/10.1016/S0925-8388(03)00575-9} {\bibfield  {journal} {\bibinfo  {journal} {Journal of Alloys and Compounds}\ }\textbf {\bibinfo {volume} {362}},\ \bibinfo {pages} {143} (\bibinfo {year} {2004}{\natexlab{a}})},\ \bibinfo {note} {proceedings of the Sixth International School and Symposium on Synchrotron Radiation in Natural
  Science (ISSRNS)}\BibitemShut {NoStop}%
\bibitem [{\citenamefont {Schmitt}\ \emph {et~al.}(2004{\natexlab{b}})\citenamefont {Schmitt}, \citenamefont {Duda}, \citenamefont {Matsubara}, \citenamefont {Mattesini}, \citenamefont {Klemm}, \citenamefont {Augustsson}, \citenamefont {Guo}, \citenamefont {Uozumi}, \citenamefont {Horn}, \citenamefont {Ahuja}, \citenamefont {Kotani},\ and\ \citenamefont {Nordgren}}]{2004PRBKotaniV6O13}%
  \BibitemOpen
  \bibfield  {author} {\bibinfo {author} {\bibfnamefont {T.}~\bibnamefont {Schmitt}}, \bibinfo {author} {\bibfnamefont {L.-C.}\ \bibnamefont {Duda}}, \bibinfo {author} {\bibfnamefont {M.}~\bibnamefont {Matsubara}}, \bibinfo {author} {\bibfnamefont {M.}~\bibnamefont {Mattesini}}, \bibinfo {author} {\bibfnamefont {M.}~\bibnamefont {Klemm}}, \bibinfo {author} {\bibfnamefont {A.}~\bibnamefont {Augustsson}}, \bibinfo {author} {\bibfnamefont {J.-H.}\ \bibnamefont {Guo}}, \bibinfo {author} {\bibfnamefont {T.}~\bibnamefont {Uozumi}}, \bibinfo {author} {\bibfnamefont {S.}~\bibnamefont {Horn}}, \bibinfo {author} {\bibfnamefont {R.}~\bibnamefont {Ahuja}}, \bibinfo {author} {\bibfnamefont {A.}~\bibnamefont {Kotani}}, \ and\ \bibinfo {author} {\bibfnamefont {J.}~\bibnamefont {Nordgren}},\ }\href {\doibase 10.1103/PhysRevB.69.125103} {\bibfield  {journal} {\bibinfo  {journal} {Phys. Rev. B}\ }\textbf {\bibinfo {volume} {69}},\ \bibinfo {pages} {125103} (\bibinfo {year} {2004}{\natexlab{b}})}\BibitemShut {NoStop}%
\bibitem [{\citenamefont {Haverkort}\ \emph {et~al.}(2005{\natexlab{b}})\citenamefont {Haverkort}, \citenamefont {Hu}, \citenamefont {Tanaka}, \citenamefont {Reichelt}, \citenamefont {Streltsov}, \citenamefont {Korotin}, \citenamefont {Anisimov}, \citenamefont {Hsieh}, \citenamefont {Lin}, \citenamefont {Chen}, \citenamefont {Khomskii},\ and\ \citenamefont {Tjeng}}]{2005PRLHaverkortVO2}%
  \BibitemOpen
  \bibfield  {author} {\bibinfo {author} {\bibfnamefont {M.~W.}\ \bibnamefont {Haverkort}}, \bibinfo {author} {\bibfnamefont {Z.}~\bibnamefont {Hu}}, \bibinfo {author} {\bibfnamefont {A.}~\bibnamefont {Tanaka}}, \bibinfo {author} {\bibfnamefont {W.}~\bibnamefont {Reichelt}}, \bibinfo {author} {\bibfnamefont {S.~V.}\ \bibnamefont {Streltsov}}, \bibinfo {author} {\bibfnamefont {M.~A.}\ \bibnamefont {Korotin}}, \bibinfo {author} {\bibfnamefont {V.~I.}\ \bibnamefont {Anisimov}}, \bibinfo {author} {\bibfnamefont {H.~H.}\ \bibnamefont {Hsieh}}, \bibinfo {author} {\bibfnamefont {H.-J.}\ \bibnamefont {Lin}}, \bibinfo {author} {\bibfnamefont {C.~T.}\ \bibnamefont {Chen}}, \bibinfo {author} {\bibfnamefont {D.~I.}\ \bibnamefont {Khomskii}}, \ and\ \bibinfo {author} {\bibfnamefont {L.~H.}\ \bibnamefont {Tjeng}},\ }\href {\doibase 10.1103/PhysRevLett.95.196404} {\bibfield  {journal} {\bibinfo  {journal} {Phys. Rev. Lett.}\ }\textbf {\bibinfo {volume} {95}},\ \bibinfo {pages} {196404} (\bibinfo {year}
  {2005}{\natexlab{b}})}\BibitemShut {NoStop}%
\bibitem [{\citenamefont {Yamaguchi}\ \emph {et~al.}(2024)\citenamefont {Yamaguchi}, \citenamefont {Higashi}, \citenamefont {Regoutz}, \citenamefont {Takahashi}, \citenamefont {Lazemi}, \citenamefont {Che}, \citenamefont {de~Groot},\ and\ \citenamefont {Hariki}}]{2024PRBHarikiVCr}%
  \BibitemOpen
  \bibfield  {author} {\bibinfo {author} {\bibfnamefont {T.}~\bibnamefont {Yamaguchi}}, \bibinfo {author} {\bibfnamefont {K.}~\bibnamefont {Higashi}}, \bibinfo {author} {\bibfnamefont {A.}~\bibnamefont {Regoutz}}, \bibinfo {author} {\bibfnamefont {Y.}~\bibnamefont {Takahashi}}, \bibinfo {author} {\bibfnamefont {M.}~\bibnamefont {Lazemi}}, \bibinfo {author} {\bibfnamefont {Q.}~\bibnamefont {Che}}, \bibinfo {author} {\bibfnamefont {F.~M.~F.}\ \bibnamefont {de~Groot}}, \ and\ \bibinfo {author} {\bibfnamefont {A.}~\bibnamefont {Hariki}},\ }\href {\doibase 10.1103/PhysRevB.109.205143} {\bibfield  {journal} {\bibinfo  {journal} {Phys. Rev. B}\ }\textbf {\bibinfo {volume} {109}},\ \bibinfo {pages} {205143} (\bibinfo {year} {2024})}\BibitemShut {NoStop}%
\bibitem [{\citenamefont {Pen}\ \emph {et~al.}(1997)\citenamefont {Pen}, \citenamefont {Tjeng}, \citenamefont {Pellegrin}, \citenamefont {de~Groot}, \citenamefont {Sawatzky}, \citenamefont {van Veenendaal},\ and\ \citenamefont {Chen}}]{1997PRBSawatzkyLiVO2}%
  \BibitemOpen
  \bibfield  {author} {\bibinfo {author} {\bibfnamefont {H.~F.}\ \bibnamefont {Pen}}, \bibinfo {author} {\bibfnamefont {L.~H.}\ \bibnamefont {Tjeng}}, \bibinfo {author} {\bibfnamefont {E.}~\bibnamefont {Pellegrin}}, \bibinfo {author} {\bibfnamefont {F.~M.~F.}\ \bibnamefont {de~Groot}}, \bibinfo {author} {\bibfnamefont {G.~A.}\ \bibnamefont {Sawatzky}}, \bibinfo {author} {\bibfnamefont {M.~A.}\ \bibnamefont {van Veenendaal}}, \ and\ \bibinfo {author} {\bibfnamefont {C.~T.}\ \bibnamefont {Chen}},\ }\href {\doibase 10.1103/PhysRevB.55.15500} {\bibfield  {journal} {\bibinfo  {journal} {Phys. Rev. B}\ }\textbf {\bibinfo {volume} {55}},\ \bibinfo {pages} {15500} (\bibinfo {year} {1997})}\BibitemShut {NoStop}%
\bibitem [{\citenamefont {Murota}\ \emph {et~al.}(2020)\citenamefont {Murota}, \citenamefont {Pachoud}, \citenamefont {Attfield}, \citenamefont {Glaum}, \citenamefont {Sutarto}, \citenamefont {Takubo}, \citenamefont {Khomskii},\ and\ \citenamefont {Mizokawa}}]{2020PRBMizokawaV2OPO4}%
  \BibitemOpen
  \bibfield  {author} {\bibinfo {author} {\bibfnamefont {K.}~\bibnamefont {Murota}}, \bibinfo {author} {\bibfnamefont {E.}~\bibnamefont {Pachoud}}, \bibinfo {author} {\bibfnamefont {J.~P.}\ \bibnamefont {Attfield}}, \bibinfo {author} {\bibfnamefont {R.}~\bibnamefont {Glaum}}, \bibinfo {author} {\bibfnamefont {R.}~\bibnamefont {Sutarto}}, \bibinfo {author} {\bibfnamefont {K.}~\bibnamefont {Takubo}}, \bibinfo {author} {\bibfnamefont {D.~I.}\ \bibnamefont {Khomskii}}, \ and\ \bibinfo {author} {\bibfnamefont {T.}~\bibnamefont {Mizokawa}},\ }\href {\doibase 10.1103/PhysRevB.101.245106} {\bibfield  {journal} {\bibinfo  {journal} {Phys. Rev. B}\ }\textbf {\bibinfo {volume} {101}},\ \bibinfo {pages} {245106} (\bibinfo {year} {2020})}\BibitemShut {NoStop}%
\bibitem [{\citenamefont {Takubo}\ \emph {et~al.}(2006)\citenamefont {Takubo}, \citenamefont {Son}, \citenamefont {Mizokawa}, \citenamefont {Ueda}, \citenamefont {Isobe}, \citenamefont {Matsushita},\ and\ \citenamefont {Ueda}}]{2006PRBTakuboV}%
  \BibitemOpen
  \bibfield  {author} {\bibinfo {author} {\bibfnamefont {K.}~\bibnamefont {Takubo}}, \bibinfo {author} {\bibfnamefont {J.-Y.}\ \bibnamefont {Son}}, \bibinfo {author} {\bibfnamefont {T.}~\bibnamefont {Mizokawa}}, \bibinfo {author} {\bibfnamefont {H.}~\bibnamefont {Ueda}}, \bibinfo {author} {\bibfnamefont {M.}~\bibnamefont {Isobe}}, \bibinfo {author} {\bibfnamefont {Y.}~\bibnamefont {Matsushita}}, \ and\ \bibinfo {author} {\bibfnamefont {Y.}~\bibnamefont {Ueda}},\ }\href {\doibase 10.1103/PhysRevB.74.155103} {\bibfield  {journal} {\bibinfo  {journal} {Phys. Rev. B}\ }\textbf {\bibinfo {volume} {74}},\ \bibinfo {pages} {155103} (\bibinfo {year} {2006})}\BibitemShut {NoStop}%
\bibitem [{\citenamefont {Zhao}\ \emph {et~al.}(2023)\citenamefont {Zhao}, \citenamefont {Haw}, \citenamefont {Wang}, \citenamefont {Cao}, \citenamefont {Lin}, \citenamefont {Chen}, \citenamefont {Sahle}, \citenamefont {Tanaka}, \citenamefont {Chen}, \citenamefont {Jin}, \citenamefont {Hu},\ and\ \citenamefont {Tjeng}}]{2023PRBTjengPbCrO3}%
  \BibitemOpen
  \bibfield  {author} {\bibinfo {author} {\bibfnamefont {J.}~\bibnamefont {Zhao}}, \bibinfo {author} {\bibfnamefont {S.-C.}\ \bibnamefont {Haw}}, \bibinfo {author} {\bibfnamefont {X.}~\bibnamefont {Wang}}, \bibinfo {author} {\bibfnamefont {L.}~\bibnamefont {Cao}}, \bibinfo {author} {\bibfnamefont {H.-J.}\ \bibnamefont {Lin}}, \bibinfo {author} {\bibfnamefont {C.-T.}\ \bibnamefont {Chen}}, \bibinfo {author} {\bibfnamefont {C.~J.}\ \bibnamefont {Sahle}}, \bibinfo {author} {\bibfnamefont {A.}~\bibnamefont {Tanaka}}, \bibinfo {author} {\bibfnamefont {J.-M.}\ \bibnamefont {Chen}}, \bibinfo {author} {\bibfnamefont {C.}~\bibnamefont {Jin}}, \bibinfo {author} {\bibfnamefont {Z.}~\bibnamefont {Hu}}, \ and\ \bibinfo {author} {\bibfnamefont {L.~H.}\ \bibnamefont {Tjeng}},\ }\href {\doibase 10.1103/PhysRevB.107.024107} {\bibfield  {journal} {\bibinfo  {journal} {Phys. Rev. B}\ }\textbf {\bibinfo {volume} {107}},\ \bibinfo {pages} {024107} (\bibinfo {year} {2023})}\BibitemShut {NoStop}%
\bibitem [{\citenamefont {Korotin}\ \emph {et~al.}(1998)\citenamefont {Korotin}, \citenamefont {Anisimov}, \citenamefont {Khomskii},\ and\ \citenamefont {Sawatzky}}]{1998PRLSawatzkyCrO2}%
  \BibitemOpen
  \bibfield  {author} {\bibinfo {author} {\bibfnamefont {M.~A.}\ \bibnamefont {Korotin}}, \bibinfo {author} {\bibfnamefont {V.~I.}\ \bibnamefont {Anisimov}}, \bibinfo {author} {\bibfnamefont {D.~I.}\ \bibnamefont {Khomskii}}, \ and\ \bibinfo {author} {\bibfnamefont {G.~A.}\ \bibnamefont {Sawatzky}},\ }\href {\doibase 10.1103/PhysRevLett.80.4305} {\bibfield  {journal} {\bibinfo  {journal} {Phys. Rev. Lett.}\ }\textbf {\bibinfo {volume} {80}},\ \bibinfo {pages} {4305} (\bibinfo {year} {1998})}\BibitemShut {NoStop}%
\bibitem [{\citenamefont {Huang}\ \emph {et~al.}(2003)\citenamefont {Huang}, \citenamefont {Tjeng}, \citenamefont {Chen}, \citenamefont {Chang}, \citenamefont {Wu}, \citenamefont {Chung}, \citenamefont {Tanaka}, \citenamefont {Guo}, \citenamefont {Lin}, \citenamefont {Shyu}, \citenamefont {Wu},\ and\ \citenamefont {Chen}}]{2003PRBTjengCrO2}%
  \BibitemOpen
  \bibfield  {author} {\bibinfo {author} {\bibfnamefont {D.~J.}\ \bibnamefont {Huang}}, \bibinfo {author} {\bibfnamefont {L.~H.}\ \bibnamefont {Tjeng}}, \bibinfo {author} {\bibfnamefont {J.}~\bibnamefont {Chen}}, \bibinfo {author} {\bibfnamefont {C.~F.}\ \bibnamefont {Chang}}, \bibinfo {author} {\bibfnamefont {W.~P.}\ \bibnamefont {Wu}}, \bibinfo {author} {\bibfnamefont {S.~C.}\ \bibnamefont {Chung}}, \bibinfo {author} {\bibfnamefont {A.}~\bibnamefont {Tanaka}}, \bibinfo {author} {\bibfnamefont {G.~Y.}\ \bibnamefont {Guo}}, \bibinfo {author} {\bibfnamefont {H.-J.}\ \bibnamefont {Lin}}, \bibinfo {author} {\bibfnamefont {S.~G.}\ \bibnamefont {Shyu}}, \bibinfo {author} {\bibfnamefont {C.~C.}\ \bibnamefont {Wu}}, \ and\ \bibinfo {author} {\bibfnamefont {C.~T.}\ \bibnamefont {Chen}},\ }\href {\doibase 10.1103/PhysRevB.67.214419} {\bibfield  {journal} {\bibinfo  {journal} {Phys. Rev. B}\ }\textbf {\bibinfo {volume} {67}},\ \bibinfo {pages} {214419} (\bibinfo {year} {2003})}\BibitemShut {NoStop}%
\bibitem [{\citenamefont {Maiti}\ and\ \citenamefont {Sarma}(1996)}]{1996PRBSarmaCr}%
  \BibitemOpen
  \bibfield  {author} {\bibinfo {author} {\bibfnamefont {K.}~\bibnamefont {Maiti}}\ and\ \bibinfo {author} {\bibfnamefont {D.~D.}\ \bibnamefont {Sarma}},\ }\href {\doibase 10.1103/PhysRevB.54.7816} {\bibfield  {journal} {\bibinfo  {journal} {Phys. Rev. B}\ }\textbf {\bibinfo {volume} {54}},\ \bibinfo {pages} {7816} (\bibinfo {year} {1996})}\BibitemShut {NoStop}%
\bibitem [{\citenamefont {Pal}\ \emph {et~al.}(2018)\citenamefont {Pal}, \citenamefont {Liu}, \citenamefont {Wen}, \citenamefont {Kareev}, \citenamefont {N'Diaye}, \citenamefont {Shafer}, \citenamefont {Arenholz},\ and\ \citenamefont {Chakhalian}}]{2018APLChakYCO}%
  \BibitemOpen
  \bibfield  {author} {\bibinfo {author} {\bibfnamefont {B.}~\bibnamefont {Pal}}, \bibinfo {author} {\bibfnamefont {X.}~\bibnamefont {Liu}}, \bibinfo {author} {\bibfnamefont {F.}~\bibnamefont {Wen}}, \bibinfo {author} {\bibfnamefont {M.}~\bibnamefont {Kareev}}, \bibinfo {author} {\bibfnamefont {A.~T.}\ \bibnamefont {N'Diaye}}, \bibinfo {author} {\bibfnamefont {P.}~\bibnamefont {Shafer}}, \bibinfo {author} {\bibfnamefont {E.}~\bibnamefont {Arenholz}}, \ and\ \bibinfo {author} {\bibfnamefont {J.}~\bibnamefont {Chakhalian}},\ }\href {\doibase 10.1063/1.5029977} {\bibfield  {journal} {\bibinfo  {journal} {Applied Physics Letters}\ }\textbf {\bibinfo {volume} {112}},\ \bibinfo {pages} {252901} (\bibinfo {year} {2018})}\BibitemShut {NoStop}%
\bibitem [{\citenamefont {Wadati}\ \emph {et~al.}(2013)\citenamefont {Wadati}, \citenamefont {Kato}, \citenamefont {Wakisaka}, \citenamefont {Sudayama}, \citenamefont {Hawthorn}, \citenamefont {Regier}, \citenamefont {Onishi}, \citenamefont {Azuma}, \citenamefont {Shimakawa}, \citenamefont {Mizokawa}, \citenamefont {Tanaka},\ and\ \citenamefont {Sawatzky}}]{2013SSCWadatiMn4}%
  \BibitemOpen
  \bibfield  {author} {\bibinfo {author} {\bibfnamefont {H.}~\bibnamefont {Wadati}}, \bibinfo {author} {\bibfnamefont {K.}~\bibnamefont {Kato}}, \bibinfo {author} {\bibfnamefont {Y.}~\bibnamefont {Wakisaka}}, \bibinfo {author} {\bibfnamefont {T.}~\bibnamefont {Sudayama}}, \bibinfo {author} {\bibfnamefont {D.}~\bibnamefont {Hawthorn}}, \bibinfo {author} {\bibfnamefont {T.}~\bibnamefont {Regier}}, \bibinfo {author} {\bibfnamefont {N.}~\bibnamefont {Onishi}}, \bibinfo {author} {\bibfnamefont {M.}~\bibnamefont {Azuma}}, \bibinfo {author} {\bibfnamefont {Y.}~\bibnamefont {Shimakawa}}, \bibinfo {author} {\bibfnamefont {T.}~\bibnamefont {Mizokawa}}, \bibinfo {author} {\bibfnamefont {A.}~\bibnamefont {Tanaka}}, \ and\ \bibinfo {author} {\bibfnamefont {G.}~\bibnamefont {Sawatzky}},\ }\href {\doibase https://doi.org/10.1016/j.ssc.2013.02.021} {\bibfield  {journal} {\bibinfo  {journal} {Solid State Communications}\ }\textbf {\bibinfo {volume} {162}},\ \bibinfo {pages} {18} (\bibinfo {year} {2013})}\BibitemShut {NoStop}%
\bibitem [{\citenamefont {Burnus}\ \emph {et~al.}(2008)\citenamefont {Burnus}, \citenamefont {Hu}, \citenamefont {Hsieh}, \citenamefont {Joly}, \citenamefont {Joy}, \citenamefont {Haverkort}, \citenamefont {Wu}, \citenamefont {Tanaka}, \citenamefont {Lin}, \citenamefont {Chen},\ and\ \citenamefont {Tjeng}}]{2008PRBBurnusLMCO}%
  \BibitemOpen
  \bibfield  {author} {\bibinfo {author} {\bibfnamefont {T.}~\bibnamefont {Burnus}}, \bibinfo {author} {\bibfnamefont {Z.}~\bibnamefont {Hu}}, \bibinfo {author} {\bibfnamefont {H.~H.}\ \bibnamefont {Hsieh}}, \bibinfo {author} {\bibfnamefont {V.~L.~J.}\ \bibnamefont {Joly}}, \bibinfo {author} {\bibfnamefont {P.~A.}\ \bibnamefont {Joy}}, \bibinfo {author} {\bibfnamefont {M.~W.}\ \bibnamefont {Haverkort}}, \bibinfo {author} {\bibfnamefont {H.}~\bibnamefont {Wu}}, \bibinfo {author} {\bibfnamefont {A.}~\bibnamefont {Tanaka}}, \bibinfo {author} {\bibfnamefont {H.-J.}\ \bibnamefont {Lin}}, \bibinfo {author} {\bibfnamefont {C.~T.}\ \bibnamefont {Chen}}, \ and\ \bibinfo {author} {\bibfnamefont {L.~H.}\ \bibnamefont {Tjeng}},\ }\href {\doibase 10.1103/PhysRevB.77.125124} {\bibfield  {journal} {\bibinfo  {journal} {Phys. Rev. B}\ }\textbf {\bibinfo {volume} {77}},\ \bibinfo {pages} {125124} (\bibinfo {year} {2008})}\BibitemShut {NoStop}%
\bibitem [{\citenamefont {Zampieri}\ \emph {et~al.}(2002)\citenamefont {Zampieri}, \citenamefont {Abbate}, \citenamefont {Prado}, \citenamefont {Caneiro},\ and\ \citenamefont {Morikawa}}]{2002PhysBAbbateMn34}%
  \BibitemOpen
  \bibfield  {author} {\bibinfo {author} {\bibfnamefont {G.}~\bibnamefont {Zampieri}}, \bibinfo {author} {\bibfnamefont {M.}~\bibnamefont {Abbate}}, \bibinfo {author} {\bibfnamefont {F.}~\bibnamefont {Prado}}, \bibinfo {author} {\bibfnamefont {A.}~\bibnamefont {Caneiro}}, \ and\ \bibinfo {author} {\bibfnamefont {E.}~\bibnamefont {Morikawa}},\ }\href {\doibase https://doi.org/10.1016/S0921-4526(02)00618-X} {\bibfield  {journal} {\bibinfo  {journal} {Physica B: Condensed Matter}\ }\textbf {\bibinfo {volume} {320}},\ \bibinfo {pages} {51} (\bibinfo {year} {2002})},\ \bibinfo {note} {proceedings of the Fifth Latin American Workshop on Magnetism, Magnetic Materials and their Applications}\BibitemShut {NoStop}%
\bibitem [{\citenamefont {Liu}\ \emph {et~al.}(2017)\citenamefont {Liu}, \citenamefont {Zhou}, \citenamefont {Zhang}, \citenamefont {Hu}, \citenamefont {Kuo}, \citenamefont {Li}, \citenamefont {Wang}, \citenamefont {Tjeng}, \citenamefont {Pi}, \citenamefont {Tanaka}, \citenamefont {Song}, \citenamefont {Wang},\ and\ \citenamefont {Zhang}}]{2017JPCCTjengLiMn2O4}%
  \BibitemOpen
  \bibfield  {author} {\bibinfo {author} {\bibfnamefont {H.}~\bibnamefont {Liu}}, \bibinfo {author} {\bibfnamefont {J.}~\bibnamefont {Zhou}}, \bibinfo {author} {\bibfnamefont {L.}~\bibnamefont {Zhang}}, \bibinfo {author} {\bibfnamefont {Z.}~\bibnamefont {Hu}}, \bibinfo {author} {\bibfnamefont {C.}~\bibnamefont {Kuo}}, \bibinfo {author} {\bibfnamefont {J.}~\bibnamefont {Li}}, \bibinfo {author} {\bibfnamefont {Y.}~\bibnamefont {Wang}}, \bibinfo {author} {\bibfnamefont {L.~H.}\ \bibnamefont {Tjeng}}, \bibinfo {author} {\bibfnamefont {T.-W.}\ \bibnamefont {Pi}}, \bibinfo {author} {\bibfnamefont {A.}~\bibnamefont {Tanaka}}, \bibinfo {author} {\bibfnamefont {L.}~\bibnamefont {Song}}, \bibinfo {author} {\bibfnamefont {J.-Q.}\ \bibnamefont {Wang}}, \ and\ \bibinfo {author} {\bibfnamefont {S.}~\bibnamefont {Zhang}},\ }\href {\doibase 10.1021/acs.jpcc.7b03875} {\bibfield  {journal} {\bibinfo  {journal} {The Journal of Physical Chemistry C}\ }\textbf {\bibinfo {volume} {121}},\ \bibinfo {pages} {16079} (\bibinfo {year}
  {2017})}\BibitemShut {NoStop}%
\bibitem [{\citenamefont {Morita}\ \emph {et~al.}(2023)\citenamefont {Morita}, \citenamefont {Itoda}, \citenamefont {Hosono}, \citenamefont {Asakura}, \citenamefont {Okubo}, \citenamefont {Takagi}, \citenamefont {Yasui}, \citenamefont {Saini},\ and\ \citenamefont {Mizokawa}}]{2023JPSJMizokawaMn4}%
  \BibitemOpen
  \bibfield  {author} {\bibinfo {author} {\bibfnamefont {K.}~\bibnamefont {Morita}}, \bibinfo {author} {\bibfnamefont {M.}~\bibnamefont {Itoda}}, \bibinfo {author} {\bibfnamefont {E.}~\bibnamefont {Hosono}}, \bibinfo {author} {\bibfnamefont {D.}~\bibnamefont {Asakura}}, \bibinfo {author} {\bibfnamefont {M.}~\bibnamefont {Okubo}}, \bibinfo {author} {\bibfnamefont {Y.}~\bibnamefont {Takagi}}, \bibinfo {author} {\bibfnamefont {A.}~\bibnamefont {Yasui}}, \bibinfo {author} {\bibfnamefont {N.~L.}\ \bibnamefont {Saini}}, \ and\ \bibinfo {author} {\bibfnamefont {T.}~\bibnamefont {Mizokawa}},\ }\href {\doibase 10.7566/JPSJ.92.014702} {\bibfield  {journal} {\bibinfo  {journal} {Journal of the Physical Society of Japan}\ }\textbf {\bibinfo {volume} {92}},\ \bibinfo {pages} {014702} (\bibinfo {year} {2023})}\BibitemShut {NoStop}%
\bibitem [{\citenamefont {Wu}\ \emph {et~al.}(2011)\citenamefont {Wu}, \citenamefont {Chang}, \citenamefont {Schumann}, \citenamefont {Hu}, \citenamefont {Cezar}, \citenamefont {Burnus}, \citenamefont {Hollmann}, \citenamefont {Brookes}, \citenamefont {Tanaka}, \citenamefont {Braden}, \citenamefont {Tjeng},\ and\ \citenamefont {Khomskii}}]{2011PRBTjengLSMO}%
  \BibitemOpen
  \bibfield  {author} {\bibinfo {author} {\bibfnamefont {H.}~\bibnamefont {Wu}}, \bibinfo {author} {\bibfnamefont {C.~F.}\ \bibnamefont {Chang}}, \bibinfo {author} {\bibfnamefont {O.}~\bibnamefont {Schumann}}, \bibinfo {author} {\bibfnamefont {Z.}~\bibnamefont {Hu}}, \bibinfo {author} {\bibfnamefont {J.~C.}\ \bibnamefont {Cezar}}, \bibinfo {author} {\bibfnamefont {T.}~\bibnamefont {Burnus}}, \bibinfo {author} {\bibfnamefont {N.}~\bibnamefont {Hollmann}}, \bibinfo {author} {\bibfnamefont {N.~B.}\ \bibnamefont {Brookes}}, \bibinfo {author} {\bibfnamefont {A.}~\bibnamefont {Tanaka}}, \bibinfo {author} {\bibfnamefont {M.}~\bibnamefont {Braden}}, \bibinfo {author} {\bibfnamefont {L.~H.}\ \bibnamefont {Tjeng}}, \ and\ \bibinfo {author} {\bibfnamefont {D.~I.}\ \bibnamefont {Khomskii}},\ }\href {\doibase 10.1103/PhysRevB.84.155126} {\bibfield  {journal} {\bibinfo  {journal} {Phys. Rev. B}\ }\textbf {\bibinfo {volume} {84}},\ \bibinfo {pages} {155126} (\bibinfo {year} {2011})}\BibitemShut {NoStop}%
\bibitem [{\citenamefont {Nguyen}\ \emph {et~al.}(2021)\citenamefont {Nguyen}, \citenamefont {Rubio-Zuazo}, \citenamefont {Castro}, \citenamefont {de~Groot}, \citenamefont {Hariharan}, \citenamefont {Elizabeth}, \citenamefont {Oura}, \citenamefont {Tseng}, \citenamefont {Lin},\ and\ \citenamefont {Chainani}}]{2021PRBdeGrootTSMO}%
  \BibitemOpen
  \bibfield  {author} {\bibinfo {author} {\bibfnamefont {T.~L.}\ \bibnamefont {Nguyen}}, \bibinfo {author} {\bibfnamefont {J.}~\bibnamefont {Rubio-Zuazo}}, \bibinfo {author} {\bibfnamefont {G.~R.}\ \bibnamefont {Castro}}, \bibinfo {author} {\bibfnamefont {F.~M.~F.}\ \bibnamefont {de~Groot}}, \bibinfo {author} {\bibfnamefont {N.}~\bibnamefont {Hariharan}}, \bibinfo {author} {\bibfnamefont {S.}~\bibnamefont {Elizabeth}}, \bibinfo {author} {\bibfnamefont {M.}~\bibnamefont {Oura}}, \bibinfo {author} {\bibfnamefont {Y.~C.}\ \bibnamefont {Tseng}}, \bibinfo {author} {\bibfnamefont {H.~J.}\ \bibnamefont {Lin}}, \ and\ \bibinfo {author} {\bibfnamefont {A.}~\bibnamefont {Chainani}},\ }\href {\doibase 10.1103/PhysRevB.103.245131} {\bibfield  {journal} {\bibinfo  {journal} {Phys. Rev. B}\ }\textbf {\bibinfo {volume} {103}},\ \bibinfo {pages} {245131} (\bibinfo {year} {2021})}\BibitemShut {NoStop}%
\bibitem [{\citenamefont {Balasubramanian}\ \emph {et~al.}(2018)\citenamefont {Balasubramanian}, \citenamefont {Joshi}, \citenamefont {Yadav}, \citenamefont {de~Groot}, \citenamefont {Singh}, \citenamefont {Ray}, \citenamefont {Gupta}, \citenamefont {Singh}, \citenamefont {Maurya}, \citenamefont {Elizabeth}, \citenamefont {Varma}, \citenamefont {Maitra},\ and\ \citenamefont {Malik}}]{2018JPCMdeGrootMnNi}%
  \BibitemOpen
  \bibfield  {author} {\bibinfo {author} {\bibfnamefont {P.}~\bibnamefont {Balasubramanian}}, \bibinfo {author} {\bibfnamefont {S.~R.}\ \bibnamefont {Joshi}}, \bibinfo {author} {\bibfnamefont {R.}~\bibnamefont {Yadav}}, \bibinfo {author} {\bibfnamefont {F.~M.~F.}\ \bibnamefont {de~Groot}}, \bibinfo {author} {\bibfnamefont {A.~K.}\ \bibnamefont {Singh}}, \bibinfo {author} {\bibfnamefont {A.}~\bibnamefont {Ray}}, \bibinfo {author} {\bibfnamefont {M.}~\bibnamefont {Gupta}}, \bibinfo {author} {\bibfnamefont {A.}~\bibnamefont {Singh}}, \bibinfo {author} {\bibfnamefont {S.}~\bibnamefont {Maurya}}, \bibinfo {author} {\bibfnamefont {S.}~\bibnamefont {Elizabeth}}, \bibinfo {author} {\bibfnamefont {S.}~\bibnamefont {Varma}}, \bibinfo {author} {\bibfnamefont {T.}~\bibnamefont {Maitra}}, \ and\ \bibinfo {author} {\bibfnamefont {V.}~\bibnamefont {Malik}},\ }\href {\doibase 10.1088/1361-648X/aae168} {\bibfield  {journal} {\bibinfo  {journal} {Journal of Physics: Condensed Matter}\ }\textbf {\bibinfo {volume} {30}},\
  \bibinfo {pages} {435603} (\bibinfo {year} {2018})}\BibitemShut {NoStop}%
\bibitem [{\citenamefont {Chen}\ \emph {et~al.}(2010)\citenamefont {Chen}, \citenamefont {Hu}, \citenamefont {Jeng}, \citenamefont {Chin}, \citenamefont {Lee}, \citenamefont {Huang}, \citenamefont {Lu}, \citenamefont {Chen}, \citenamefont {Haw}, \citenamefont {Chou}, \citenamefont {Lin}, \citenamefont {Shen}, \citenamefont {Liu}, \citenamefont {Tanaka}, \citenamefont {Tjeng},\ and\ \citenamefont {Chen}}]{2010PRBTjengDyMnO3}%
  \BibitemOpen
  \bibfield  {author} {\bibinfo {author} {\bibfnamefont {J.~M.}\ \bibnamefont {Chen}}, \bibinfo {author} {\bibfnamefont {Z.}~\bibnamefont {Hu}}, \bibinfo {author} {\bibfnamefont {H.~T.}\ \bibnamefont {Jeng}}, \bibinfo {author} {\bibfnamefont {Y.~Y.}\ \bibnamefont {Chin}}, \bibinfo {author} {\bibfnamefont {J.~M.}\ \bibnamefont {Lee}}, \bibinfo {author} {\bibfnamefont {S.~W.}\ \bibnamefont {Huang}}, \bibinfo {author} {\bibfnamefont {K.~T.}\ \bibnamefont {Lu}}, \bibinfo {author} {\bibfnamefont {C.~K.}\ \bibnamefont {Chen}}, \bibinfo {author} {\bibfnamefont {S.~C.}\ \bibnamefont {Haw}}, \bibinfo {author} {\bibfnamefont {T.~L.}\ \bibnamefont {Chou}}, \bibinfo {author} {\bibfnamefont {H.-J.}\ \bibnamefont {Lin}}, \bibinfo {author} {\bibfnamefont {C.~C.}\ \bibnamefont {Shen}}, \bibinfo {author} {\bibfnamefont {R.~S.}\ \bibnamefont {Liu}}, \bibinfo {author} {\bibfnamefont {A.}~\bibnamefont {Tanaka}}, \bibinfo {author} {\bibfnamefont {L.~H.}\ \bibnamefont {Tjeng}}, \ and\ \bibinfo {author} {\bibfnamefont {C.~T.}\
  \bibnamefont {Chen}},\ }\href {\doibase 10.1103/PhysRevB.81.201102} {\bibfield  {journal} {\bibinfo  {journal} {Phys. Rev. B}\ }\textbf {\bibinfo {volume} {81}},\ \bibinfo {pages} {201102} (\bibinfo {year} {2010})}\BibitemShut {NoStop}%
\bibitem [{\citenamefont {Fujimori}\ \emph {et~al.}(1990)\citenamefont {Fujimori}, \citenamefont {Kimizuka}, \citenamefont {Akahane}, \citenamefont {Chiba}, \citenamefont {Kimura}, \citenamefont {Minami}, \citenamefont {Siratori}, \citenamefont {Taniguchi}, \citenamefont {Ogawa},\ and\ \citenamefont {Suga}}]{1990PRBFujimoriMnO}%
  \BibitemOpen
  \bibfield  {author} {\bibinfo {author} {\bibfnamefont {A.}~\bibnamefont {Fujimori}}, \bibinfo {author} {\bibfnamefont {N.}~\bibnamefont {Kimizuka}}, \bibinfo {author} {\bibfnamefont {T.}~\bibnamefont {Akahane}}, \bibinfo {author} {\bibfnamefont {T.}~\bibnamefont {Chiba}}, \bibinfo {author} {\bibfnamefont {S.}~\bibnamefont {Kimura}}, \bibinfo {author} {\bibfnamefont {F.}~\bibnamefont {Minami}}, \bibinfo {author} {\bibfnamefont {K.}~\bibnamefont {Siratori}}, \bibinfo {author} {\bibfnamefont {M.}~\bibnamefont {Taniguchi}}, \bibinfo {author} {\bibfnamefont {S.}~\bibnamefont {Ogawa}}, \ and\ \bibinfo {author} {\bibfnamefont {S.}~\bibnamefont {Suga}},\ }\href {\doibase 10.1103/PhysRevB.42.7580} {\bibfield  {journal} {\bibinfo  {journal} {Phys. Rev. B}\ }\textbf {\bibinfo {volume} {42}},\ \bibinfo {pages} {7580} (\bibinfo {year} {1990})}\BibitemShut {NoStop}%
\bibitem [{\citenamefont {Okada}\ and\ \citenamefont {Kotani}(1992{\natexlab{a}})}]{1992JPSJKotaniFeMnXPS}%
  \BibitemOpen
  \bibfield  {author} {\bibinfo {author} {\bibfnamefont {K.}~\bibnamefont {Okada}}\ and\ \bibinfo {author} {\bibfnamefont {A.}~\bibnamefont {Kotani}},\ }\href {\doibase 10.1143/JPSJ.61.4619} {\bibfield  {journal} {\bibinfo  {journal} {Journal of the Physical Society of Japan}\ }\textbf {\bibinfo {volume} {61}},\ \bibinfo {pages} {4619} (\bibinfo {year} {1992}{\natexlab{a}})}\BibitemShut {NoStop}%
\bibitem [{\citenamefont {Taguchi}\ \emph {et~al.}(1997)\citenamefont {Taguchi}, \citenamefont {Uozumi},\ and\ \citenamefont {Kotani}}]{1997JPSJKotaniMnXPS}%
  \BibitemOpen
  \bibfield  {author} {\bibinfo {author} {\bibfnamefont {M.}~\bibnamefont {Taguchi}}, \bibinfo {author} {\bibfnamefont {T.}~\bibnamefont {Uozumi}}, \ and\ \bibinfo {author} {\bibfnamefont {A.}~\bibnamefont {Kotani}},\ }\href {\doibase 10.1143/JPSJ.66.247} {\bibfield  {journal} {\bibinfo  {journal} {Journal of the Physical Society of Japan}\ }\textbf {\bibinfo {volume} {66}},\ \bibinfo {pages} {247} (\bibinfo {year} {1997})}\BibitemShut {NoStop}%
\bibitem [{\citenamefont {Shoji}\ \emph {et~al.}(2003)\citenamefont {Shoji}, \citenamefont {Taguchi}, \citenamefont {Hirai}, \citenamefont {Iwazumi}, \citenamefont {Kotani}, \citenamefont {Nanao},\ and\ \citenamefont {Isozumi}}]{2003JPSJKotaniMnKEdge}%
  \BibitemOpen
  \bibfield  {author} {\bibinfo {author} {\bibfnamefont {H.}~\bibnamefont {Shoji}}, \bibinfo {author} {\bibfnamefont {M.}~\bibnamefont {Taguchi}}, \bibinfo {author} {\bibfnamefont {E.}~\bibnamefont {Hirai}}, \bibinfo {author} {\bibfnamefont {T.}~\bibnamefont {Iwazumi}}, \bibinfo {author} {\bibfnamefont {A.}~\bibnamefont {Kotani}}, \bibinfo {author} {\bibfnamefont {S.}~\bibnamefont {Nanao}}, \ and\ \bibinfo {author} {\bibfnamefont {Y.}~\bibnamefont {Isozumi}},\ }\href {\doibase 10.1143/JPSJ.72.1560} {\bibfield  {journal} {\bibinfo  {journal} {Journal of the Physical Society of Japan}\ }\textbf {\bibinfo {volume} {72}},\ \bibinfo {pages} {1560} (\bibinfo {year} {2003})}\BibitemShut {NoStop}%
\bibitem [{\citenamefont {van Elp}\ \emph {et~al.}(1991{\natexlab{a}})\citenamefont {van Elp}, \citenamefont {Potze}, \citenamefont {Eskes}, \citenamefont {Berger},\ and\ \citenamefont {Sawatzky}}]{1991PRBvanElpMnO}%
  \BibitemOpen
  \bibfield  {author} {\bibinfo {author} {\bibfnamefont {J.}~\bibnamefont {van Elp}}, \bibinfo {author} {\bibfnamefont {R.~H.}\ \bibnamefont {Potze}}, \bibinfo {author} {\bibfnamefont {H.}~\bibnamefont {Eskes}}, \bibinfo {author} {\bibfnamefont {R.}~\bibnamefont {Berger}}, \ and\ \bibinfo {author} {\bibfnamefont {G.~A.}\ \bibnamefont {Sawatzky}},\ }\href {\doibase 10.1103/PhysRevB.44.1530} {\bibfield  {journal} {\bibinfo  {journal} {Phys. Rev. B}\ }\textbf {\bibinfo {volume} {44}},\ \bibinfo {pages} {1530} (\bibinfo {year} {1991}{\natexlab{a}})}\BibitemShut {NoStop}%
\bibitem [{\citenamefont {Maignan}\ \emph {et~al.}(2020)\citenamefont {Maignan}, \citenamefont {Peng}, \citenamefont {Komarek}, \citenamefont {Kuo}, \citenamefont {Chang}, \citenamefont {Wang}, \citenamefont {Hu}, \citenamefont {Chen}, \citenamefont {Tanaka}, \citenamefont {Tjeng}, \citenamefont {Lebedev},\ and\ \citenamefont {Martin}}]{2020ChemMatTjengMn3WO6}%
  \BibitemOpen
  \bibfield  {author} {\bibinfo {author} {\bibfnamefont {A.}~\bibnamefont {Maignan}}, \bibinfo {author} {\bibfnamefont {W.}~\bibnamefont {Peng}}, \bibinfo {author} {\bibfnamefont {A.~C.}\ \bibnamefont {Komarek}}, \bibinfo {author} {\bibfnamefont {C.-Y.}\ \bibnamefont {Kuo}}, \bibinfo {author} {\bibfnamefont {C.-F.}\ \bibnamefont {Chang}}, \bibinfo {author} {\bibfnamefont {X.}~\bibnamefont {Wang}}, \bibinfo {author} {\bibfnamefont {Z.}~\bibnamefont {Hu}}, \bibinfo {author} {\bibfnamefont {C.-T.}\ \bibnamefont {Chen}}, \bibinfo {author} {\bibfnamefont {A.}~\bibnamefont {Tanaka}}, \bibinfo {author} {\bibfnamefont {L.~H.}\ \bibnamefont {Tjeng}}, \bibinfo {author} {\bibfnamefont {O.~I.}\ \bibnamefont {Lebedev}}, \ and\ \bibinfo {author} {\bibfnamefont {C.}~\bibnamefont {Martin}},\ }\href {\doibase 10.1021/acs.chemmater.0c01303} {\bibfield  {journal} {\bibinfo  {journal} {Chemistry of Materials}\ }\textbf {\bibinfo {volume} {32}},\ \bibinfo {pages} {5664} (\bibinfo {year} {2020})}\BibitemShut {NoStop}%
\bibitem [{\citenamefont {Hollmann}\ \emph {et~al.}(2010)\citenamefont {Hollmann}, \citenamefont {Hu}, \citenamefont {Willers}, \citenamefont {Bohat\'y}, \citenamefont {Becker}, \citenamefont {Tanaka}, \citenamefont {Hsieh}, \citenamefont {Lin}, \citenamefont {Chen},\ and\ \citenamefont {Tjeng}}]{2010PRBTjengMnCoWO4}%
  \BibitemOpen
  \bibfield  {author} {\bibinfo {author} {\bibfnamefont {N.}~\bibnamefont {Hollmann}}, \bibinfo {author} {\bibfnamefont {Z.}~\bibnamefont {Hu}}, \bibinfo {author} {\bibfnamefont {T.}~\bibnamefont {Willers}}, \bibinfo {author} {\bibfnamefont {L.}~\bibnamefont {Bohat\'y}}, \bibinfo {author} {\bibfnamefont {P.}~\bibnamefont {Becker}}, \bibinfo {author} {\bibfnamefont {A.}~\bibnamefont {Tanaka}}, \bibinfo {author} {\bibfnamefont {H.~H.}\ \bibnamefont {Hsieh}}, \bibinfo {author} {\bibfnamefont {H.-J.}\ \bibnamefont {Lin}}, \bibinfo {author} {\bibfnamefont {C.~T.}\ \bibnamefont {Chen}}, \ and\ \bibinfo {author} {\bibfnamefont {L.~H.}\ \bibnamefont {Tjeng}},\ }\href {\doibase 10.1103/PhysRevB.82.184429} {\bibfield  {journal} {\bibinfo  {journal} {Phys. Rev. B}\ }\textbf {\bibinfo {volume} {82}},\ \bibinfo {pages} {184429} (\bibinfo {year} {2010})}\BibitemShut {NoStop}%
\bibitem [{\citenamefont {Tsuyama}\ \emph {et~al.}(2015)\citenamefont {Tsuyama}, \citenamefont {Matsuda}, \citenamefont {Chakraverty}, \citenamefont {Okamoto}, \citenamefont {Ikenaga}, \citenamefont {Tanaka}, \citenamefont {Mizokawa}, \citenamefont {Hwang}, \citenamefont {Tokura},\ and\ \citenamefont {Wadati}}]{2015PRBWadatiBaFeO3}%
  \BibitemOpen
  \bibfield  {author} {\bibinfo {author} {\bibfnamefont {T.}~\bibnamefont {Tsuyama}}, \bibinfo {author} {\bibfnamefont {T.}~\bibnamefont {Matsuda}}, \bibinfo {author} {\bibfnamefont {S.}~\bibnamefont {Chakraverty}}, \bibinfo {author} {\bibfnamefont {J.}~\bibnamefont {Okamoto}}, \bibinfo {author} {\bibfnamefont {E.}~\bibnamefont {Ikenaga}}, \bibinfo {author} {\bibfnamefont {A.}~\bibnamefont {Tanaka}}, \bibinfo {author} {\bibfnamefont {T.}~\bibnamefont {Mizokawa}}, \bibinfo {author} {\bibfnamefont {H.~Y.}\ \bibnamefont {Hwang}}, \bibinfo {author} {\bibfnamefont {Y.}~\bibnamefont {Tokura}}, \ and\ \bibinfo {author} {\bibfnamefont {H.}~\bibnamefont {Wadati}},\ }\href {\doibase 10.1103/PhysRevB.91.115101} {\bibfield  {journal} {\bibinfo  {journal} {Phys. Rev. B}\ }\textbf {\bibinfo {volume} {91}},\ \bibinfo {pages} {115101} (\bibinfo {year} {2015})}\BibitemShut {NoStop}%
\bibitem [{\citenamefont {Wang}\ \emph {et~al.}(2021)\citenamefont {Wang}, \citenamefont {Hu}, \citenamefont {Agrestini}, \citenamefont {Herrero-Martín}, \citenamefont {Valvidares}, \citenamefont {Sankar}, \citenamefont {Chou}, \citenamefont {Chu}, \citenamefont {Tanaka}, \citenamefont {Tjeng},\ and\ \citenamefont {Pellegrin}}]{2021JMMMTjengCoFe2O4}%
  \BibitemOpen
  \bibfield  {author} {\bibinfo {author} {\bibfnamefont {X.}~\bibnamefont {Wang}}, \bibinfo {author} {\bibfnamefont {Z.}~\bibnamefont {Hu}}, \bibinfo {author} {\bibfnamefont {S.}~\bibnamefont {Agrestini}}, \bibinfo {author} {\bibfnamefont {J.}~\bibnamefont {Herrero-Martín}}, \bibinfo {author} {\bibfnamefont {M.}~\bibnamefont {Valvidares}}, \bibinfo {author} {\bibfnamefont {R.}~\bibnamefont {Sankar}}, \bibinfo {author} {\bibfnamefont {F.-C.}\ \bibnamefont {Chou}}, \bibinfo {author} {\bibfnamefont {Y.-H.}\ \bibnamefont {Chu}}, \bibinfo {author} {\bibfnamefont {A.}~\bibnamefont {Tanaka}}, \bibinfo {author} {\bibfnamefont {L.~H.}\ \bibnamefont {Tjeng}}, \ and\ \bibinfo {author} {\bibfnamefont {E.}~\bibnamefont {Pellegrin}},\ }\href {\doibase https://doi.org/10.1016/j.jmmm.2021.167940} {\bibfield  {journal} {\bibinfo  {journal} {Journal of Magnetism and Magnetic Materials}\ }\textbf {\bibinfo {volume} {530}},\ \bibinfo {pages} {167940} (\bibinfo {year} {2021})}\BibitemShut {NoStop}%
\bibitem [{\citenamefont {Asakura}\ \emph {et~al.}(2018)\citenamefont {Asakura}, \citenamefont {Nanba}, \citenamefont {Makinose}, \citenamefont {Matsuda}, \citenamefont {Glans}, \citenamefont {Guo},\ and\ \citenamefont {Hosono}}]{2018CPCHosonoLPFO}%
  \BibitemOpen
  \bibfield  {author} {\bibinfo {author} {\bibfnamefont {D.}~\bibnamefont {Asakura}}, \bibinfo {author} {\bibfnamefont {Y.}~\bibnamefont {Nanba}}, \bibinfo {author} {\bibfnamefont {Y.}~\bibnamefont {Makinose}}, \bibinfo {author} {\bibfnamefont {H.}~\bibnamefont {Matsuda}}, \bibinfo {author} {\bibfnamefont {P.-A.}\ \bibnamefont {Glans}}, \bibinfo {author} {\bibfnamefont {J.}~\bibnamefont {Guo}}, \ and\ \bibinfo {author} {\bibfnamefont {E.}~\bibnamefont {Hosono}},\ }\href {\doibase https://doi.org/10.1002/cphc.201800038} {\bibfield  {journal} {\bibinfo  {journal} {ChemPhysChem}\ }\textbf {\bibinfo {volume} {19}},\ \bibinfo {pages} {988} (\bibinfo {year} {2018})}\BibitemShut {NoStop}%
\bibitem [{\citenamefont {Fujimori}\ \emph {et~al.}(1986)\citenamefont {Fujimori}, \citenamefont {Saeki}, \citenamefont {Kimizuka}, \citenamefont {Taniguchi},\ and\ \citenamefont {Suga}}]{1986PRBFujimoriFe2O3}%
  \BibitemOpen
  \bibfield  {author} {\bibinfo {author} {\bibfnamefont {A.}~\bibnamefont {Fujimori}}, \bibinfo {author} {\bibfnamefont {M.}~\bibnamefont {Saeki}}, \bibinfo {author} {\bibfnamefont {N.}~\bibnamefont {Kimizuka}}, \bibinfo {author} {\bibfnamefont {M.}~\bibnamefont {Taniguchi}}, \ and\ \bibinfo {author} {\bibfnamefont {S.}~\bibnamefont {Suga}},\ }\href {\doibase 10.1103/PhysRevB.34.7318} {\bibfield  {journal} {\bibinfo  {journal} {Phys. Rev. B}\ }\textbf {\bibinfo {volume} {34}},\ \bibinfo {pages} {7318} (\bibinfo {year} {1986})}\BibitemShut {NoStop}%
\bibitem [{\citenamefont {Ellis}\ \emph {et~al.}(2022)\citenamefont {Ellis}, \citenamefont {Wang}, \citenamefont {Wong}, \citenamefont {Cooper}, \citenamefont {Schulz}, \citenamefont {Chuang}, \citenamefont {Piekner}, \citenamefont {Grave}, \citenamefont {Schleuning}, \citenamefont {Friedrich}, \citenamefont {de~Groot},\ and\ \citenamefont {Rothschild}}]{2022PRBdeGrootFe2O3}%
  \BibitemOpen
  \bibfield  {author} {\bibinfo {author} {\bibfnamefont {D.~S.}\ \bibnamefont {Ellis}}, \bibinfo {author} {\bibfnamefont {R.-P.}\ \bibnamefont {Wang}}, \bibinfo {author} {\bibfnamefont {D.}~\bibnamefont {Wong}}, \bibinfo {author} {\bibfnamefont {J.~K.}\ \bibnamefont {Cooper}}, \bibinfo {author} {\bibfnamefont {C.}~\bibnamefont {Schulz}}, \bibinfo {author} {\bibfnamefont {Y.-D.}\ \bibnamefont {Chuang}}, \bibinfo {author} {\bibfnamefont {Y.}~\bibnamefont {Piekner}}, \bibinfo {author} {\bibfnamefont {D.~A.}\ \bibnamefont {Grave}}, \bibinfo {author} {\bibfnamefont {M.}~\bibnamefont {Schleuning}}, \bibinfo {author} {\bibfnamefont {D.}~\bibnamefont {Friedrich}}, \bibinfo {author} {\bibfnamefont {F.~M.~F.}\ \bibnamefont {de~Groot}}, \ and\ \bibinfo {author} {\bibfnamefont {A.}~\bibnamefont {Rothschild}},\ }\href {\doibase 10.1103/PhysRevB.105.075101} {\bibfield  {journal} {\bibinfo  {journal} {Phys. Rev. B}\ }\textbf {\bibinfo {volume} {105}},\ \bibinfo {pages} {075101} (\bibinfo {year} {2022})}\BibitemShut
  {NoStop}%
\bibitem [{\citenamefont {Ghiasi}\ \emph {et~al.}(2019)\citenamefont {Ghiasi}, \citenamefont {Hariki}, \citenamefont {Winder}, \citenamefont {Kune\ifmmode~\check{s}\else \v{s}\fi{}}, \citenamefont {Regoutz}, \citenamefont {Lee}, \citenamefont {Hu}, \citenamefont {Rueff},\ and\ \citenamefont {de~Groot}}]{2019PRBHariki1sXPS}%
  \BibitemOpen
  \bibfield  {author} {\bibinfo {author} {\bibfnamefont {M.}~\bibnamefont {Ghiasi}}, \bibinfo {author} {\bibfnamefont {A.}~\bibnamefont {Hariki}}, \bibinfo {author} {\bibfnamefont {M.}~\bibnamefont {Winder}}, \bibinfo {author} {\bibfnamefont {J.}~\bibnamefont {Kune\ifmmode~\check{s}\else \v{s}\fi{}}}, \bibinfo {author} {\bibfnamefont {A.}~\bibnamefont {Regoutz}}, \bibinfo {author} {\bibfnamefont {T.-L.}\ \bibnamefont {Lee}}, \bibinfo {author} {\bibfnamefont {Y.}~\bibnamefont {Hu}}, \bibinfo {author} {\bibfnamefont {J.-P.}\ \bibnamefont {Rueff}}, \ and\ \bibinfo {author} {\bibfnamefont {F.~M.~F.}\ \bibnamefont {de~Groot}},\ }\href {\doibase 10.1103/PhysRevB.100.075146} {\bibfield  {journal} {\bibinfo  {journal} {Phys. Rev. B}\ }\textbf {\bibinfo {volume} {100}},\ \bibinfo {pages} {075146} (\bibinfo {year} {2019})}\BibitemShut {NoStop}%
\bibitem [{\citenamefont {Chang}\ \emph {et~al.}(2016)\citenamefont {Chang}, \citenamefont {Hu}, \citenamefont {Klein}, \citenamefont {Liu}, \citenamefont {Sutarto}, \citenamefont {Tanaka}, \citenamefont {Cezar}, \citenamefont {Brookes}, \citenamefont {Lin}, \citenamefont {Hsieh}, \citenamefont {Chen}, \citenamefont {Rata},\ and\ \citenamefont {Tjeng}}]{2016PRXTjengFe3O4}%
  \BibitemOpen
  \bibfield  {author} {\bibinfo {author} {\bibfnamefont {C.~F.}\ \bibnamefont {Chang}}, \bibinfo {author} {\bibfnamefont {Z.}~\bibnamefont {Hu}}, \bibinfo {author} {\bibfnamefont {S.}~\bibnamefont {Klein}}, \bibinfo {author} {\bibfnamefont {X.~H.}\ \bibnamefont {Liu}}, \bibinfo {author} {\bibfnamefont {R.}~\bibnamefont {Sutarto}}, \bibinfo {author} {\bibfnamefont {A.}~\bibnamefont {Tanaka}}, \bibinfo {author} {\bibfnamefont {J.~C.}\ \bibnamefont {Cezar}}, \bibinfo {author} {\bibfnamefont {N.~B.}\ \bibnamefont {Brookes}}, \bibinfo {author} {\bibfnamefont {H.-J.}\ \bibnamefont {Lin}}, \bibinfo {author} {\bibfnamefont {H.~H.}\ \bibnamefont {Hsieh}}, \bibinfo {author} {\bibfnamefont {C.~T.}\ \bibnamefont {Chen}}, \bibinfo {author} {\bibfnamefont {A.~D.}\ \bibnamefont {Rata}}, \ and\ \bibinfo {author} {\bibfnamefont {L.~H.}\ \bibnamefont {Tjeng}},\ }\href {\doibase 10.1103/PhysRevX.6.041011} {\bibfield  {journal} {\bibinfo  {journal} {Phys. Rev. X}\ }\textbf {\bibinfo {volume} {6}},\ \bibinfo {pages} {041011}
  (\bibinfo {year} {2016})}\BibitemShut {NoStop}%
\bibitem [{\citenamefont {Kuo}\ \emph {et~al.}(2014)\citenamefont {Kuo}, \citenamefont {Drees}, \citenamefont {Fern\'andez-D\'{\i}az}, \citenamefont {Zhao}, \citenamefont {Vasylechko}, \citenamefont {Sheptyakov}, \citenamefont {Bell}, \citenamefont {Pi}, \citenamefont {Lin}, \citenamefont {Wu}, \citenamefont {Pellegrin}, \citenamefont {Valvidares}, \citenamefont {Li}, \citenamefont {Adler}, \citenamefont {Todorova}, \citenamefont {K\"uchler}, \citenamefont {Steppke}, \citenamefont {Tjeng}, \citenamefont {Hu},\ and\ \citenamefont {Komarek}}]{2014PRLTjengSmFeO3}%
  \BibitemOpen
  \bibfield  {author} {\bibinfo {author} {\bibfnamefont {C.-Y.}\ \bibnamefont {Kuo}}, \bibinfo {author} {\bibfnamefont {Y.}~\bibnamefont {Drees}}, \bibinfo {author} {\bibfnamefont {M.~T.}\ \bibnamefont {Fern\'andez-D\'{\i}az}}, \bibinfo {author} {\bibfnamefont {L.}~\bibnamefont {Zhao}}, \bibinfo {author} {\bibfnamefont {L.}~\bibnamefont {Vasylechko}}, \bibinfo {author} {\bibfnamefont {D.}~\bibnamefont {Sheptyakov}}, \bibinfo {author} {\bibfnamefont {A.~M.~T.}\ \bibnamefont {Bell}}, \bibinfo {author} {\bibfnamefont {T.~W.}\ \bibnamefont {Pi}}, \bibinfo {author} {\bibfnamefont {H.-J.}\ \bibnamefont {Lin}}, \bibinfo {author} {\bibfnamefont {M.-K.}\ \bibnamefont {Wu}}, \bibinfo {author} {\bibfnamefont {E.}~\bibnamefont {Pellegrin}}, \bibinfo {author} {\bibfnamefont {S.~M.}\ \bibnamefont {Valvidares}}, \bibinfo {author} {\bibfnamefont {Z.~W.}\ \bibnamefont {Li}}, \bibinfo {author} {\bibfnamefont {P.}~\bibnamefont {Adler}}, \bibinfo {author} {\bibfnamefont {A.}~\bibnamefont {Todorova}}, \bibinfo {author}
  {\bibfnamefont {R.}~\bibnamefont {K\"uchler}}, \bibinfo {author} {\bibfnamefont {A.}~\bibnamefont {Steppke}}, \bibinfo {author} {\bibfnamefont {L.~H.}\ \bibnamefont {Tjeng}}, \bibinfo {author} {\bibfnamefont {Z.}~\bibnamefont {Hu}}, \ and\ \bibinfo {author} {\bibfnamefont {A.~C.}\ \bibnamefont {Komarek}},\ }\href {\doibase 10.1103/PhysRevLett.113.217203} {\bibfield  {journal} {\bibinfo  {journal} {Phys. Rev. Lett.}\ }\textbf {\bibinfo {volume} {113}},\ \bibinfo {pages} {217203} (\bibinfo {year} {2014})}\BibitemShut {NoStop}%
\bibitem [{\citenamefont {Hollmann}\ \emph {et~al.}(2011)\citenamefont {Hollmann}, \citenamefont {Valldor}, \citenamefont {Wu}, \citenamefont {Hu}, \citenamefont {Qureshi}, \citenamefont {Willers}, \citenamefont {Chin}, \citenamefont {Cezar}, \citenamefont {Tanaka}, \citenamefont {Brookes},\ and\ \citenamefont {Tjeng}}]{2011PRBTjengCBFO}%
  \BibitemOpen
  \bibfield  {author} {\bibinfo {author} {\bibfnamefont {N.}~\bibnamefont {Hollmann}}, \bibinfo {author} {\bibfnamefont {M.}~\bibnamefont {Valldor}}, \bibinfo {author} {\bibfnamefont {H.}~\bibnamefont {Wu}}, \bibinfo {author} {\bibfnamefont {Z.}~\bibnamefont {Hu}}, \bibinfo {author} {\bibfnamefont {N.}~\bibnamefont {Qureshi}}, \bibinfo {author} {\bibfnamefont {T.}~\bibnamefont {Willers}}, \bibinfo {author} {\bibfnamefont {Y.-Y.}\ \bibnamefont {Chin}}, \bibinfo {author} {\bibfnamefont {J.~C.}\ \bibnamefont {Cezar}}, \bibinfo {author} {\bibfnamefont {A.}~\bibnamefont {Tanaka}}, \bibinfo {author} {\bibfnamefont {N.~B.}\ \bibnamefont {Brookes}}, \ and\ \bibinfo {author} {\bibfnamefont {L.~H.}\ \bibnamefont {Tjeng}},\ }\href {\doibase 10.1103/PhysRevB.83.180405} {\bibfield  {journal} {\bibinfo  {journal} {Phys. Rev. B}\ }\textbf {\bibinfo {volume} {83}},\ \bibinfo {pages} {180405} (\bibinfo {year} {2011})}\BibitemShut {NoStop}%
\bibitem [{\citenamefont {Fujimori}\ \emph {et~al.}(1987)\citenamefont {Fujimori}, \citenamefont {Kimizuka}, \citenamefont {Taniguchi},\ and\ \citenamefont {Suga}}]{1987PRBFujimoriFeO}%
  \BibitemOpen
  \bibfield  {author} {\bibinfo {author} {\bibfnamefont {A.}~\bibnamefont {Fujimori}}, \bibinfo {author} {\bibfnamefont {N.}~\bibnamefont {Kimizuka}}, \bibinfo {author} {\bibfnamefont {M.}~\bibnamefont {Taniguchi}}, \ and\ \bibinfo {author} {\bibfnamefont {S.}~\bibnamefont {Suga}},\ }\href {\doibase 10.1103/PhysRevB.36.6691} {\bibfield  {journal} {\bibinfo  {journal} {Phys. Rev. B}\ }\textbf {\bibinfo {volume} {36}},\ \bibinfo {pages} {6691} (\bibinfo {year} {1987})}\BibitemShut {NoStop}%
\bibitem [{\citenamefont {Haupricht}\ \emph {et~al.}(2010)\citenamefont {Haupricht}, \citenamefont {Sutarto}, \citenamefont {Haverkort}, \citenamefont {Ott}, \citenamefont {Tanaka}, \citenamefont {Hsieh}, \citenamefont {Lin}, \citenamefont {Chen}, \citenamefont {Hu},\ and\ \citenamefont {Tjeng}}]{2010PRBTjengMgOFe}%
  \BibitemOpen
  \bibfield  {author} {\bibinfo {author} {\bibfnamefont {T.}~\bibnamefont {Haupricht}}, \bibinfo {author} {\bibfnamefont {R.}~\bibnamefont {Sutarto}}, \bibinfo {author} {\bibfnamefont {M.~W.}\ \bibnamefont {Haverkort}}, \bibinfo {author} {\bibfnamefont {H.}~\bibnamefont {Ott}}, \bibinfo {author} {\bibfnamefont {A.}~\bibnamefont {Tanaka}}, \bibinfo {author} {\bibfnamefont {H.~H.}\ \bibnamefont {Hsieh}}, \bibinfo {author} {\bibfnamefont {H.-J.}\ \bibnamefont {Lin}}, \bibinfo {author} {\bibfnamefont {C.~T.}\ \bibnamefont {Chen}}, \bibinfo {author} {\bibfnamefont {Z.}~\bibnamefont {Hu}}, \ and\ \bibinfo {author} {\bibfnamefont {L.~H.}\ \bibnamefont {Tjeng}},\ }\href {\doibase 10.1103/PhysRevB.82.035120} {\bibfield  {journal} {\bibinfo  {journal} {Phys. Rev. B}\ }\textbf {\bibinfo {volume} {82}},\ \bibinfo {pages} {035120} (\bibinfo {year} {2010})}\BibitemShut {NoStop}%
\bibitem [{\citenamefont {Altendorf}\ \emph {et~al.}(2023)\citenamefont {Altendorf}, \citenamefont {Takegami}, \citenamefont {Mel\'endez-Sans}, \citenamefont {Chang}, \citenamefont {Yoshimura}, \citenamefont {Tsuei}, \citenamefont {Tanaka}, \citenamefont {Schmidt},\ and\ \citenamefont {Tjeng}}]{2023PRBTjengFeWO4}%
  \BibitemOpen
  \bibfield  {author} {\bibinfo {author} {\bibfnamefont {S.~G.}\ \bibnamefont {Altendorf}}, \bibinfo {author} {\bibfnamefont {D.}~\bibnamefont {Takegami}}, \bibinfo {author} {\bibfnamefont {A.}~\bibnamefont {Mel\'endez-Sans}}, \bibinfo {author} {\bibfnamefont {C.~F.}\ \bibnamefont {Chang}}, \bibinfo {author} {\bibfnamefont {M.}~\bibnamefont {Yoshimura}}, \bibinfo {author} {\bibfnamefont {K.~D.}\ \bibnamefont {Tsuei}}, \bibinfo {author} {\bibfnamefont {A.}~\bibnamefont {Tanaka}}, \bibinfo {author} {\bibfnamefont {M.}~\bibnamefont {Schmidt}}, \ and\ \bibinfo {author} {\bibfnamefont {L.~H.}\ \bibnamefont {Tjeng}},\ }\href {\doibase 10.1103/PhysRevB.108.085119} {\bibfield  {journal} {\bibinfo  {journal} {Phys. Rev. B}\ }\textbf {\bibinfo {volume} {108}},\ \bibinfo {pages} {085119} (\bibinfo {year} {2023})}\BibitemShut {NoStop}%
\bibitem [{\citenamefont {Chin}\ \emph {et~al.}(2021)\citenamefont {Chin}, \citenamefont {Hu}, \citenamefont {Shimakawa}, \citenamefont {Yang}, \citenamefont {Long}, \citenamefont {Tanaka}, \citenamefont {Tjeng}, \citenamefont {Lin},\ and\ \citenamefont {Chen}}]{20221PRBTjengCo4}%
  \BibitemOpen
  \bibfield  {author} {\bibinfo {author} {\bibfnamefont {Y.-Y.}\ \bibnamefont {Chin}}, \bibinfo {author} {\bibfnamefont {Z.}~\bibnamefont {Hu}}, \bibinfo {author} {\bibfnamefont {Y.}~\bibnamefont {Shimakawa}}, \bibinfo {author} {\bibfnamefont {J.}~\bibnamefont {Yang}}, \bibinfo {author} {\bibfnamefont {Y.}~\bibnamefont {Long}}, \bibinfo {author} {\bibfnamefont {A.}~\bibnamefont {Tanaka}}, \bibinfo {author} {\bibfnamefont {L.~H.}\ \bibnamefont {Tjeng}}, \bibinfo {author} {\bibfnamefont {H.-J.}\ \bibnamefont {Lin}}, \ and\ \bibinfo {author} {\bibfnamefont {C.-T.}\ \bibnamefont {Chen}},\ }\href {\doibase 10.1103/PhysRevB.103.115149} {\bibfield  {journal} {\bibinfo  {journal} {Phys. Rev. B}\ }\textbf {\bibinfo {volume} {103}},\ \bibinfo {pages} {115149} (\bibinfo {year} {2021})}\BibitemShut {NoStop}%
\bibitem [{\citenamefont {Takegami}\ \emph {et~al.}(2023{\natexlab{a}})\citenamefont {Takegami}, \citenamefont {Hu}, \citenamefont {Falke}, \citenamefont {Meléndez-Sans}, \citenamefont {Liu}, \citenamefont {Chang}, \citenamefont {Kuo}, \citenamefont {Chen}, \citenamefont {Guo}, \citenamefont {Komarek}, \citenamefont {Tanaka}, \citenamefont {Hébert},\ and\ \citenamefont {Tjeng}}]{2023ZAACTjengBa2CoO4}%
  \BibitemOpen
  \bibfield  {author} {\bibinfo {author} {\bibfnamefont {D.}~\bibnamefont {Takegami}}, \bibinfo {author} {\bibfnamefont {Z.}~\bibnamefont {Hu}}, \bibinfo {author} {\bibfnamefont {J.}~\bibnamefont {Falke}}, \bibinfo {author} {\bibfnamefont {A.}~\bibnamefont {Meléndez-Sans}}, \bibinfo {author} {\bibfnamefont {C.-E.}\ \bibnamefont {Liu}}, \bibinfo {author} {\bibfnamefont {C.-F.}\ \bibnamefont {Chang}}, \bibinfo {author} {\bibfnamefont {C.-Y.}\ \bibnamefont {Kuo}}, \bibinfo {author} {\bibfnamefont {C.-T.}\ \bibnamefont {Chen}}, \bibinfo {author} {\bibfnamefont {H.}~\bibnamefont {Guo}}, \bibinfo {author} {\bibfnamefont {A.}~\bibnamefont {Komarek}}, \bibinfo {author} {\bibfnamefont {A.}~\bibnamefont {Tanaka}}, \bibinfo {author} {\bibfnamefont {S.}~\bibnamefont {Hébert}}, \ and\ \bibinfo {author} {\bibfnamefont {L.~H.}\ \bibnamefont {Tjeng}},\ }\href {\doibase https://doi.org/10.1002/zaac.202300077} {\bibfield  {journal} {\bibinfo  {journal} {Zeitschrift für anorganische und allgemeine Chemie}\ }\textbf {\bibinfo
  {volume} {649}},\ \bibinfo {pages} {e202300077} (\bibinfo {year} {2023}{\natexlab{a}})}\BibitemShut {NoStop}%
\bibitem [{\citenamefont {Chin}\ \emph {et~al.}(2019)\citenamefont {Chin}, \citenamefont {Hu}, \citenamefont {Lin}, \citenamefont {Agrestini}, \citenamefont {Weinen}, \citenamefont {Martin}, \citenamefont {H\'ebert}, \citenamefont {Maignan}, \citenamefont {Tanaka}, \citenamefont {Cezar}, \citenamefont {Brookes}, \citenamefont {Liao}, \citenamefont {Tsuei}, \citenamefont {Chen}, \citenamefont {Khomskii},\ and\ \citenamefont {Tjeng}}]{2019PRBTjengBaCoO3}%
  \BibitemOpen
  \bibfield  {author} {\bibinfo {author} {\bibfnamefont {Y.~Y.}\ \bibnamefont {Chin}}, \bibinfo {author} {\bibfnamefont {Z.}~\bibnamefont {Hu}}, \bibinfo {author} {\bibfnamefont {H.-J.}\ \bibnamefont {Lin}}, \bibinfo {author} {\bibfnamefont {S.}~\bibnamefont {Agrestini}}, \bibinfo {author} {\bibfnamefont {J.}~\bibnamefont {Weinen}}, \bibinfo {author} {\bibfnamefont {C.}~\bibnamefont {Martin}}, \bibinfo {author} {\bibfnamefont {S.}~\bibnamefont {H\'ebert}}, \bibinfo {author} {\bibfnamefont {A.}~\bibnamefont {Maignan}}, \bibinfo {author} {\bibfnamefont {A.}~\bibnamefont {Tanaka}}, \bibinfo {author} {\bibfnamefont {J.~C.}\ \bibnamefont {Cezar}}, \bibinfo {author} {\bibfnamefont {N.~B.}\ \bibnamefont {Brookes}}, \bibinfo {author} {\bibfnamefont {Y.-F.}\ \bibnamefont {Liao}}, \bibinfo {author} {\bibfnamefont {K.-D.}\ \bibnamefont {Tsuei}}, \bibinfo {author} {\bibfnamefont {C.~T.}\ \bibnamefont {Chen}}, \bibinfo {author} {\bibfnamefont {D.~I.}\ \bibnamefont {Khomskii}}, \ and\ \bibinfo {author} {\bibfnamefont
  {L.~H.}\ \bibnamefont {Tjeng}},\ }\href {\doibase 10.1103/PhysRevB.100.205139} {\bibfield  {journal} {\bibinfo  {journal} {Phys. Rev. B}\ }\textbf {\bibinfo {volume} {100}},\ \bibinfo {pages} {205139} (\bibinfo {year} {2019})}\BibitemShut {NoStop}%
\bibitem [{\citenamefont {Potze}\ \emph {et~al.}(1995)\citenamefont {Potze}, \citenamefont {Sawatzky},\ and\ \citenamefont {Abbate}}]{1995PRBSawatzkyCo4}%
  \BibitemOpen
  \bibfield  {author} {\bibinfo {author} {\bibfnamefont {R.~H.}\ \bibnamefont {Potze}}, \bibinfo {author} {\bibfnamefont {G.~A.}\ \bibnamefont {Sawatzky}}, \ and\ \bibinfo {author} {\bibfnamefont {M.}~\bibnamefont {Abbate}},\ }\href {\doibase 10.1103/PhysRevB.51.11501} {\bibfield  {journal} {\bibinfo  {journal} {Phys. Rev. B}\ }\textbf {\bibinfo {volume} {51}},\ \bibinfo {pages} {11501} (\bibinfo {year} {1995})}\BibitemShut {NoStop}%
\bibitem [{\citenamefont {Lin}\ \emph {et~al.}(2010)\citenamefont {Lin}, \citenamefont {Chin}, \citenamefont {Hu}, \citenamefont {Shu}, \citenamefont {Chou}, \citenamefont {Ohta}, \citenamefont {Yoshimura}, \citenamefont {H\'ebert}, \citenamefont {Maignan}, \citenamefont {Tanaka}, \citenamefont {Tjeng},\ and\ \citenamefont {Chen}}]{2010PRBTjengNaCoO2}%
  \BibitemOpen
  \bibfield  {author} {\bibinfo {author} {\bibfnamefont {H.-J.}\ \bibnamefont {Lin}}, \bibinfo {author} {\bibfnamefont {Y.~Y.}\ \bibnamefont {Chin}}, \bibinfo {author} {\bibfnamefont {Z.}~\bibnamefont {Hu}}, \bibinfo {author} {\bibfnamefont {G.~J.}\ \bibnamefont {Shu}}, \bibinfo {author} {\bibfnamefont {F.~C.}\ \bibnamefont {Chou}}, \bibinfo {author} {\bibfnamefont {H.}~\bibnamefont {Ohta}}, \bibinfo {author} {\bibfnamefont {K.}~\bibnamefont {Yoshimura}}, \bibinfo {author} {\bibfnamefont {S.}~\bibnamefont {H\'ebert}}, \bibinfo {author} {\bibfnamefont {A.}~\bibnamefont {Maignan}}, \bibinfo {author} {\bibfnamefont {A.}~\bibnamefont {Tanaka}}, \bibinfo {author} {\bibfnamefont {L.~H.}\ \bibnamefont {Tjeng}}, \ and\ \bibinfo {author} {\bibfnamefont {C.~T.}\ \bibnamefont {Chen}},\ }\href {\doibase 10.1103/PhysRevB.81.115138} {\bibfield  {journal} {\bibinfo  {journal} {Phys. Rev. B}\ }\textbf {\bibinfo {volume} {81}},\ \bibinfo {pages} {115138} (\bibinfo {year} {2010})}\BibitemShut {NoStop}%
\bibitem [{\citenamefont {Hu}\ \emph {et~al.}(2004)\citenamefont {Hu}, \citenamefont {Wu}, \citenamefont {Haverkort}, \citenamefont {Hsieh}, \citenamefont {Lin}, \citenamefont {Lorenz}, \citenamefont {Baier}, \citenamefont {Reichl}, \citenamefont {Bonn}, \citenamefont {Felser}, \citenamefont {Tanaka}, \citenamefont {Chen},\ and\ \citenamefont {Tjeng}}]{2004PRLTjengSr2CoO3Cl}%
  \BibitemOpen
  \bibfield  {author} {\bibinfo {author} {\bibfnamefont {Z.}~\bibnamefont {Hu}}, \bibinfo {author} {\bibfnamefont {H.}~\bibnamefont {Wu}}, \bibinfo {author} {\bibfnamefont {M.~W.}\ \bibnamefont {Haverkort}}, \bibinfo {author} {\bibfnamefont {H.~H.}\ \bibnamefont {Hsieh}}, \bibinfo {author} {\bibfnamefont {H.~J.}\ \bibnamefont {Lin}}, \bibinfo {author} {\bibfnamefont {T.}~\bibnamefont {Lorenz}}, \bibinfo {author} {\bibfnamefont {J.}~\bibnamefont {Baier}}, \bibinfo {author} {\bibfnamefont {A.}~\bibnamefont {Reichl}}, \bibinfo {author} {\bibfnamefont {I.}~\bibnamefont {Bonn}}, \bibinfo {author} {\bibfnamefont {C.}~\bibnamefont {Felser}}, \bibinfo {author} {\bibfnamefont {A.}~\bibnamefont {Tanaka}}, \bibinfo {author} {\bibfnamefont {C.~T.}\ \bibnamefont {Chen}}, \ and\ \bibinfo {author} {\bibfnamefont {L.~H.}\ \bibnamefont {Tjeng}},\ }\href {\doibase 10.1103/PhysRevLett.92.207402} {\bibfield  {journal} {\bibinfo  {journal} {Phys. Rev. Lett.}\ }\textbf {\bibinfo {volume} {92}},\ \bibinfo {pages} {207402} (\bibinfo
  {year} {2004})}\BibitemShut {NoStop}%
\bibitem [{\citenamefont {Haverkort}\ \emph {et~al.}(2006)\citenamefont {Haverkort}, \citenamefont {Hu}, \citenamefont {Cezar}, \citenamefont {Burnus}, \citenamefont {Hartmann}, \citenamefont {Reuther}, \citenamefont {Zobel}, \citenamefont {Lorenz}, \citenamefont {Tanaka}, \citenamefont {Brookes}, \citenamefont {Hsieh}, \citenamefont {Lin}, \citenamefont {Chen},\ and\ \citenamefont {Tjeng}}]{2006PRLHaverkortLCO}%
  \BibitemOpen
  \bibfield  {author} {\bibinfo {author} {\bibfnamefont {M.~W.}\ \bibnamefont {Haverkort}}, \bibinfo {author} {\bibfnamefont {Z.}~\bibnamefont {Hu}}, \bibinfo {author} {\bibfnamefont {J.~C.}\ \bibnamefont {Cezar}}, \bibinfo {author} {\bibfnamefont {T.}~\bibnamefont {Burnus}}, \bibinfo {author} {\bibfnamefont {H.}~\bibnamefont {Hartmann}}, \bibinfo {author} {\bibfnamefont {M.}~\bibnamefont {Reuther}}, \bibinfo {author} {\bibfnamefont {C.}~\bibnamefont {Zobel}}, \bibinfo {author} {\bibfnamefont {T.}~\bibnamefont {Lorenz}}, \bibinfo {author} {\bibfnamefont {A.}~\bibnamefont {Tanaka}}, \bibinfo {author} {\bibfnamefont {N.~B.}\ \bibnamefont {Brookes}}, \bibinfo {author} {\bibfnamefont {H.~H.}\ \bibnamefont {Hsieh}}, \bibinfo {author} {\bibfnamefont {H.-J.}\ \bibnamefont {Lin}}, \bibinfo {author} {\bibfnamefont {C.~T.}\ \bibnamefont {Chen}}, \ and\ \bibinfo {author} {\bibfnamefont {L.~H.}\ \bibnamefont {Tjeng}},\ }\href {\doibase 10.1103/PhysRevLett.97.176405} {\bibfield  {journal} {\bibinfo  {journal} {Phys. Rev.
  Lett.}\ }\textbf {\bibinfo {volume} {97}},\ \bibinfo {pages} {176405} (\bibinfo {year} {2006})}\BibitemShut {NoStop}%
\bibitem [{\citenamefont {Takegami}\ \emph {et~al.}(2023{\natexlab{b}})\citenamefont {Takegami}, \citenamefont {Tanaka}, \citenamefont {Agrestini}, \citenamefont {Hu}, \citenamefont {Weinen}, \citenamefont {Rotter}, \citenamefont {Sch\"u\ss{}ler-Langeheine}, \citenamefont {Willers}, \citenamefont {Koethe}, \citenamefont {Lorenz}, \citenamefont {Liao}, \citenamefont {Tsuei}, \citenamefont {Lin}, \citenamefont {Chen},\ and\ \citenamefont {Tjeng}}]{2023PRXTjengLCO}%
  \BibitemOpen
  \bibfield  {author} {\bibinfo {author} {\bibfnamefont {D.}~\bibnamefont {Takegami}}, \bibinfo {author} {\bibfnamefont {A.}~\bibnamefont {Tanaka}}, \bibinfo {author} {\bibfnamefont {S.}~\bibnamefont {Agrestini}}, \bibinfo {author} {\bibfnamefont {Z.}~\bibnamefont {Hu}}, \bibinfo {author} {\bibfnamefont {J.}~\bibnamefont {Weinen}}, \bibinfo {author} {\bibfnamefont {M.}~\bibnamefont {Rotter}}, \bibinfo {author} {\bibfnamefont {C.}~\bibnamefont {Sch\"u\ss{}ler-Langeheine}}, \bibinfo {author} {\bibfnamefont {T.}~\bibnamefont {Willers}}, \bibinfo {author} {\bibfnamefont {T.~C.}\ \bibnamefont {Koethe}}, \bibinfo {author} {\bibfnamefont {T.}~\bibnamefont {Lorenz}}, \bibinfo {author} {\bibfnamefont {Y.~F.}\ \bibnamefont {Liao}}, \bibinfo {author} {\bibfnamefont {K.~D.}\ \bibnamefont {Tsuei}}, \bibinfo {author} {\bibfnamefont {H.-J.}\ \bibnamefont {Lin}}, \bibinfo {author} {\bibfnamefont {C.~T.}\ \bibnamefont {Chen}}, \ and\ \bibinfo {author} {\bibfnamefont {L.~H.}\ \bibnamefont {Tjeng}},\ }\href {\doibase
  10.1103/PhysRevX.13.011037} {\bibfield  {journal} {\bibinfo  {journal} {Phys. Rev. X}\ }\textbf {\bibinfo {volume} {13}},\ \bibinfo {pages} {011037} (\bibinfo {year} {2023}{\natexlab{b}})}\BibitemShut {NoStop}%
\bibitem [{\citenamefont {Sudayama}\ \emph {et~al.}(2011)\citenamefont {Sudayama}, \citenamefont {Wakisaka}, \citenamefont {Mizokawa}, \citenamefont {Wadati}, \citenamefont {Sawatzky}, \citenamefont {Hawthorn}, \citenamefont {Regier}, \citenamefont {Oka}, \citenamefont {Azuma},\ and\ \citenamefont {Shimakawa}}]{2011PRBShimakawaBiCoO3}%
  \BibitemOpen
  \bibfield  {author} {\bibinfo {author} {\bibfnamefont {T.}~\bibnamefont {Sudayama}}, \bibinfo {author} {\bibfnamefont {Y.}~\bibnamefont {Wakisaka}}, \bibinfo {author} {\bibfnamefont {T.}~\bibnamefont {Mizokawa}}, \bibinfo {author} {\bibfnamefont {H.}~\bibnamefont {Wadati}}, \bibinfo {author} {\bibfnamefont {G.~A.}\ \bibnamefont {Sawatzky}}, \bibinfo {author} {\bibfnamefont {D.~G.}\ \bibnamefont {Hawthorn}}, \bibinfo {author} {\bibfnamefont {T.~Z.}\ \bibnamefont {Regier}}, \bibinfo {author} {\bibfnamefont {K.}~\bibnamefont {Oka}}, \bibinfo {author} {\bibfnamefont {M.}~\bibnamefont {Azuma}}, \ and\ \bibinfo {author} {\bibfnamefont {Y.}~\bibnamefont {Shimakawa}},\ }\href {\doibase 10.1103/PhysRevB.83.235105} {\bibfield  {journal} {\bibinfo  {journal} {Phys. Rev. B}\ }\textbf {\bibinfo {volume} {83}},\ \bibinfo {pages} {235105} (\bibinfo {year} {2011})}\BibitemShut {NoStop}%
\bibitem [{\citenamefont {van Elp}\ \emph {et~al.}(1991{\natexlab{b}})\citenamefont {van Elp}, \citenamefont {Wieland}, \citenamefont {Eskes}, \citenamefont {Kuiper}, \citenamefont {Sawatzky}, \citenamefont {de~Groot},\ and\ \citenamefont {Turner}}]{1991PRBvanElpCo}%
  \BibitemOpen
  \bibfield  {author} {\bibinfo {author} {\bibfnamefont {J.}~\bibnamefont {van Elp}}, \bibinfo {author} {\bibfnamefont {J.~L.}\ \bibnamefont {Wieland}}, \bibinfo {author} {\bibfnamefont {H.}~\bibnamefont {Eskes}}, \bibinfo {author} {\bibfnamefont {P.}~\bibnamefont {Kuiper}}, \bibinfo {author} {\bibfnamefont {G.~A.}\ \bibnamefont {Sawatzky}}, \bibinfo {author} {\bibfnamefont {F.~M.~F.}\ \bibnamefont {de~Groot}}, \ and\ \bibinfo {author} {\bibfnamefont {T.~S.}\ \bibnamefont {Turner}},\ }\href {\doibase 10.1103/PhysRevB.44.6090} {\bibfield  {journal} {\bibinfo  {journal} {Phys. Rev. B}\ }\textbf {\bibinfo {volume} {44}},\ \bibinfo {pages} {6090} (\bibinfo {year} {1991}{\natexlab{b}})}\BibitemShut {NoStop}%
\bibitem [{\citenamefont {Agrestini}\ \emph {et~al.}(2017)\citenamefont {Agrestini}, \citenamefont {Kuo}, \citenamefont {Mikhailova}, \citenamefont {Chen}, \citenamefont {Ohresser}, \citenamefont {Pi}, \citenamefont {Guo}, \citenamefont {Komarek}, \citenamefont {Tanaka}, \citenamefont {Hu},\ and\ \citenamefont {Tjeng}}]{2017PRBTjengCo3}%
  \BibitemOpen
  \bibfield  {author} {\bibinfo {author} {\bibfnamefont {S.}~\bibnamefont {Agrestini}}, \bibinfo {author} {\bibfnamefont {C.-Y.}\ \bibnamefont {Kuo}}, \bibinfo {author} {\bibfnamefont {D.}~\bibnamefont {Mikhailova}}, \bibinfo {author} {\bibfnamefont {K.}~\bibnamefont {Chen}}, \bibinfo {author} {\bibfnamefont {P.}~\bibnamefont {Ohresser}}, \bibinfo {author} {\bibfnamefont {T.~W.}\ \bibnamefont {Pi}}, \bibinfo {author} {\bibfnamefont {H.}~\bibnamefont {Guo}}, \bibinfo {author} {\bibfnamefont {A.~C.}\ \bibnamefont {Komarek}}, \bibinfo {author} {\bibfnamefont {A.}~\bibnamefont {Tanaka}}, \bibinfo {author} {\bibfnamefont {Z.}~\bibnamefont {Hu}}, \ and\ \bibinfo {author} {\bibfnamefont {L.~H.}\ \bibnamefont {Tjeng}},\ }\href {\doibase 10.1103/PhysRevB.95.245131} {\bibfield  {journal} {\bibinfo  {journal} {Phys. Rev. B}\ }\textbf {\bibinfo {volume} {95}},\ \bibinfo {pages} {245131} (\bibinfo {year} {2017})}\BibitemShut {NoStop}%
\bibitem [{\citenamefont {Hu}\ \emph {et~al.}(2012)\citenamefont {Hu}, \citenamefont {Wu}, \citenamefont {Koethe}, \citenamefont {Barilo}, \citenamefont {Shiryaev}, \citenamefont {Bychkov}, \citenamefont {Schüßler-Langeheine}, \citenamefont {Lorenz}, \citenamefont {Tanaka}, \citenamefont {Hsieh}, \citenamefont {Lin}, \citenamefont {Chen}, \citenamefont {Brookes}, \citenamefont {Agrestini}, \citenamefont {Chin}, \citenamefont {Rotter},\ and\ \citenamefont {Tjeng}}]{2012NJPTjengGdBaCoO}%
  \BibitemOpen
  \bibfield  {author} {\bibinfo {author} {\bibfnamefont {Z.}~\bibnamefont {Hu}}, \bibinfo {author} {\bibfnamefont {H.}~\bibnamefont {Wu}}, \bibinfo {author} {\bibfnamefont {T.~C.}\ \bibnamefont {Koethe}}, \bibinfo {author} {\bibfnamefont {S.~N.}\ \bibnamefont {Barilo}}, \bibinfo {author} {\bibfnamefont {S.~V.}\ \bibnamefont {Shiryaev}}, \bibinfo {author} {\bibfnamefont {G.~L.}\ \bibnamefont {Bychkov}}, \bibinfo {author} {\bibfnamefont {C.}~\bibnamefont {Schüßler-Langeheine}}, \bibinfo {author} {\bibfnamefont {T.}~\bibnamefont {Lorenz}}, \bibinfo {author} {\bibfnamefont {A.}~\bibnamefont {Tanaka}}, \bibinfo {author} {\bibfnamefont {H.~H.}\ \bibnamefont {Hsieh}}, \bibinfo {author} {\bibfnamefont {H.-J.}\ \bibnamefont {Lin}}, \bibinfo {author} {\bibfnamefont {C.~T.}\ \bibnamefont {Chen}}, \bibinfo {author} {\bibfnamefont {N.~B.}\ \bibnamefont {Brookes}}, \bibinfo {author} {\bibfnamefont {S.}~\bibnamefont {Agrestini}}, \bibinfo {author} {\bibfnamefont {Y.-Y.}\ \bibnamefont {Chin}}, \bibinfo {author}
  {\bibfnamefont {M.}~\bibnamefont {Rotter}}, \ and\ \bibinfo {author} {\bibfnamefont {L.~H.}\ \bibnamefont {Tjeng}},\ }\href {\doibase 10.1088/1367-2630/14/12/123025} {\bibfield  {journal} {\bibinfo  {journal} {New Journal of Physics}\ }\textbf {\bibinfo {volume} {14}},\ \bibinfo {pages} {123025} (\bibinfo {year} {2012})}\BibitemShut {NoStop}%
\bibitem [{\citenamefont {Burnus}\ \emph {et~al.}(2006)\citenamefont {Burnus}, \citenamefont {Hu}, \citenamefont {Haverkort}, \citenamefont {Cezar}, \citenamefont {Flahaut}, \citenamefont {Hardy}, \citenamefont {Maignan}, \citenamefont {Brookes}, \citenamefont {Tanaka}, \citenamefont {Hsieh}, \citenamefont {Lin}, \citenamefont {Chen},\ and\ \citenamefont {Tjeng}}]{2006PRBTjengCCO}%
  \BibitemOpen
  \bibfield  {author} {\bibinfo {author} {\bibfnamefont {T.}~\bibnamefont {Burnus}}, \bibinfo {author} {\bibfnamefont {Z.}~\bibnamefont {Hu}}, \bibinfo {author} {\bibfnamefont {M.~W.}\ \bibnamefont {Haverkort}}, \bibinfo {author} {\bibfnamefont {J.~C.}\ \bibnamefont {Cezar}}, \bibinfo {author} {\bibfnamefont {D.}~\bibnamefont {Flahaut}}, \bibinfo {author} {\bibfnamefont {V.}~\bibnamefont {Hardy}}, \bibinfo {author} {\bibfnamefont {A.}~\bibnamefont {Maignan}}, \bibinfo {author} {\bibfnamefont {N.~B.}\ \bibnamefont {Brookes}}, \bibinfo {author} {\bibfnamefont {A.}~\bibnamefont {Tanaka}}, \bibinfo {author} {\bibfnamefont {H.~H.}\ \bibnamefont {Hsieh}}, \bibinfo {author} {\bibfnamefont {H.-J.}\ \bibnamefont {Lin}}, \bibinfo {author} {\bibfnamefont {C.~T.}\ \bibnamefont {Chen}}, \ and\ \bibinfo {author} {\bibfnamefont {L.~H.}\ \bibnamefont {Tjeng}},\ }\href {\doibase 10.1103/PhysRevB.74.245111} {\bibfield  {journal} {\bibinfo  {journal} {Phys. Rev. B}\ }\textbf {\bibinfo {volume} {74}},\ \bibinfo {pages} {245111}
  (\bibinfo {year} {2006})}\BibitemShut {NoStop}%
\bibitem [{\citenamefont {Wang}\ \emph {et~al.}(2022)\citenamefont {Wang}, \citenamefont {Huang}, \citenamefont {Hariki}, \citenamefont {Okamoto}, \citenamefont {Huang}, \citenamefont {Singh}, \citenamefont {Huang}, \citenamefont {Nagel}, \citenamefont {Schuppler}, \citenamefont {Haarman}, \citenamefont {Liu},\ and\ \citenamefont {de~Groot}}]{2022JPCCdeGrootCo3O4}%
  \BibitemOpen
  \bibfield  {author} {\bibinfo {author} {\bibfnamefont {R.-P.}\ \bibnamefont {Wang}}, \bibinfo {author} {\bibfnamefont {M.-J.}\ \bibnamefont {Huang}}, \bibinfo {author} {\bibfnamefont {A.}~\bibnamefont {Hariki}}, \bibinfo {author} {\bibfnamefont {J.}~\bibnamefont {Okamoto}}, \bibinfo {author} {\bibfnamefont {H.-Y.}\ \bibnamefont {Huang}}, \bibinfo {author} {\bibfnamefont {A.}~\bibnamefont {Singh}}, \bibinfo {author} {\bibfnamefont {D.-J.}\ \bibnamefont {Huang}}, \bibinfo {author} {\bibfnamefont {P.}~\bibnamefont {Nagel}}, \bibinfo {author} {\bibfnamefont {S.}~\bibnamefont {Schuppler}}, \bibinfo {author} {\bibfnamefont {T.}~\bibnamefont {Haarman}}, \bibinfo {author} {\bibfnamefont {B.}~\bibnamefont {Liu}}, \ and\ \bibinfo {author} {\bibfnamefont {F.~M.~F.}\ \bibnamefont {de~Groot}},\ }\href {\doibase 10.1021/acs.jpcc.2c01521} {\bibfield  {journal} {\bibinfo  {journal} {The Journal of Physical Chemistry C}\ }\textbf {\bibinfo {volume} {126}},\ \bibinfo {pages} {8752} (\bibinfo {year} {2022})}\BibitemShut
  {NoStop}%
\bibitem [{\citenamefont {Hollmann}\ \emph {et~al.}(2014)\citenamefont {Hollmann}, \citenamefont {Agrestini}, \citenamefont {Hu}, \citenamefont {He}, \citenamefont {Schmidt}, \citenamefont {Kuo}, \citenamefont {Rotter}, \citenamefont {Nugroho}, \citenamefont {Sessi}, \citenamefont {Tanaka}, \citenamefont {Brookes},\ and\ \citenamefont {Tjeng}}]{2014PRBTjengCoV2O6}%
  \BibitemOpen
  \bibfield  {author} {\bibinfo {author} {\bibfnamefont {N.}~\bibnamefont {Hollmann}}, \bibinfo {author} {\bibfnamefont {S.}~\bibnamefont {Agrestini}}, \bibinfo {author} {\bibfnamefont {Z.}~\bibnamefont {Hu}}, \bibinfo {author} {\bibfnamefont {Z.}~\bibnamefont {He}}, \bibinfo {author} {\bibfnamefont {M.}~\bibnamefont {Schmidt}}, \bibinfo {author} {\bibfnamefont {C.-Y.}\ \bibnamefont {Kuo}}, \bibinfo {author} {\bibfnamefont {M.}~\bibnamefont {Rotter}}, \bibinfo {author} {\bibfnamefont {A.~A.}\ \bibnamefont {Nugroho}}, \bibinfo {author} {\bibfnamefont {V.}~\bibnamefont {Sessi}}, \bibinfo {author} {\bibfnamefont {A.}~\bibnamefont {Tanaka}}, \bibinfo {author} {\bibfnamefont {N.~B.}\ \bibnamefont {Brookes}}, \ and\ \bibinfo {author} {\bibfnamefont {L.~H.}\ \bibnamefont {Tjeng}},\ }\href {\doibase 10.1103/PhysRevB.89.201101} {\bibfield  {journal} {\bibinfo  {journal} {Phys. Rev. B}\ }\textbf {\bibinfo {volume} {89}},\ \bibinfo {pages} {201101} (\bibinfo {year} {2014})}\BibitemShut {NoStop}%
\bibitem [{\citenamefont {Csiszar}\ \emph {et~al.}(2005)\citenamefont {Csiszar}, \citenamefont {Haverkort}, \citenamefont {Hu}, \citenamefont {Tanaka}, \citenamefont {Hsieh}, \citenamefont {Lin}, \citenamefont {Chen}, \citenamefont {Hibma},\ and\ \citenamefont {Tjeng}}]{2005PRLTjengCoO}%
  \BibitemOpen
  \bibfield  {author} {\bibinfo {author} {\bibfnamefont {S.~I.}\ \bibnamefont {Csiszar}}, \bibinfo {author} {\bibfnamefont {M.~W.}\ \bibnamefont {Haverkort}}, \bibinfo {author} {\bibfnamefont {Z.}~\bibnamefont {Hu}}, \bibinfo {author} {\bibfnamefont {A.}~\bibnamefont {Tanaka}}, \bibinfo {author} {\bibfnamefont {H.~H.}\ \bibnamefont {Hsieh}}, \bibinfo {author} {\bibfnamefont {H.-J.}\ \bibnamefont {Lin}}, \bibinfo {author} {\bibfnamefont {C.~T.}\ \bibnamefont {Chen}}, \bibinfo {author} {\bibfnamefont {T.}~\bibnamefont {Hibma}}, \ and\ \bibinfo {author} {\bibfnamefont {L.~H.}\ \bibnamefont {Tjeng}},\ }\href {\doibase 10.1103/PhysRevLett.95.187205} {\bibfield  {journal} {\bibinfo  {journal} {Phys. Rev. Lett.}\ }\textbf {\bibinfo {volume} {95}},\ \bibinfo {pages} {187205} (\bibinfo {year} {2005})}\BibitemShut {NoStop}%
\bibitem [{\citenamefont {Okada}\ and\ \citenamefont {Kotani}(1992{\natexlab{b}})}]{1992JPSJKotaniCoO}%
  \BibitemOpen
  \bibfield  {author} {\bibinfo {author} {\bibfnamefont {K.}~\bibnamefont {Okada}}\ and\ \bibinfo {author} {\bibfnamefont {A.}~\bibnamefont {Kotani}},\ }\href {\doibase 10.1143/JPSJ.61.449} {\bibfield  {journal} {\bibinfo  {journal} {Journal of the Physical Society of Japan}\ }\textbf {\bibinfo {volume} {61}},\ \bibinfo {pages} {449} (\bibinfo {year} {1992}{\natexlab{b}})}\BibitemShut {NoStop}%
\bibitem [{\citenamefont {Parmigiani}\ and\ \citenamefont {Sangaletti}(1993)}]{1993CPLParmigiani}%
  \BibitemOpen
  \bibfield  {author} {\bibinfo {author} {\bibfnamefont {F.}~\bibnamefont {Parmigiani}}\ and\ \bibinfo {author} {\bibfnamefont {L.}~\bibnamefont {Sangaletti}},\ }\href {\doibase https://doi.org/10.1016/0009-2614(93)89170-M} {\bibfield  {journal} {\bibinfo  {journal} {Chemical Physics Letters}\ }\textbf {\bibinfo {volume} {213}},\ \bibinfo {pages} {613} (\bibinfo {year} {1993})}\BibitemShut {NoStop}%
\bibitem [{\citenamefont {Huang}\ \emph {et~al.}(2023)\citenamefont {Huang}, \citenamefont {Chang}, \citenamefont {Huang}, \citenamefont {Li}, \citenamefont {Komarek}, \citenamefont {Tjeng}, \citenamefont {Orikasa}, \citenamefont {Pao}, \citenamefont {Chan}, \citenamefont {Chen}, \citenamefont {Haw}, \citenamefont {Zhou}, \citenamefont {Wang}, \citenamefont {Lin}, \citenamefont {Chen}, \citenamefont {Dong}, \citenamefont {Kuo}, \citenamefont {Wang}, \citenamefont {Hu},\ and\ \citenamefont {Zhang}}]{2023NatCommTjengLiNiO2}%
  \BibitemOpen
  \bibfield  {author} {\bibinfo {author} {\bibfnamefont {H.}~\bibnamefont {Huang}}, \bibinfo {author} {\bibfnamefont {Y.-C.}\ \bibnamefont {Chang}}, \bibinfo {author} {\bibfnamefont {Y.-C.}\ \bibnamefont {Huang}}, \bibinfo {author} {\bibfnamefont {L.}~\bibnamefont {Li}}, \bibinfo {author} {\bibfnamefont {A.~C.}\ \bibnamefont {Komarek}}, \bibinfo {author} {\bibfnamefont {L.~H.}\ \bibnamefont {Tjeng}}, \bibinfo {author} {\bibfnamefont {Y.}~\bibnamefont {Orikasa}}, \bibinfo {author} {\bibfnamefont {C.-W.}\ \bibnamefont {Pao}}, \bibinfo {author} {\bibfnamefont {T.-S.}\ \bibnamefont {Chan}}, \bibinfo {author} {\bibfnamefont {J.-M.}\ \bibnamefont {Chen}}, \bibinfo {author} {\bibfnamefont {S.-C.}\ \bibnamefont {Haw}}, \bibinfo {author} {\bibfnamefont {J.}~\bibnamefont {Zhou}}, \bibinfo {author} {\bibfnamefont {Y.}~\bibnamefont {Wang}}, \bibinfo {author} {\bibfnamefont {H.-J.}\ \bibnamefont {Lin}}, \bibinfo {author} {\bibfnamefont {C.-T.}\ \bibnamefont {Chen}}, \bibinfo {author} {\bibfnamefont {C.-L.}\ \bibnamefont
  {Dong}}, \bibinfo {author} {\bibfnamefont {C.-Y.}\ \bibnamefont {Kuo}}, \bibinfo {author} {\bibfnamefont {J.-Q.}\ \bibnamefont {Wang}}, \bibinfo {author} {\bibfnamefont {Z.}~\bibnamefont {Hu}}, \ and\ \bibinfo {author} {\bibfnamefont {L.}~\bibnamefont {Zhang}},\ }\href@noop {} {\bibfield  {journal} {\bibinfo  {journal} {Nature Communications}\ }\textbf {\bibinfo {volume} {14}},\ \bibinfo {pages} {2112} (\bibinfo {year} {2023})}\BibitemShut {NoStop}%
\bibitem [{\citenamefont {Mizokawa}\ \emph {et~al.}(1996)\citenamefont {Mizokawa}, \citenamefont {Fujimori}, \citenamefont {Arima}, \citenamefont {Tokura}, \citenamefont {Mõri},\ and\ \citenamefont {Akimitsu}}]{1996JELSPECMizokawaPNO}%
  \BibitemOpen
  \bibfield  {author} {\bibinfo {author} {\bibfnamefont {T.}~\bibnamefont {Mizokawa}}, \bibinfo {author} {\bibfnamefont {A.}~\bibnamefont {Fujimori}}, \bibinfo {author} {\bibfnamefont {T.}~\bibnamefont {Arima}}, \bibinfo {author} {\bibfnamefont {Y.}~\bibnamefont {Tokura}}, \bibinfo {author} {\bibfnamefont {N.}~\bibnamefont {Mõri}}, \ and\ \bibinfo {author} {\bibfnamefont {J.}~\bibnamefont {Akimitsu}},\ }\href {\doibase https://doi.org/10.1016/S0368-2048(96)80059-4} {\bibfield  {journal} {\bibinfo  {journal} {Journal of Electron Spectroscopy and Related Phenomena}\ }\textbf {\bibinfo {volume} {78}},\ \bibinfo {pages} {191} (\bibinfo {year} {1996})}\BibitemShut {NoStop}%
\bibitem [{\citenamefont {Abbate}\ \emph {et~al.}(2002{\natexlab{b}})\citenamefont {Abbate}, \citenamefont {Zampieri}, \citenamefont {Prado}, \citenamefont {Caneiro}, \citenamefont {Gonzalez-Calbet},\ and\ \citenamefont {Vallet-Regi}}]{2002PRBAbbateLaNiO3}%
  \BibitemOpen
  \bibfield  {author} {\bibinfo {author} {\bibfnamefont {M.}~\bibnamefont {Abbate}}, \bibinfo {author} {\bibfnamefont {G.}~\bibnamefont {Zampieri}}, \bibinfo {author} {\bibfnamefont {F.}~\bibnamefont {Prado}}, \bibinfo {author} {\bibfnamefont {A.}~\bibnamefont {Caneiro}}, \bibinfo {author} {\bibfnamefont {J.~M.}\ \bibnamefont {Gonzalez-Calbet}}, \ and\ \bibinfo {author} {\bibfnamefont {M.}~\bibnamefont {Vallet-Regi}},\ }\href {\doibase 10.1103/PhysRevB.65.155101} {\bibfield  {journal} {\bibinfo  {journal} {Phys. Rev. B}\ }\textbf {\bibinfo {volume} {65}},\ \bibinfo {pages} {155101} (\bibinfo {year} {2002}{\natexlab{b}})}\BibitemShut {NoStop}%
\bibitem [{\citenamefont {van Elp}\ \emph {et~al.}(1992)\citenamefont {van Elp}, \citenamefont {Eskes}, \citenamefont {Kuiper},\ and\ \citenamefont {Sawatzky}}]{1992PRBvanElpNi}%
  \BibitemOpen
  \bibfield  {author} {\bibinfo {author} {\bibfnamefont {J.}~\bibnamefont {van Elp}}, \bibinfo {author} {\bibfnamefont {H.}~\bibnamefont {Eskes}}, \bibinfo {author} {\bibfnamefont {P.}~\bibnamefont {Kuiper}}, \ and\ \bibinfo {author} {\bibfnamefont {G.~A.}\ \bibnamefont {Sawatzky}},\ }\href {\doibase 10.1103/PhysRevB.45.1612} {\bibfield  {journal} {\bibinfo  {journal} {Phys. Rev. B}\ }\textbf {\bibinfo {volume} {45}},\ \bibinfo {pages} {1612} (\bibinfo {year} {1992})}\BibitemShut {NoStop}%
\bibitem [{\citenamefont {Maiti}\ \emph {et~al.}(1999)\citenamefont {Maiti}, \citenamefont {Mahadevan},\ and\ \citenamefont {Sarma}}]{1999PRBSarmaNi}%
  \BibitemOpen
  \bibfield  {author} {\bibinfo {author} {\bibfnamefont {K.}~\bibnamefont {Maiti}}, \bibinfo {author} {\bibfnamefont {P.}~\bibnamefont {Mahadevan}}, \ and\ \bibinfo {author} {\bibfnamefont {D.~D.}\ \bibnamefont {Sarma}},\ }\href {\doibase 10.1103/PhysRevB.59.12457} {\bibfield  {journal} {\bibinfo  {journal} {Phys. Rev. B}\ }\textbf {\bibinfo {volume} {59}},\ \bibinfo {pages} {12457} (\bibinfo {year} {1999})}\BibitemShut {NoStop}%
\bibitem [{\citenamefont {Kuo}\ \emph {et~al.}(2017)\citenamefont {Kuo}, \citenamefont {Haupricht}, \citenamefont {Weinen}, \citenamefont {Wu}, \citenamefont {Tsuei}, \citenamefont {Haverkort}, \citenamefont {Tanaka},\ and\ \citenamefont {Tjeng}}]{2017EPJSTTjengNiOReview}%
  \BibitemOpen
  \bibfield  {author} {\bibinfo {author} {\bibfnamefont {C.~Y.}\ \bibnamefont {Kuo}}, \bibinfo {author} {\bibfnamefont {T.}~\bibnamefont {Haupricht}}, \bibinfo {author} {\bibfnamefont {J.}~\bibnamefont {Weinen}}, \bibinfo {author} {\bibfnamefont {H.}~\bibnamefont {Wu}}, \bibinfo {author} {\bibfnamefont {K.~D.}\ \bibnamefont {Tsuei}}, \bibinfo {author} {\bibfnamefont {M.}~\bibnamefont {Haverkort}}, \bibinfo {author} {\bibfnamefont {A.}~\bibnamefont {Tanaka}}, \ and\ \bibinfo {author} {\bibfnamefont {L.}~\bibnamefont {Tjeng}},\ }\href@noop {} {\bibfield  {journal} {\bibinfo  {journal} {The European Physical Journal Special Topics}\ }\textbf {\bibinfo {volume} {226}},\ \bibinfo {pages} {2445} (\bibinfo {year} {2017})}\BibitemShut {NoStop}%
\bibitem [{\citenamefont {van~der Laan}\ \emph {et~al.}(1986)\citenamefont {van~der Laan}, \citenamefont {Zaanen}, \citenamefont {Sawatzky}, \citenamefont {Karnatak},\ and\ \citenamefont {Esteva}}]{1986PRBvanderlaanNi}%
  \BibitemOpen
  \bibfield  {author} {\bibinfo {author} {\bibfnamefont {G.}~\bibnamefont {van~der Laan}}, \bibinfo {author} {\bibfnamefont {J.}~\bibnamefont {Zaanen}}, \bibinfo {author} {\bibfnamefont {G.~A.}\ \bibnamefont {Sawatzky}}, \bibinfo {author} {\bibfnamefont {R.}~\bibnamefont {Karnatak}}, \ and\ \bibinfo {author} {\bibfnamefont {J.-M.}\ \bibnamefont {Esteva}},\ }\href {\doibase 10.1103/PhysRevB.33.4253} {\bibfield  {journal} {\bibinfo  {journal} {Phys. Rev. B}\ }\textbf {\bibinfo {volume} {33}},\ \bibinfo {pages} {4253} (\bibinfo {year} {1986})}\BibitemShut {NoStop}%
\bibitem [{\citenamefont {Matsubara}\ \emph {et~al.}(2005)\citenamefont {Matsubara}, \citenamefont {Uozumi}, \citenamefont {Kotani},\ and\ \citenamefont {Claude~Parlebas}}]{2005JPSJKotaniNiO}%
  \BibitemOpen
  \bibfield  {author} {\bibinfo {author} {\bibfnamefont {M.}~\bibnamefont {Matsubara}}, \bibinfo {author} {\bibfnamefont {T.}~\bibnamefont {Uozumi}}, \bibinfo {author} {\bibfnamefont {A.}~\bibnamefont {Kotani}}, \ and\ \bibinfo {author} {\bibfnamefont {J.}~\bibnamefont {Claude~Parlebas}},\ }\href {\doibase 10.1143/JPSJ.74.2052} {\bibfield  {journal} {\bibinfo  {journal} {Journal of the Physical Society of Japan}\ }\textbf {\bibinfo {volume} {74}},\ \bibinfo {pages} {2052} (\bibinfo {year} {2005})}\BibitemShut {NoStop}%
\bibitem [{\citenamefont {Altieri}\ \emph {et~al.}(2000)\citenamefont {Altieri}, \citenamefont {Tjeng}, \citenamefont {Tanaka},\ and\ \citenamefont {Sawatzky}}]{2000PRBTjengMgONi}%
  \BibitemOpen
  \bibfield  {author} {\bibinfo {author} {\bibfnamefont {S.}~\bibnamefont {Altieri}}, \bibinfo {author} {\bibfnamefont {L.~H.}\ \bibnamefont {Tjeng}}, \bibinfo {author} {\bibfnamefont {A.}~\bibnamefont {Tanaka}}, \ and\ \bibinfo {author} {\bibfnamefont {G.~A.}\ \bibnamefont {Sawatzky}},\ }\href {\doibase 10.1103/PhysRevB.61.13403} {\bibfield  {journal} {\bibinfo  {journal} {Phys. Rev. B}\ }\textbf {\bibinfo {volume} {61}},\ \bibinfo {pages} {13403} (\bibinfo {year} {2000})}\BibitemShut {NoStop}%
\bibitem [{\citenamefont {Mizokawa}\ \emph {et~al.}(1998)\citenamefont {Mizokawa}, \citenamefont {Fujimori}, \citenamefont {Namatame}, \citenamefont {Takeda},\ and\ \citenamefont {Takano}}]{1998PRBFujimoriLaCuO3}%
  \BibitemOpen
  \bibfield  {author} {\bibinfo {author} {\bibfnamefont {T.}~\bibnamefont {Mizokawa}}, \bibinfo {author} {\bibfnamefont {A.}~\bibnamefont {Fujimori}}, \bibinfo {author} {\bibfnamefont {H.}~\bibnamefont {Namatame}}, \bibinfo {author} {\bibfnamefont {Y.}~\bibnamefont {Takeda}}, \ and\ \bibinfo {author} {\bibfnamefont {M.}~\bibnamefont {Takano}},\ }\href {\doibase 10.1103/PhysRevB.57.9550} {\bibfield  {journal} {\bibinfo  {journal} {Phys. Rev. B}\ }\textbf {\bibinfo {volume} {57}},\ \bibinfo {pages} {9550} (\bibinfo {year} {1998})}\BibitemShut {NoStop}%
\bibitem [{\citenamefont {Okada}\ and\ \citenamefont {Kotani}(1999)}]{1999JPSJKotaniLaCuO3}%
  \BibitemOpen
  \bibfield  {author} {\bibinfo {author} {\bibfnamefont {K.}~\bibnamefont {Okada}}\ and\ \bibinfo {author} {\bibfnamefont {A.}~\bibnamefont {Kotani}},\ }\href {\doibase 10.1143/JPSJ.68.666} {\bibfield  {journal} {\bibinfo  {journal} {Journal of the Physical Society of Japan}\ }\textbf {\bibinfo {volume} {68}},\ \bibinfo {pages} {666} (\bibinfo {year} {1999})}\BibitemShut {NoStop}%
\bibitem [{\citenamefont {Choudhury}\ \emph {et~al.}(2015)\citenamefont {Choudhury}, \citenamefont {Rivero}, \citenamefont {Meyers}, \citenamefont {Liu}, \citenamefont {Cao}, \citenamefont {Middey}, \citenamefont {Whitaker}, \citenamefont {Barraza-Lopez}, \citenamefont {Freeland}, \citenamefont {Greenblatt},\ and\ \citenamefont {Chakhalian}}]{2015PRBChakCu3}%
  \BibitemOpen
  \bibfield  {author} {\bibinfo {author} {\bibfnamefont {D.}~\bibnamefont {Choudhury}}, \bibinfo {author} {\bibfnamefont {P.}~\bibnamefont {Rivero}}, \bibinfo {author} {\bibfnamefont {D.}~\bibnamefont {Meyers}}, \bibinfo {author} {\bibfnamefont {X.}~\bibnamefont {Liu}}, \bibinfo {author} {\bibfnamefont {Y.}~\bibnamefont {Cao}}, \bibinfo {author} {\bibfnamefont {S.}~\bibnamefont {Middey}}, \bibinfo {author} {\bibfnamefont {M.~J.}\ \bibnamefont {Whitaker}}, \bibinfo {author} {\bibfnamefont {S.}~\bibnamefont {Barraza-Lopez}}, \bibinfo {author} {\bibfnamefont {J.~W.}\ \bibnamefont {Freeland}}, \bibinfo {author} {\bibfnamefont {M.}~\bibnamefont {Greenblatt}}, \ and\ \bibinfo {author} {\bibfnamefont {J.}~\bibnamefont {Chakhalian}},\ }\href {\doibase 10.1103/PhysRevB.92.201108} {\bibfield  {journal} {\bibinfo  {journal} {Phys. Rev. B}\ }\textbf {\bibinfo {volume} {92}},\ \bibinfo {pages} {201108} (\bibinfo {year} {2015})}\BibitemShut {NoStop}%
\bibitem [{\citenamefont {Ghijsen}\ \emph {et~al.}(1988)\citenamefont {Ghijsen}, \citenamefont {Tjeng}, \citenamefont {van Elp}, \citenamefont {Eskes}, \citenamefont {Westerink}, \citenamefont {Sawatzky},\ and\ \citenamefont {Czyzyk}}]{1988PRBTjengCuOxides}%
  \BibitemOpen
  \bibfield  {author} {\bibinfo {author} {\bibfnamefont {J.}~\bibnamefont {Ghijsen}}, \bibinfo {author} {\bibfnamefont {L.~H.}\ \bibnamefont {Tjeng}}, \bibinfo {author} {\bibfnamefont {J.}~\bibnamefont {van Elp}}, \bibinfo {author} {\bibfnamefont {H.}~\bibnamefont {Eskes}}, \bibinfo {author} {\bibfnamefont {J.}~\bibnamefont {Westerink}}, \bibinfo {author} {\bibfnamefont {G.~A.}\ \bibnamefont {Sawatzky}}, \ and\ \bibinfo {author} {\bibfnamefont {M.~T.}\ \bibnamefont {Czyzyk}},\ }\href {\doibase 10.1103/PhysRevB.38.11322} {\bibfield  {journal} {\bibinfo  {journal} {Phys. Rev. B}\ }\textbf {\bibinfo {volume} {38}},\ \bibinfo {pages} {11322} (\bibinfo {year} {1988})}\BibitemShut {NoStop}%
\bibitem [{\citenamefont {Tanaka}\ \emph {et~al.}(1991)\citenamefont {Tanaka}, \citenamefont {Okada},\ and\ \citenamefont {Kotani}}]{1991JPSJKotaniCuOLCO}%
  \BibitemOpen
  \bibfield  {author} {\bibinfo {author} {\bibfnamefont {S.}~\bibnamefont {Tanaka}}, \bibinfo {author} {\bibfnamefont {K.}~\bibnamefont {Okada}}, \ and\ \bibinfo {author} {\bibfnamefont {A.}~\bibnamefont {Kotani}},\ }\href {\doibase 10.1143/JPSJ.60.3893} {\bibfield  {journal} {\bibinfo  {journal} {Journal of the Physical Society of Japan}\ }\textbf {\bibinfo {volume} {60}},\ \bibinfo {pages} {3893} (\bibinfo {year} {1991})}\BibitemShut {NoStop}%
\bibitem [{\citenamefont {Maiti}\ \emph {et~al.}(1997)\citenamefont {Maiti}, \citenamefont {Sarma}, \citenamefont {Mizokawa},\ and\ \citenamefont {Fujimori}}]{1997EPLFujimoriCu}%
  \BibitemOpen
  \bibfield  {author} {\bibinfo {author} {\bibfnamefont {K.}~\bibnamefont {Maiti}}, \bibinfo {author} {\bibfnamefont {D.~D.}\ \bibnamefont {Sarma}}, \bibinfo {author} {\bibfnamefont {T.}~\bibnamefont {Mizokawa}}, \ and\ \bibinfo {author} {\bibfnamefont {A.}~\bibnamefont {Fujimori}},\ }\href {\doibase 10.1209/epl/i1997-00157-x} {\bibfield  {journal} {\bibinfo  {journal} {Europhysics Letters}\ }\textbf {\bibinfo {volume} {37}},\ \bibinfo {pages} {359} (\bibinfo {year} {1997})}\BibitemShut {NoStop}%
\bibitem [{\citenamefont {Hollmann}\ \emph {et~al.}(2013)\citenamefont {Hollmann}, \citenamefont {Hu}, \citenamefont {Maignan}, \citenamefont {G\"unther}, \citenamefont {Jang}, \citenamefont {Tanaka}, \citenamefont {Lin}, \citenamefont {Chen}, \citenamefont {Thalmeier},\ and\ \citenamefont {Tjeng}}]{2013PRBTjengCCRO}%
  \BibitemOpen
  \bibfield  {author} {\bibinfo {author} {\bibfnamefont {N.}~\bibnamefont {Hollmann}}, \bibinfo {author} {\bibfnamefont {Z.}~\bibnamefont {Hu}}, \bibinfo {author} {\bibfnamefont {A.}~\bibnamefont {Maignan}}, \bibinfo {author} {\bibfnamefont {A.}~\bibnamefont {G\"unther}}, \bibinfo {author} {\bibfnamefont {L.-Y.}\ \bibnamefont {Jang}}, \bibinfo {author} {\bibfnamefont {A.}~\bibnamefont {Tanaka}}, \bibinfo {author} {\bibfnamefont {H.-J.}\ \bibnamefont {Lin}}, \bibinfo {author} {\bibfnamefont {C.~T.}\ \bibnamefont {Chen}}, \bibinfo {author} {\bibfnamefont {P.}~\bibnamefont {Thalmeier}}, \ and\ \bibinfo {author} {\bibfnamefont {L.~H.}\ \bibnamefont {Tjeng}},\ }\href {\doibase 10.1103/PhysRevB.87.155122} {\bibfield  {journal} {\bibinfo  {journal} {Phys. Rev. B}\ }\textbf {\bibinfo {volume} {87}},\ \bibinfo {pages} {155122} (\bibinfo {year} {2013})}\BibitemShut {NoStop}%
\bibitem [{\citenamefont {H\"am\"al\"ainen}\ \emph {et~al.}(2000)\citenamefont {H\"am\"al\"ainen}, \citenamefont {Hill}, \citenamefont {Huotari}, \citenamefont {Kao}, \citenamefont {Berman}, \citenamefont {Kotani}, \citenamefont {Id\'e}, \citenamefont {Peng},\ and\ \citenamefont {Greene}}]{2000PRBKotaniNCO}%
  \BibitemOpen
  \bibfield  {author} {\bibinfo {author} {\bibfnamefont {K.}~\bibnamefont {H\"am\"al\"ainen}}, \bibinfo {author} {\bibfnamefont {J.~P.}\ \bibnamefont {Hill}}, \bibinfo {author} {\bibfnamefont {S.}~\bibnamefont {Huotari}}, \bibinfo {author} {\bibfnamefont {C.-C.}\ \bibnamefont {Kao}}, \bibinfo {author} {\bibfnamefont {L.~E.}\ \bibnamefont {Berman}}, \bibinfo {author} {\bibfnamefont {A.}~\bibnamefont {Kotani}}, \bibinfo {author} {\bibfnamefont {T.}~\bibnamefont {Id\'e}}, \bibinfo {author} {\bibfnamefont {J.~L.}\ \bibnamefont {Peng}}, \ and\ \bibinfo {author} {\bibfnamefont {R.~L.}\ \bibnamefont {Greene}},\ }\href {\doibase 10.1103/PhysRevB.61.1836} {\bibfield  {journal} {\bibinfo  {journal} {Phys. Rev. B}\ }\textbf {\bibinfo {volume} {61}},\ \bibinfo {pages} {1836} (\bibinfo {year} {2000})}\BibitemShut {NoStop}%
\bibitem [{\citenamefont {Id\'{e}}\ and\ \citenamefont {Kotani}(2000)}]{2000JPSJKotaniNCO}%
  \BibitemOpen
  \bibfield  {author} {\bibinfo {author} {\bibfnamefont {T.}~\bibnamefont {Id\'{e}}}\ and\ \bibinfo {author} {\bibfnamefont {A.}~\bibnamefont {Kotani}},\ }\href {\doibase 10.1143/JPSJ.69.3107} {\bibfield  {journal} {\bibinfo  {journal} {Journal of the Physical Society of Japan}\ }\textbf {\bibinfo {volume} {69}},\ \bibinfo {pages} {3107} (\bibinfo {year} {2000})}\BibitemShut {NoStop}%
\bibitem [{\citenamefont {Okada}\ \emph {et~al.}(1996)\citenamefont {Okada}, \citenamefont {Kotani}, \citenamefont {Maiti},\ and\ \citenamefont {Sarma}}]{1996JPSJKotaniSCO}%
  \BibitemOpen
  \bibfield  {author} {\bibinfo {author} {\bibfnamefont {K.}~\bibnamefont {Okada}}, \bibinfo {author} {\bibfnamefont {A.}~\bibnamefont {Kotani}}, \bibinfo {author} {\bibfnamefont {K.}~\bibnamefont {Maiti}}, \ and\ \bibinfo {author} {\bibfnamefont {D.}~\bibnamefont {Sarma}},\ }\href {\doibase 10.1143/JPSJ.65.1844} {\bibfield  {journal} {\bibinfo  {journal} {Journal of the Physical Society of Japan}\ }\textbf {\bibinfo {volume} {65}},\ \bibinfo {pages} {1844} (\bibinfo {year} {1996})}\BibitemShut {NoStop}%
\end{thebibliography}%

\end{document}